\documentclass[rmp,twocolumn]{revtex4}

\usepackage{epsfig}
\usepackage{epsf}
\usepackage{amssymb}
\usepackage[dvips]{color}
\usepackage{lscape}
\usepackage{longtable}
\usepackage{rotating}

\def\gtwid{\mathrel{\raise.3ex\hbox{$>$\kern-.75em\lower1ex\hbox{$\sim$}}}}
\def\ltwid{\mathrel{\raise.3ex\hbox{$<$\kern-.75em\lower1ex\hbox{$\sim$}}}}

\def\agt{\mathrel{\raise.3ex\hbox{$>$\kern-.75em\lower1ex\hbox{$\sim$}}}}
\def\alt{\mathrel{\raise.3ex\hbox{$<$\kern-.75em\lower1ex\hbox{$\sim$}}}}

\bibpunct{(}{)}{,}{a}{}{,}

\def\be{\begin{equation}}
\newcommand{\ee}{\end{equation}}
\newcommand{\ba}{\begin{eqnarray}}
\newcommand{\ea}{\end{eqnarray}}

\def\x{{\mbox{\boldmath$x$}}}
\def\u{{\mbox{\boldmath$u$}}}

\def\eps{{\epsilon}}

\renewcommand{\cite}{\citet}

\newcommand{\aleq}{\mbox{\ 
\raisebox{-.9ex}{$\stackrel{\textstyle<}{\sim}$}\ }}

\newcommand{\bea}{\begin{eqnarray}}
\newcommand{\eea}{\end{eqnarray}}
\newcommand{\bean}{\begin{eqnarray}}
\newcommand{\eean}{\end{eqnarray}}

\bibpunct{(}{)}{,}{a}{}{,}

\def\be{\begin{equation}}
\def\ee{\end{equation}}
\def\ba{\begin{eqnarray}}
\def\ea{\end{eqnarray}}

\def\Re{\mbox{Re}}
\def\Ra{\mbox{Ra}}
\def\Ro{\mbox{Ro}}

\def\Pr{\mbox{Pr}}
\def\Nu{\mbox{Nu}}

\def\x{{\mbox{\boldmath$x$}}}
\def\u{{\mbox{\boldmath$u$}}}

\def\bea{\begin{eqnarray}}
\def\eea{\end{eqnarray}}
\def\bean{\begin{eqnarray}}
\def\eean{\end{eqnarray}}

\begin{document}

\title{Heat transfer \& large-scale dynamics in turbulent Rayleigh-B\'enard
 convection}

\markboth{Rayleigh-B\'enard convection}{Ahlers, Grossmann, Lohse}

\author{
Guenter Ahlers$^1$, Siegfried Grossmann$^2$, and Detlef Lohse$^3$}
\affiliation{
$^1$Department of Physics and iQCD, University of California, 
Santa Barbara, CA 93106, USA\\
$^2$Fachbereich Physik, 
Philipps-Universit\"at Marburg, D-35032 Marburg, Germany\\
$^3$Physics of Fluids group,
Department of Science and Technology, 
J.\  M.\ Burgers Centre for
Fluid Dynamics, and Impact-Institute, 
University of Twente, 7500 AE Enschede, The Netherlands}


\begin{abstract}
The progress in our understanding of several aspects of turbulent Rayleigh-B\'enard convection 
is reviewed. The focus is on the question of
how the Nusselt number and the Reynolds 
number depend on the Rayleigh number Ra and the Prandtl number Pr, and on 
how the thicknesses of the
thermal and the kinetic boundary layers scale with Ra and Pr. 
Non-Oberbeck-Boussinesq effects and the 
dynamics of the large-scale convection-roll are addressed as well. The review ends with a list of challenges for future research 
on the turbulent Rayleigh-B\'enard system. 
\end{abstract}
\date{\today}

\maketitle

\section{Introduction}

Rayleigh-B\'enard (RB) convection 
  -- the buoyancy driven
flow of a fluid heated from below 
and cooled from above --  
is a
classical problem in fluid
dynamics. It played a crucial role in the development of 
stability theory in hydrodynamics (\cite{cha81,dra81}) and had
been paradigmatic in pattern formation and in the study
of spatial-temporal chaos (\cite{bod00,get98}). 
From an applied viewpoint, thermally driven flows are of utmost
importance. Examples are thermal 
convection in the atmosphere (see e.g.\ \cite{har01}), 
in the oceans (see e.g.\ \cite{mar99}) 
(including thermohaline convection, see e.g.\ \cite{rah00}), 
in buildings (see e.g.\
\cite{hun99}), 
in process technology, or in 
metal-production processes (see e.g.\ \cite{bre88}). In the geophysical and astrophysical context,
we mention convection in Earth's mantle (see e.g.\ \cite{mck74}), 
in Earth's outer core (see e.g.\ \cite{car94}), and 
in stars including our sun (see e.g.\ \cite{cat03}). 
Convection has been associated with the generation and reversal
of Earth's magnetic field, see e.g.\ \cite{gla95}.

Even if one restricts oneself to thermally driven flows in a closed
box, there are so many aspects that not all of them can be addressed
in this single review. We shall focus on  developed turbulence
when spatial coherence throughout the cell is lost 
and only on the
{\it large scale dynamics} of the flow and aspects intimately  
connected with it such as the boundary layer structures. 
The scaling of the spectra of velocity 
and temperature fluctuations, or of the corresponding
structure functions, will not be addressed. 
These issues had been discussed in the review  by \cite{sig94}, but
meanwhile considerable progress has been achieved, in particular
on the question of whether and where in the flow to expect
Bolgiano-Obukhov scaling (\cite{bol59,obu59,my75}) 
of the structure functions, see e.g.\ \cite{cal02,sun06,kun08}.

The  question to be asked about the Rayleigh-B\'enard problem is as follows:
For a given fluid in a closed container of height $L$ 
heated from below
and cooled from
above, what are the flow properties inside the container and in addition,
what is the heat transfer from bottom to top? Here spatially and 
temporally constant temperatures are assumed at the bottom and top. In section \ref{nurenu} we will discuss to what degree this
assumption can be justified in reality (\cite{cha02,ver04,bro05}). 

The problem is further simplified by the so-called Oberbeck-Boussinesq
(OB) approximation (\cite{obe79,bou03,ll87}) in 
which the fluid density $\rho$ is assumed to depend linearly 
on the
temperature, 
\be
\rho(T) = \rho (T_0) (1 - \beta (T-T_0)),
\label{beta-def}
\ee
with $\beta$ being the thermal expansion coefficient. In addition, it is
assumed that 
the material properties
of the fluid such as $\beta$, 
the viscosity $\nu$, and the thermal diffusivity $\kappa$ do not depend
on temperature. The  governing equations of the RB problem 
are then the Oberbeck-Boussinesq equations (\cite{ll87}) 
\begin{eqnarray}
\partial_t u_i  + u_j \partial_j u_i &=& -\partial_i p
+\nu \partial_j^2 u_i +\beta g
\delta_{i3} \theta,
\label{ob-u}
\\
\partial_t \theta  + u_j \partial_j
  \theta &=&  \kappa \partial_j^2 \theta  
\label{ob-theta}
\end{eqnarray}
for the velocity field
$\u (\x , t ) $,
the kinematic pressure
field $p(\x , t)$, 
 and the temperature field $\theta (\x , t)$ relative to some reference
temperature. 
Here and in the following we assume summation over double indices;
$\delta_{ij}$ is the Kronecker symbol. 
The Oberbeck-Boussinesq
equations are assisted by 
continuity $\partial_i u_i = 0$ and 
the
boundary conditions $\u = 0$ for the velocities 
at all walls, 
$\theta (z=-L/2) = \Delta/2$
for the temperature at the bottom plate, and $\theta (z=L/2) = -\Delta/2$
for the temperature at the top plate.  At the side-walls the 
condition of no lateral heat flow is imposed. The limitations of the Oberbeck-Boussinesq 
approximations 
will be discussed in section \ref{sec-nob}.

Within the OB approximation and for a given cell geometry, 
the system is determined by only two dimensionless control parameters,
namely the Rayleigh number and the Prandtl number
\be
\Ra = {\beta g L^3 \Delta \over \kappa \nu},\qquad
\Pr ={\nu \over \kappa}.
\label{ra-pr}
\ee
The cell geometry is described by its symmetry and one or more aspect ratios $\Gamma$. 
For a cylindrical cell $\Gamma \equiv d/L$, where $d$ is the cell diameter.

The key response of the system to the imposed $\Ra$  is the 
heat flux $H$ from bottom to top. 
The dimensionless heat flux $\Nu = H/ (\Lambda \Delta L^{-1}$ is the
Nusselt number. Here $\Lambda = c_p \rho \kappa$ is the thermal 
conductivity. Within the Oberbeck-Boussinesq approximation one obtains
for incompressible flow
\be 
\Nu ={\left< u_z \theta\right>_A 
-\kappa \partial_3 \left< \theta \right>_A
 \over \kappa
 \Delta  L^{-1}}. 
\label{nu}
 \ee
Here
 $\left<.\right>_A$ denotes the average over (any) horizontal plane and over time. 
Correspondingly,
 $\left<.\right>_V$ used below denotes the volume and time average.

 Another key system response is the extent of turbulence, best expressed
in terms of a characteristic velocity amplitude $U$, nondimensionalized by $\nu L^{-1}$ to define a Reynolds number
\be
\Re = {U \over \nu L^{-1}}.
 \label{re}
\ee
As we shall see in section \ref{nurere}, there are various reasonable possibilities to choose a velocity, e.g.\ the components or the magnitude of the velocity field at different positions, local or averaged amplitudes, turnover times or frequency peaks in the thermal spectrum, etc. In some parameter ranges these amplitudes differ not only in magnitude but even show different dependences on $\Ra$ and $\Pr$ (\cite{bro07c,sug08}). Mostly we shall restrict ourselves to that Reynolds number which is associated with the large-scale flow (LSC), also
called  
the ``wind of turbulence" $U$ (\cite{nie01,xia03,sun05}).
There is discussion in the literature whether the LSC evolves out
of the well-known cellular structures at small Ra or whether it does not.
On
the one hand \cite{kri81} performed experiments from which they
concluded that the LSC is not a simple reminder and continuation
of the roll structure observed just after the onset of convection. On the
other hand we are not aware that their observations have been
confirmed. Even explicit search for such a mode as \cite{kri81}
 report was not
successful, see the summarizing review by  \cite{bus03} 
and \cite{har05}. These authors conclude that the LSC at
large Ra indeed is a reminder of the low Ra structures.
%
%
The dynamics of the large scale wind, its azimuthal oscillation, diffusion, reorientation, cessation,
and possible breakdown at very large $\Ra$ 
will be discussed in detail in section \ref{sec:global}.

The key question to ask is: How do Nu and Re depend on Ra and Pr?
The experimental situation 
 will be the subject of sections \ref{nurenu} for
Nu(Ra,Pr) and \ref{nurere} for Re(Ra,Pr).
Results from numerical simulations
are reported in section \ref{nuredns}.
However,  first (section \ref{theories})  we will summarize 
older theories (\ref{nureold}) and then (subsection \ref{nuregl})
the Grossmann-Lohse (GL) theory (\cite{gro00,gro01,gro02,gro04}).
In subsection \ref{nureup} we will discuss theories about a possible 
asymptotic
regime at very large $\Ra$ and strict upper bounds for $\Nu$.

\begin{figure}
\begin{center}
\includegraphics*[height=5cm]{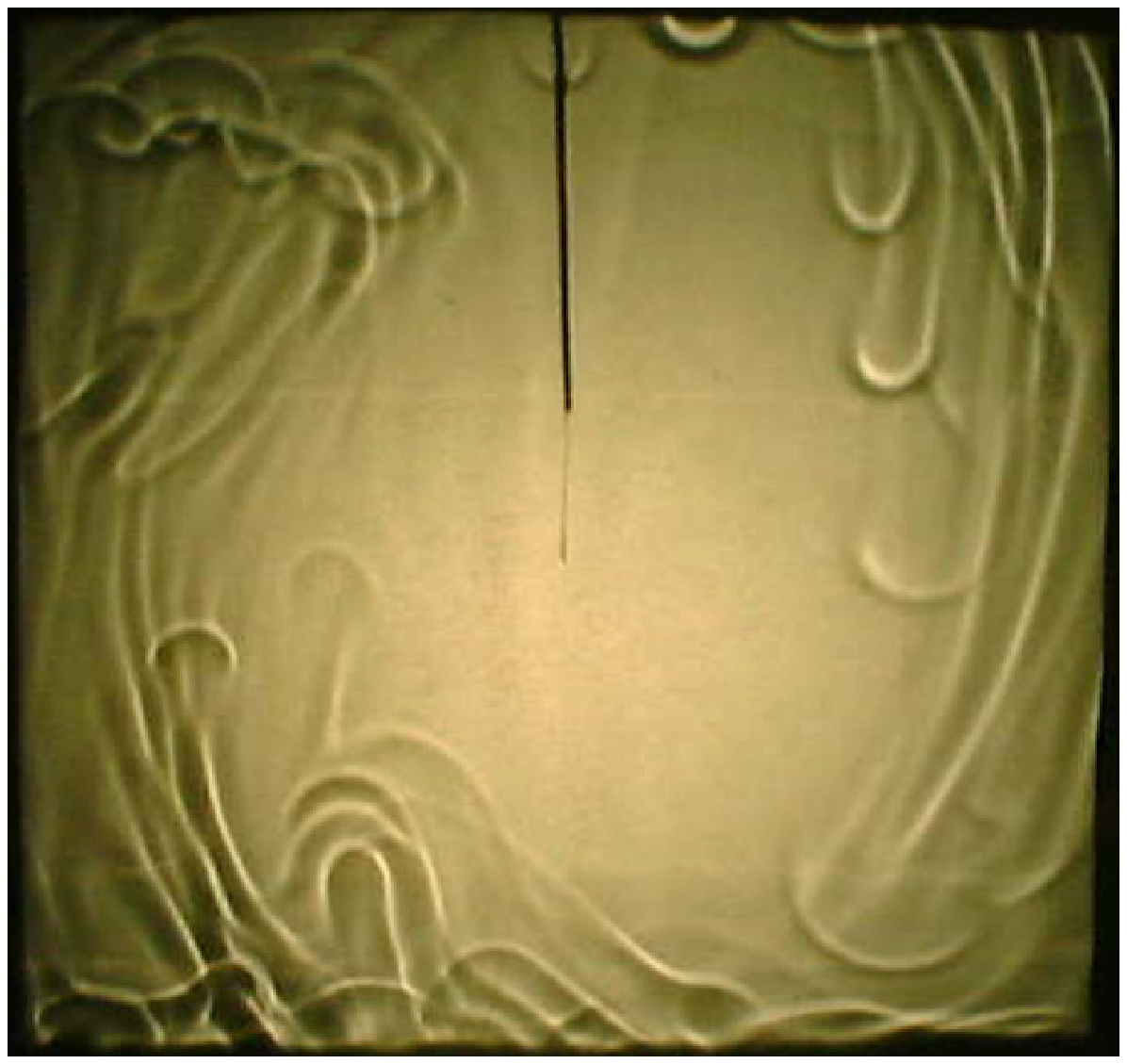}
\vspace*{0.5cm}
\includegraphics*[height=4cm]{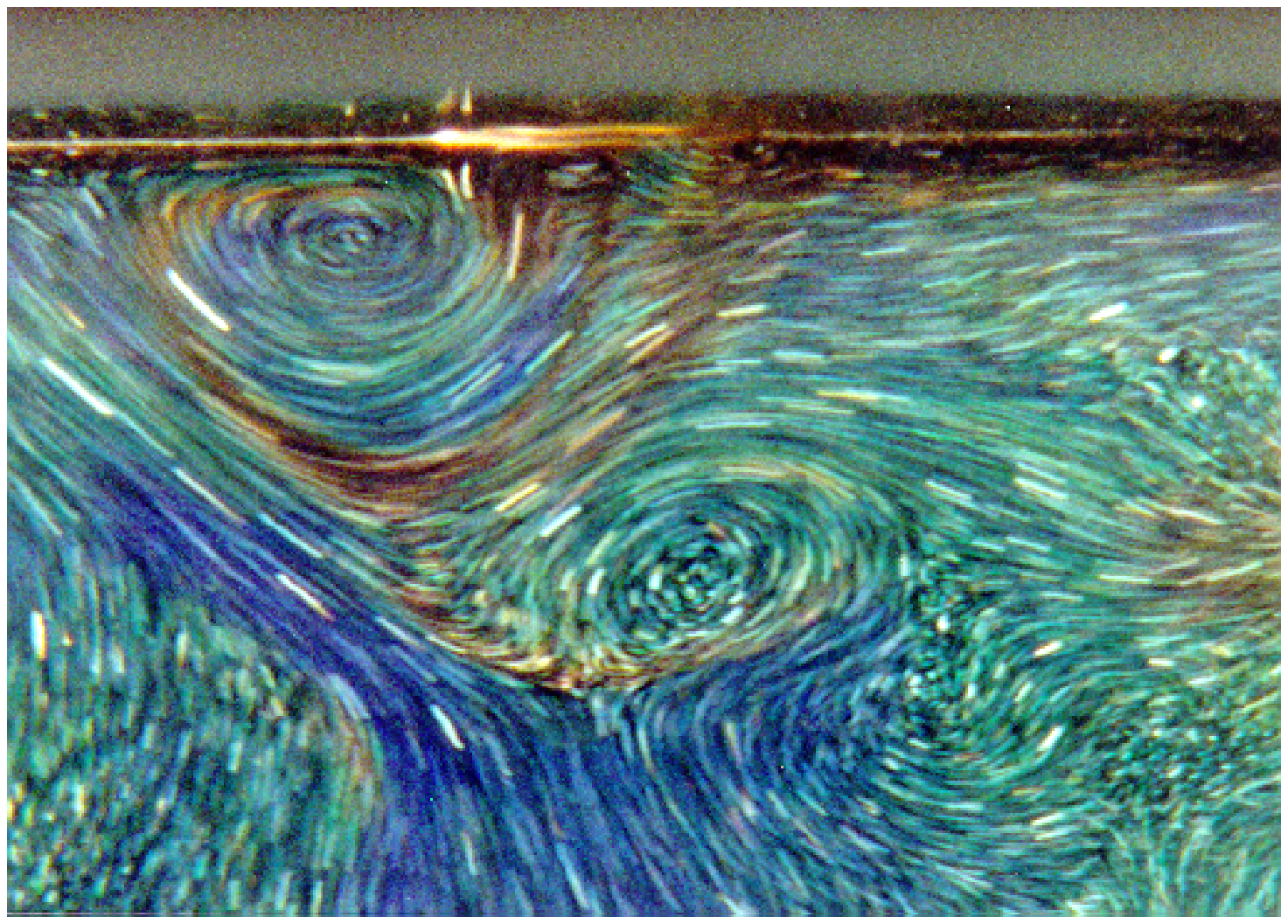}
\vspace*{0.5cm}
\includegraphics*[height=5cm]{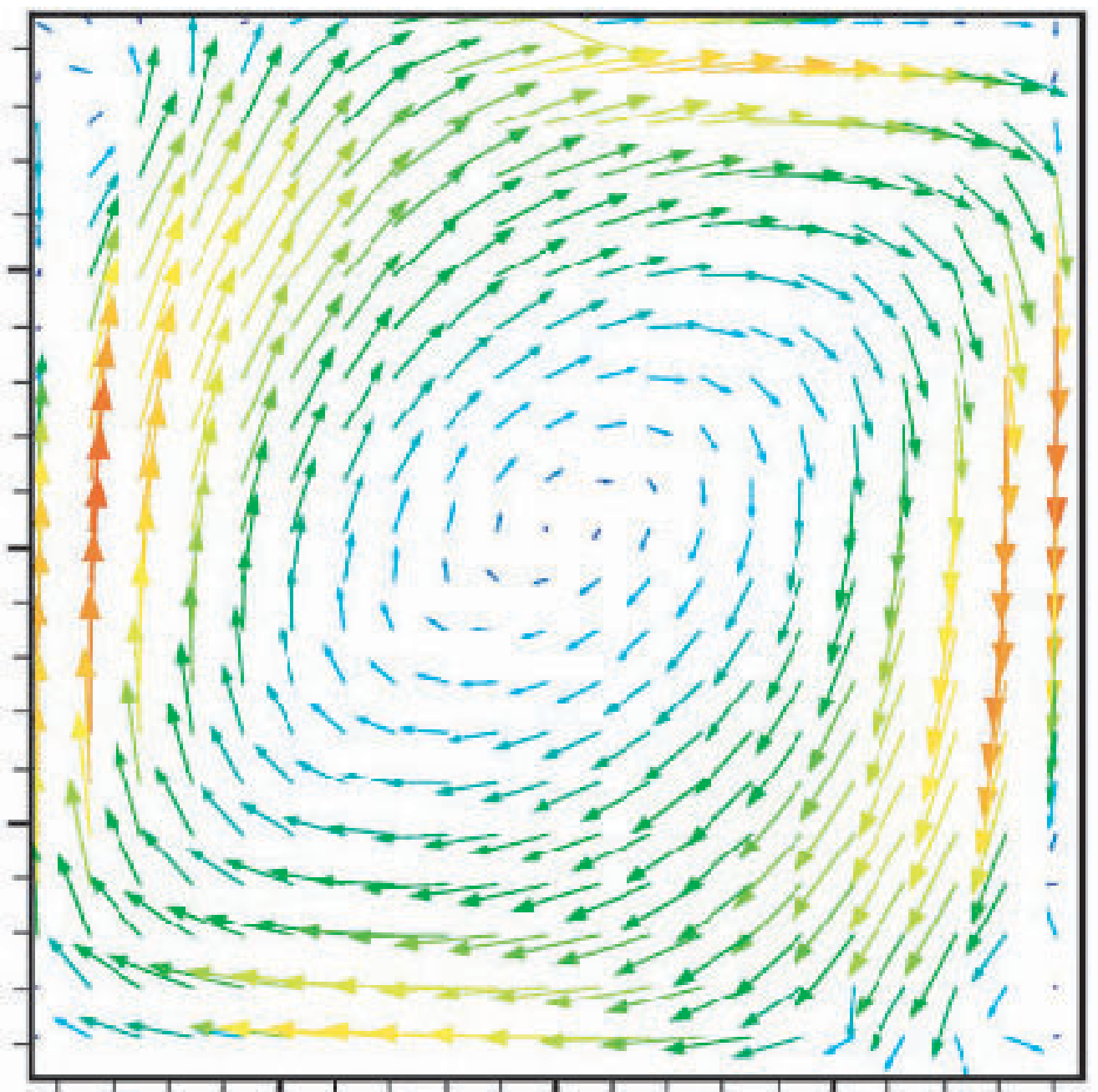}
\caption{Upper: Shadowgraph visualization of rising and falling
plumes at $\Ra = 6.8 \cdot 10^8$, 
$\Pr = 596$ (dipropylene glycol) in a $\Gamma = 1$ cell. 
From \cite{sha03}. 
Middle: Streak picture of temperature sensitive liquid crystal spheres
taken near the top cold surface in a $\Gamma = 1$ cell at
$\Ra = 2.6\cdot 10^9$ and $\Pr = 5.4$ (water), in order to visualize
plume detachment.
The view shows an area of 6.5cm by
4cm. From \cite{du00}.
Bottom: Time-averaged velocity-vector map in the plane of the LSC at $\Ra = 7.0\times 10^9$ (adapted from \cite{sun05}).
}
\label{plumes} 
\end{center}
\end{figure}

Section \ref{sec:bl} is devoted to the structure and width
of the thermal and kinetic boundary
layers (BLs). The thermal BLs  play a crucial role in determining the heat transfer, and the kinetic BLs provide viscous dissipation of the LSC. 
Another important feature is the thermal plumes 
(\cite{zoc90,kad01,xi04,fun04,zho07})
that detach from the thermal boundary layers; they contribute to the  
driving of the flow. In order to give an idea of the importance and organization
of these plumes and their shapes (for large Pr) 
we show their shadowgraph visualization in the top image of 
figure \ref{plumes}, taken from \cite{sha03}. 
The middle image of figure \ref{plumes} shows a streak picture of the temperature distribution
close to the upper plate, including a detaching plume for medium Pr. The lower
 image is a velocity-vector map of the LSC.

As already mentioned, section \ref{sec-nob}
 is devoted to non-Oberbeck-Boussinesq
effects and section \ref{sec:global}
 deals with the global wind dynamics. In section \ref{future} we outline 
some major issues in Rayleigh-B\'enard convection for future research.

\section{Theories of global properties: The Nusselt and Reynolds number}
\label{theories}

\subsection{Older theories for Nu(Ra,Pr) and Re(Ra,Pr)}\label{nureold}

For a detailed discussion of the theories developed prior
to the review by \cite{sig94}  we refer to that paper
and to \cite{cha81}.
These theories predicted power laws
\bea
Nu & \sim & Ra^{\gamma_{Nu}} Pr^{\alpha_{Nu}},\label{eq:Nupow} \\
Re & \sim & Ra^{\gamma_{Re}} Pr^{\alpha_{Re}} \label{eq:Repow}
\eea
for the dependences of Nu and Re on Ra and Pr.  A  summary of  predicted  exponents
is given in table \ref{tab-old-theories}. Early experiments were of limited precision, and were consistent with power-law dependences over their limited ranges of Ra and Pr. 

{
\tiny
{
\begin{table*}
 \begin{tabular}{|c|c|c|c|c|c|c|}
 \hline
         Reference
       & Pr \& Ra range
       & $\gamma_{Nu}$
       & $\alpha_{Nu}$
       & $\gamma_{Re}$
       & $\gamma_{Re_{fluct}}$
       & $\alpha_{Re}$
 \\
\hline
         \cite{dav22a,dav22b}
       & Ra small
       & $1/4$
       & $$
       & $$
       & $$
       & $$
\\
         \cite{mal54}
       & 
       & $1/3$
       & $$
       & $$
       & $$
       & $$
\\
         \cite{kra62}
       & Ra ultimate, 
       & $1/2$
       & $1/2$
       & $1/2$
       & $$
       & $-1/2$
\\
       & Pr$<0.15$
       & 
       & 
       & 
       & 
       & 
\\
        
       & Ra ultimate, 
       & $1/2$
       & $-1/4$
       & $1/2$
       & $$
       & $-3/4$
\\
       &  $0.15< \Pr \aleq 1$
       & 
       & 
       & 
       & 
       & 
\\
         \cite{spi71}
       & Ra ultimate
       & $1/2$
       & $1/2$
       & $1/2$
       & $$
       & $-1/2$
\\
         \cite{cas89}
       & 
       & $2/7$
       & $$
       & $1/2$
       & $3/7$
       & $$
\\
         \cite{shr90}
       & Pr$>1$
       & $2/7$
       & $-1/7$
       & $3/7$
       & $$
       & $-5/7$
\\
         \cite{yak92}
       & 
       & $5/19$
       & $$
       & $$
       & $8/19$
       & $$
\\
         \cite{zal98}
       & 
       & $2/7$
       & $$
       & $$
       & $$
       & $$
\\
         \cite{cio97}
       & Pr$<1$
       & $2/7$
       & $2/7$
       & $$
       & $3/7$
       & $-4/7$
\\
\hline
 \end{tabular}
\caption[]{
Power-law exponents for Nu and Re as functions
of Ra and Pr predicted by theories developed prior to the review by \cite{sig94}.
The exponents are defined by Eqs.~(\ref{eq:Nupow}) and (\ref{eq:Repow}).
Whereas Re is based on the
large-scale wind-velocity, $Re_{fluct}$ is based on the velocity fluctuations.
}
\label{tab-old-theories}
\end{table*}

}}

The conceptually easiest early theory is Malkus' marginal-stability theory of 1954. It assumed that the thermal BL thickness
adjusts itself so as to yield a  critical BL Rayleigh number.
This  immediately gave $\gamma_{Nu} = 1/3$. 
After the experiments by \cite{chu73,thr75} and the later ground-breaking
Chicago experiments in cryogenic helium
(\cite{hes87,san89,cas89,wu90,pro91}) 
had suggested a smaller power-law exponent, 
the Chicago group developed the mixing-zone
model (\cite{cas89}) which later was extended by \cite{cio97} to
include the Prandtl-number dependences. The central result 
was $\gamma_{Nu} = 2/7$. 
The same scaling exponent could
also be obtained from the BL theory of \cite{shr90}, assuming a
turbulent boundary layer. The assumptions of that theory
are however very different from those of the mixing-layer theory,
leading to very different power-law exponents for the dependences
on the Prandtl number, see table \ref{tab-old-theories}. 

As we will see later, the assumption of a fully developed turbulent
BL is far from being fulfilled in the parameter regime of the 
Chicago experiments. That can already be seen from an estimate
of the coherence length $\ell$ of the RB flow. 
Taking the data from \cite{pro91} for
the scaling of the velocity fluctuations and of the crossover frequency
to the viscous subrange, \cite{gnlo93a} obtained $\ell/L \approx 
50 \Ra^{-0.32}$. For the $\Gamma = 1/2$ cell of \cite{pro91}
this implies that only at 
$\Ra\approx 10^8$ the coherence length becomes
 about 1/3 of the lateral cell width and 1/6 of its height, a pre-requisite
for independent fluctuations to develop in the bulk. Estimates based
on $\ell \approx 10 \eta$, where $\eta$ is the (locally or globally defined)
Kolmogorov scale, give similar
results. 
The transition to turbulence in the BL is correspondingly 
expected only at much
large Ra, namely, at $\Ra \approx 10^{14}$ (at the edge of the achievable
regime in the Chicago experiments), 
as we will see in the next
subsection.

In any case, at ``large enough'' Rayleigh number a transition should
occur towards  
an ultimate Rayleigh-number regime. Such a regime was first suggested by
\cite{kra62}. 
\cite{spi71} hypothesized that in that regime the heat flux and
the turbulence intensity are independent of the 
kinematic viscosity and the thermal diffusivity, which leads to 
$\gamma_{Nu} = 1/2$ (for more details, see section \ref{nureup}). 
Though in those days (1971 and before) 
no measured power-law exponent was even close to that value, that paper
has been extremely influential, perhaps also because
 from a mathematical point of view
no lower strict upper bound than $\gamma_{Nu} = 1/2$
could be proven to exist for finite Pr (see \cite{doe96}).

As will be shown in sections \ref{nurenu} and \ref{nurere}, 
the experiments of the last decade
reveal the limitations of most of these older theories.

\subsection{Grossmann-Lohse theory for Nu(Ra,Pr) and Re(Ra,Pr)}\label{nuregl}
Given the increasing richness and precision 
of experimental and numerical 
data for Nu(Ra,Pr) (section \ref{nurenu})  and Re(Ra,Pr) (section
\ref{nurere}), 
it became clear near the end of the last decade  that none of the theories developed
up to then could offer a unifying view, accounting for all data. 
In particular, the predicted Prandtl-number dependences of Nu 
(\cite{shr90,cio97})
are in
disagreement with measured and calculated data. 
Therefore, in a series of four papers, \cite{gro00,gro01,gro02,gro04} tried
to develop a unifying theory to account for
Nu(Ra,Pr) and Re(Ra,Pr) over wide parameter ranges. 

\begin{figure}
\begin{center}
\includegraphics*[height=5cm]{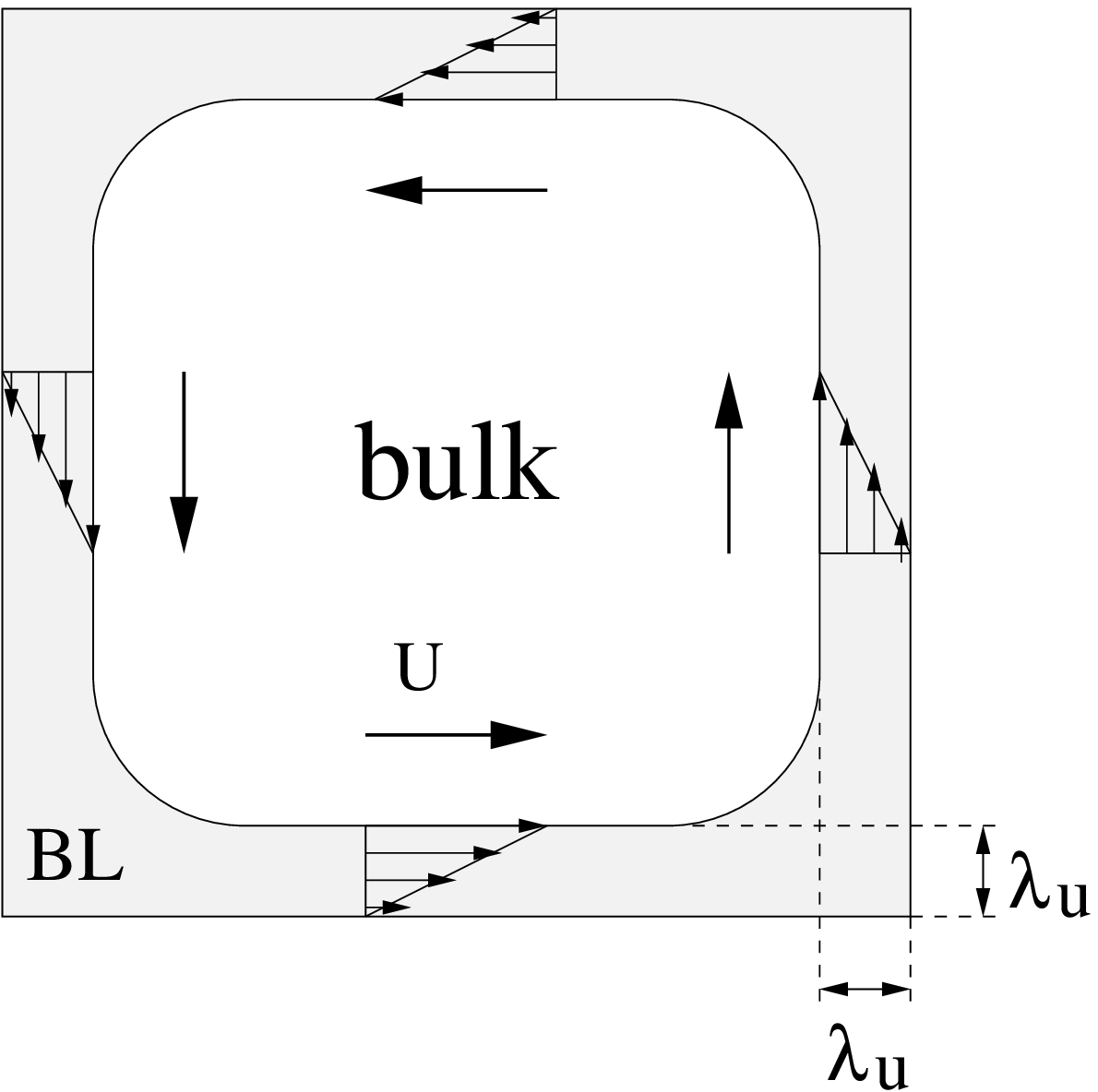}
\vspace{0.5cm}
\includegraphics*[height=5cm]{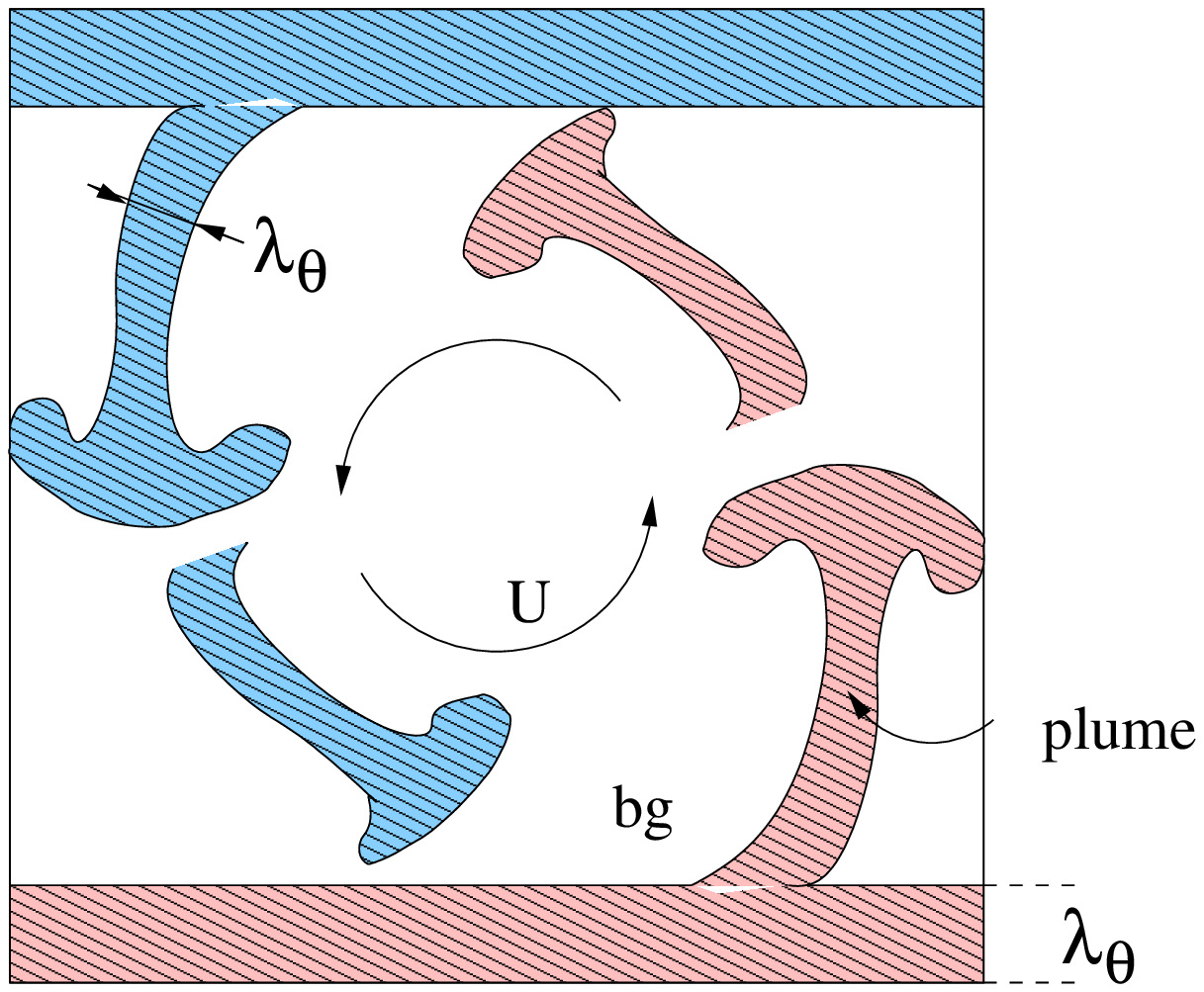}
\caption{
Sketch of the splitting of the kinetic (upper) and thermal (lower)
dissipation rates on which the GL theory is based.
In both figures the large-scale convection-roll with typical velocity amplitude 
U is sketched. The typical width of the kinetic BL is $\lambda_u$,
whereas the typical thermal BL thicknesses and the plume thicknesses
are $\lambda_\theta$. Outside the BL/plume region is the background flow "bg". 
}
\label{sketch-splitting} 
\end{center}
\end{figure}

The backbone of the theory is a set
of  two exact relations for the
kinetic and thermal energy-dissipation rates $\eps_u$ and $\eps_\theta$ respectively, namely
\begin{eqnarray}
\eps_u &\equiv& 
\left< \nu (\partial_i u_j (\x , t))^2\right>_V
=
{\nu^3\over L^4 } (\Nu-1) \Ra \Pr^{-2},\label{eps_u}\\
\eps_\theta &\equiv& 
\left< \kappa (\partial_i \theta (\x , t))^2\right>_V
=
\kappa {\Delta^2 \over L^2} \Nu.  \label{eps_theta} 
\end{eqnarray}
These relations can easily be derived from the Boussinesq equations
and the corresponding boundary conditions (see e.g.\ \cite{shr90}),
assuming only statistical stationarity. 
The central idea of the theory now is to split the 
volume averages of both the kinetic and the thermal dissipation rate 
into respective bulk and   boundary layer (or rather 
 boundary layer-like) contributions,
\begin{eqnarray}
\eps_u &=& \eps_{u,BL} + \eps_{u,bulk}, \label{split1}\\
\eps_\theta &=& \eps_{\theta ,BL} + \eps_{\theta , bulk}.\label{split2}
\end{eqnarray}
The motivation for this splitting is that the physics of the 
bulk and the BL (or BL-like) contributions to the dissipation rates is fundamentally different and thus the corresponding dissipation rate contributions must be modeled in different ways. The phrase ``BL-like'' indicates that 
from a scaling point of view
we consider the detaching thermal plumes 
as parts of the thermal BLs.
 Thus instead of BL and bulk we could also use the labels 
pl (plume) and bg (background) for the two parts of the dissipation rates.
A sketch of the splitting is shown in
figure \ref{sketch-splitting}.
Rather than eq.\ (\ref{split2}) one therefore could also write
\be 
\eps_\theta = \eps_{\theta ,pl} + \eps_{\theta , bg}, 
\label{split3}
\ee
signalling the contributions from the BL and the plumes (pl)
on the one
hand and from the background (bg) on the other. 

Two further assumptions of the GL theory are indicated as well in figure
\ref{sketch-splitting}, namely that there exists a large-scale
wind with only {\it one} typical velocity scale $U$ (defining a Reynolds
number $\Re=UL/\nu$), and that the kinetic BLs are 
(scalingwise) characterized by
a single effective thickness $\lambda_u$ regardless of the position along
the plates and walls in the flow. As we shall see in section \ref{nurere} 
for the velocity scales and in section \ref{sec:bl} for the BL thicknesses, 
both assumptions are  simplifications. In
particular,  even the scaling of the kinetic BL thickness with Ra may
 be different at the side-walls as compared to the top and bottom plates
(see \cite{xin96,xin97,qiu98,lui98}). Nevertheless, for the sake of simplicity
and in view of Occam's razor, 
-- and consistent with the most recent experimental results for the
BLs by \cite{sun08} --
these simplifications have been used.

Accepting the splitting 
(\ref{split1}) and 
(\ref{split2}) (or (\ref{split3})),
the immediate consequence is that there are four main regimes in parameter
space:
Regime I in which both $\eps_u$ and $\eps_\theta$ are dominated
by the BL/plume contribution,
regime II in which  $\eps_u$ is dominated by
$\eps_{u,bulk}$ and $\eps_\theta$ by 
$\eps_{\theta, BL}$, 
regime III in which  $\eps_u$ is dominated by
$\eps_{u,BL}$ and $\eps_\theta$ by 
$\eps_{\theta, bulk}$, 
and finally
regime IV in which both $\eps_u$ and $\eps_\theta$ are dominated
by their bulk contributions. It remains to be determined where in
Ra-Pr parameter space the crossovers between the different regimes are 
located.

\begin{figure}
\begin{center}
\includegraphics*[width=7cm]{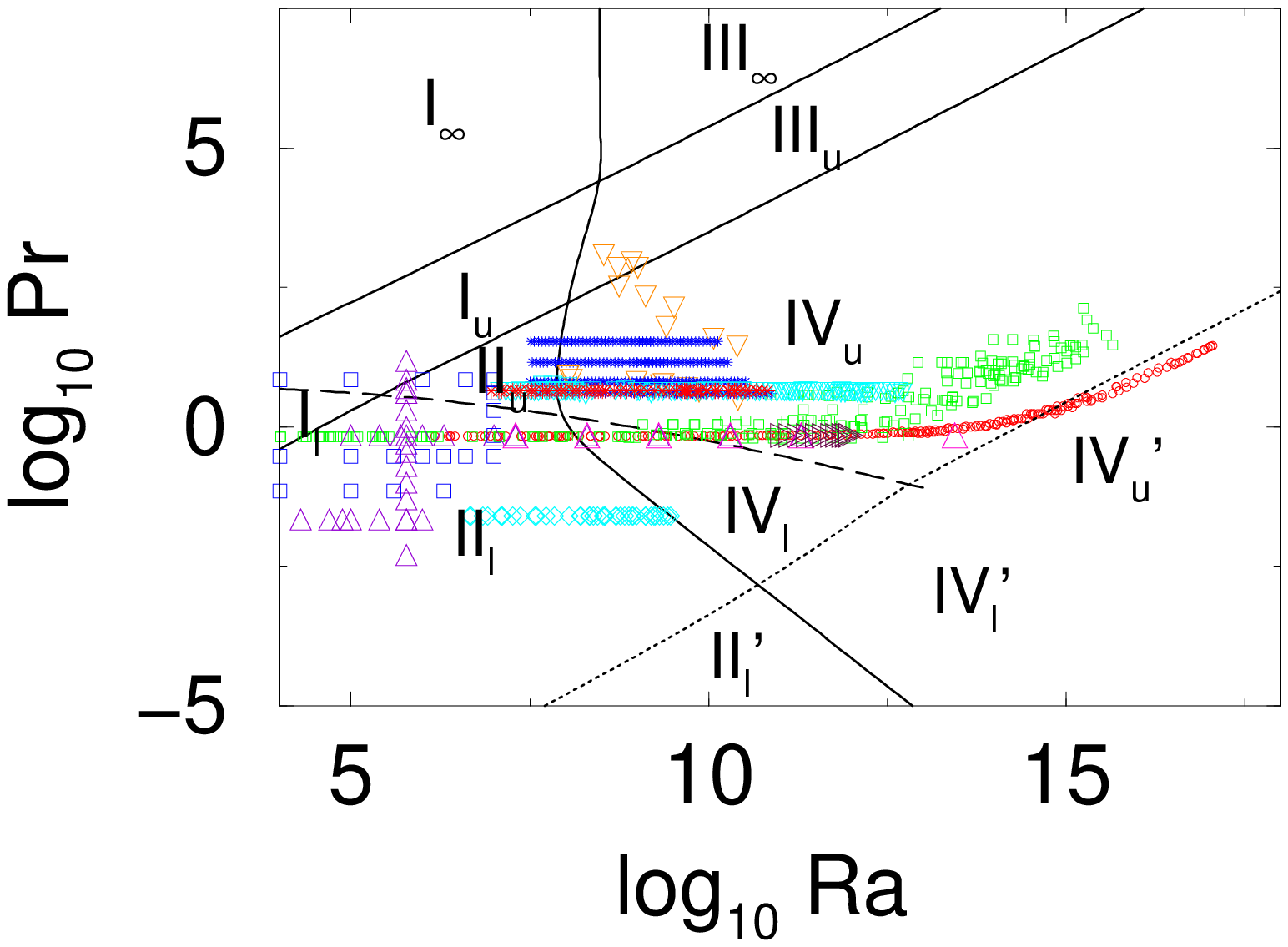}
\hspace*{0.6cm}
\includegraphics*[width=7cm]{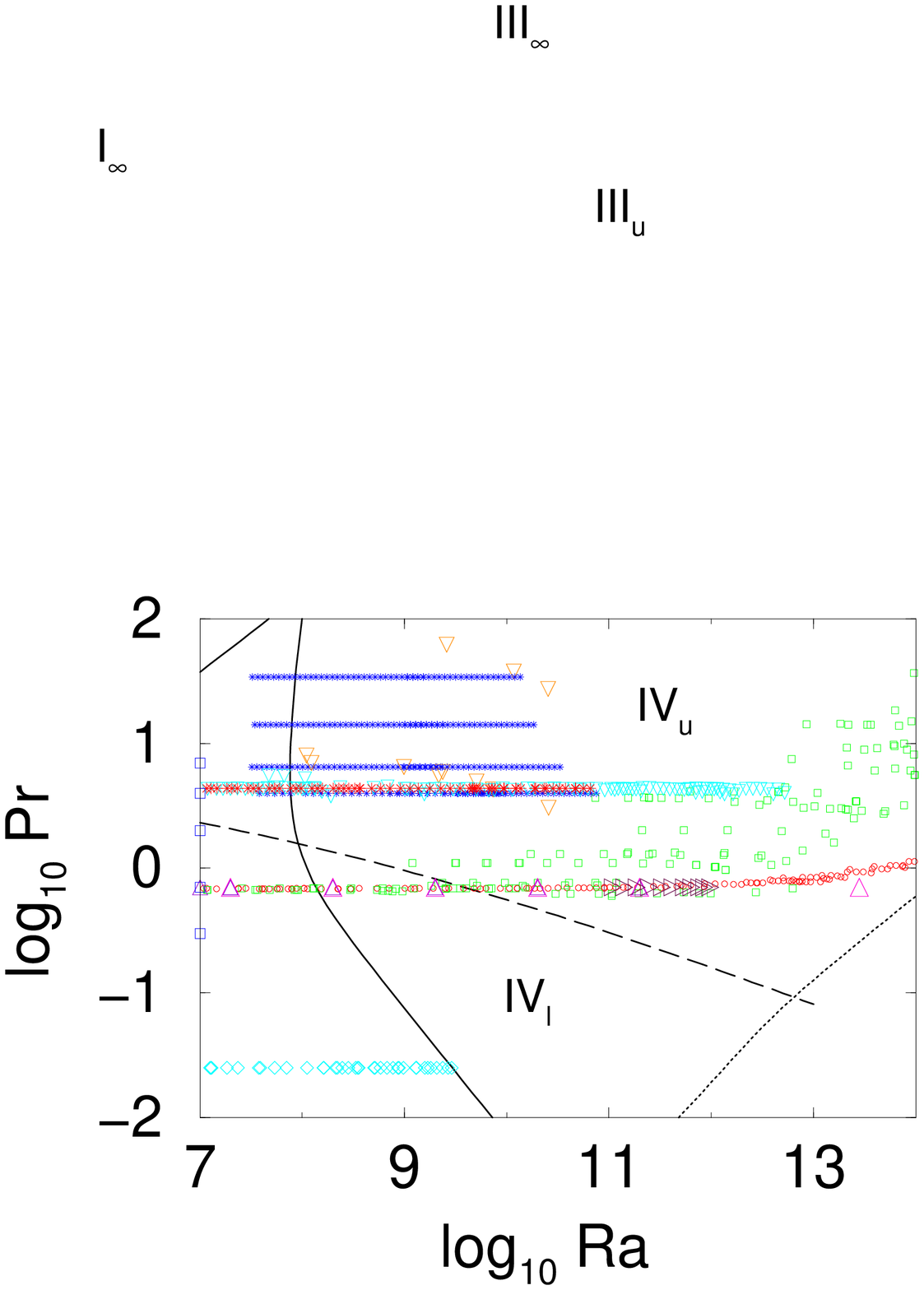}
\caption{Upper: 
Phase diagram in the $Ra-Pr$ plane according to
\cite{gro00,gro01,gro02,gro04}: The upper solid line means
$Re = Re_c$; the lower nearly parallel
solid line corresponds to $\eps_{u,BL} = \eps_{u,bulk}$;
the curved solid line is $\eps_{\theta,BL} = \eps_{\theta,bulk}$;
and along the long-dashed line $\lambda_u = \lambda_\theta$, i.e., $2aNu = \sqrt{Re}$. 
The dotted line indicates where the laminar kinetic BL is expected to become 
turbulent, based on a critical shear 
Reynolds number $Re_s^*\approx 420 $ of the kinetic BL, cf. \cite{ll87}.
Data points where $Nu$ has
been measured or numerically calculated have been included
(for aspect ratios $\Gamma \approx 1/2 ~\mathrm{to}~ 1$):
green squares:   \cite{cha97};
cyan diamonds:  \cite{cio97};
red circles:  \cite{nie00};
blue stars:  \cite{ahl01};
red stars: \cite{fun05,nik05};
brown triangles down:  \cite{xia02};
cyan triangles down: \cite{sun05a};
maroon triangles right: \cite{pui07};
violet triangles up:   \cite{ver99} (numerical simulations);
blue squares:   \cite{ker00} (numerical simulations);
magenta 
triangles up:   \cite{ama05,ver08} (numerical simulations).
Note that some of the large Ra data probably are influenced by NOB effects.
Lower: an enlargement of the upper figure. 
}
\label{pd} 
\end{center}
\end{figure}

The next step is to model the individual contributions to the dissipation
rates. We start with the bulk contributions.
The turbulence in the bulk is driven by the large-scale wind $U$.
The corresponding time scale therefore is $L/U$, and from 
Kolmogorov's energy-cascade picture (see e.g.\ \cite{fri95})
the bulk energy dissipation-rate scalingwise becomes{\footnote{Note that the
 Bolgiano-Obukhov length scale does not enter here.}}
\be
\eps_{u,bulk} 
 \sim  {U^3\over L} = {\nu^3\over L^4} Re^3.
\label{eps_u_bulk}
\ee
This seems justified because the turbulence in the bulk is
more or less homogeneous and isotropic (\cite{sun06,zho08}). 
The same reasoning can be applied to the temperature equation, see
again \cite{fri95}.
The bulk thermal dissipation-rate 
then becomes
\be
\eps_{\theta,bulk} 
 \sim {U\Delta^2 \over L} = \kappa {\Delta^2 \over L^2}
Pr Re.
\label{eps_u_theta}
\ee
The scaling of the boundary-layer contributions to the dissipation rates
are estimated from their definitions as BL averages 
$
\eps_{u, BL} =
\nu  \left< (\partial_i u_j (\x \in BL,t))^2\right>_{V}
$ 
and 
$
\eps_{\theta, BL} =
\kappa  \left< (\partial_i \theta (\x \in BL,t))^2\right>_{V}
$, 
namely
\be
\eps_{u,BL} \sim \nu {U^2 \over \lambda_u^2 } \cdot {\lambda_u \over L}
\label{eps_u_bl}
\ee
and
\be
\eps_{\theta,BL} \sim \kappa {\Delta^2 \over \lambda_\theta^2 } 
\cdot {\lambda_\theta \over L}. 
\label{eps_theta_bl}
\ee
As detailed by \cite{gro04}, the kinetic and thermal BL thicknesses 
$\lambda_u$ and 
$\lambda_\theta$  are obtained from the Prandtl-Blasius
BL theory (\cite{pra04,bla08,mek61,sch00,cow01}):
\be
{\lambda_u \over L} = a Re^{-1/2},
\label{lambda_u}
\ee
where $a$ is a dimensionless prefactor of order one, and 
\bea
{\lambda_\theta \over L} & \sim & Re^{-1/2} Pr^{-1/2} 
\ \quad \hbox{for Pr} \ll 1,
\label{lambda_theta_small}\\
& \sim & Re^{-1/2} Pr^{-1/3} 
\ \quad \hbox{for Pr} \gg 1.
\label{lambda_theta_large}
\label{lambda_theta}
\eea
Note that scalingwise {\it laminar} BL theory is applied which seems
 justified because
of the low prevailing boundary Reynolds numbers. Further below it will be estimated when this
assumption breaks down for increasing $Re$. In the small $\Pr$ regimes
(eq.\ (\ref{lambda_theta_small})) (label 
``l'' stands for  ``lower'' in Fig.~\ref{pd})
the kinetic BL is nested in the thermal one, 
$\lambda_u < \lambda_\theta$, whereas in 
 the large Pr regimes 
(eq.\ (\ref{lambda_theta_large})) (``u'' for ``upper''  in Fig.~\ref{pd})
the thermal BL is nested in the kinetic one, 
$\lambda_\theta < \lambda_u$. 
The transition from one regime to the other
is modelled ``by hand'' through a crossover function 
$f(x_\theta)=(1+x_{\theta}^4)^{-1/4}$ of the variable 
$x_\theta = \lambda_u / \lambda_\theta = 2 a Nu / Re^{1/2}$, see \cite{gro01}. 
Note that in the crossover function $f(\lambda_u / \lambda_\theta)$ 
the thermal BL thickness
$\lambda_\theta$ has been replaced by $L/(2\Nu)$.
Finally, when Re becomes very small the expression (\ref{lambda_u}) for the kinetic BL 
thickness diverges, while the physical $\lambda_u$ 
is limited instead by an outer length
scale of the order of the cell height L. This saturation is happening
at some small but a priori unknown Reynolds number $Re_c$. The transition
towards the saturation regime is again modeled ``by hand'' with the 
crossover function 
$g(x_L) = x_L (1+ x_L^4)^{-1/4}$ with  
$x_L = \lambda_u (Re)/\lambda_u (Re_c)  = 
\sqrt{Re_c/Re}$ (again see \cite{gro01} for details). 

When putting the splitting and modeling assumptions 
together with the two exact relations (\ref{eps_u}) and 
(\ref{eps_theta}), one finally 
obtains two implicit equations for Nu(Ra,Pr) and Re(Ra,Pr) with
six free parameters $a$, $Re_c$, and $c_i$, $i = 1,2,3,4$:
\be
(Nu-1) Ra Pr^{-2} = c_1 {Re^2 \over g (\sqrt{Re_c/Re})} + c_2 Re^3,
\label{eq13}
\ee
\bea
Nu -1 &=& c_3 Re^{1/2} Pr^{1/2} 
\left[ 
f\left(  {2 a Nu \over \sqrt{Re_c}} g\left(
{\sqrt{Re_c \over Re}}
\right) \right)
\right]^{1/2} \nonumber \\
&+& c_4 Pr Re
f\left(   {2 aNu \over \sqrt{Re_c}} g\left(
{\sqrt{Re_c \over Re}}
\right) \right).
\label{eq14}
\eea
The $-1$ on the lhs of eq.\ (\ref{eq14}) stems from the contribution of the molecular transport, which survives when the Peclet number $Pe \equiv Re Pr = U L / \kappa$ tends to zero, $Pe \rightarrow 0$, cf.\ \cite{gro08}. This happens if either the velocity field decreases, $u_i \rightarrow 0$, or if the thermal diffusivity becomes large, $\kappa \rightarrow \infty$. In either case the time averaged Oberbeck-Boussinesq equation \ref{ob-theta} takes the form 
$\partial_j^2 \theta = 0$, whose solution with the proper boundary conditions is $\theta = -\Delta L^{-1} z$. Inserting this solution into the Nusselt number definition (\ref{nu}) gives $\lim_{Pe \rightarrow 0} Nu = 1$. Of course the $-1$ does not matter much in the turbulent regime with large Nu.

The six parameters in eqs.\ (\ref{eq13}) and (\ref{eq14})
were adjusted so as to provide a fit to 155 data points for Nu(Ra,Pr)
from \cite{ahl01}. These data were in the range 
$3\cdot 10^7 \le \Ra \le 3 \cdot 10^{9}$ and $4 \le \Pr \le 34$ for 
a $\Gamma = 1$ cylindrical cell. 
As elaborated by \cite{gro02}, in order to fix the parameter
$a$ one also needs to know Re for (at least)
one pair (Ra,Pr); \cite{gro02} 
took that value from \cite{qiu01a}. The final results were
$a=0.482$, $c_1 = 8.7$,
$c_2 = 1.45$,
$c_3 = 0.46$,  
$c_4 = 0.013$, and 
$Re_c=1.0$.
With this set  the data of \cite{ahl01} were described very well.
Later these data were adjusted for  
side-wall and plate corrections. 
However, the agreement with them as well as with additional data (\cite{fun05}) for 
Ra up to $3\cdot 10^{10}$ and $Pr = 4.38$ (see figure \ref{fig:Nred_of_R}) is still excellent (see subsection \ref{sec:Pr=4}). 
For the Nu(Ra,Pr) and Re(Ra,Pr) predicted with these parameters over wide ranges of Ra and Pr we
refer to the figures given by \cite{gro01} and \cite{gro02}. 
We mention that in principle one expects an aspect-ratio dependence of
the $c_i$, since the relative contributions of BL and bulk change with
aspect ratio. However, from experiment it is known that the $\Gamma$ dependence of $\Nu(\Ra,\Pr)$ is very weak in the explored range of Ra and Pr  (see subsection~\ref{sec:Gamma}). 

After the determination of these six parameters, Nu(Ra,Pr) and
Re(Ra,Pr) are given for all Ra and Pr by equations (\ref{eq13}) and (\ref{eq14}). In addition,
also the Ra-Pr parameter-space structure with all transitions from one regime to 
another is determined. The corresponding phase diagram is reproduced in figure \ref{pd}.

 \begin{table*}
 \begin{tabular}{|c|c|c|c|c|}
 \hline
          regime 
       &  dominance of
       &  BLs
       & $Nu$
       & $Re$
\\
\hline
         $I_l $ 
       & $\eps_{u,BL}$, $\eps_{\theta,BL}$
       & $\lambda_u < \lambda_\theta$
       & $  Ra^{1/4} Pr^{1/8} $
       & $  Ra^{1/2} Pr^{-3/4} $
\\         $I_u$ 
       & 
       & $\lambda_u > \lambda_\theta$
       & $  Ra^{1/4} Pr^{-1/12} $
       & $  Ra^{1/2} Pr^{-5/6} $
\\         $I_\infty^<$ 
       & 
       & $\lambda_u = L/4 > \lambda_\theta$
       & $  Ra^{1/3}  $
       & $   Ra^{2/3} Pr^{-1} $
\\         $I_\infty^>$ 
       & 
       & $\lambda_u = L/4 > \lambda_\theta$
       & $  Ra^{1/5}  $
       & $   Ra^{3/5} Pr^{-1} $
\\
\hline
         $II_l$
       & $\eps_{u,bulk}$, $\eps_{\theta,BL}$
       & $\lambda_u < \lambda_\theta$
       & $  Ra^{1/5} Pr^{1/5} $
       & $  Ra^{2/5} Pr^{-3/5} $
\\
         $II_u$
       & 
       & $\lambda_u > \lambda_\theta$
       & $  Ra^{1/5}  $
       & $  Ra^{2/5} Pr^{-2/3} $
\\
\hline
         $III_u$
       &  $\eps_{u,BL}$, $\eps_{\theta,bulk}$
       &  $\lambda_u > \lambda_\theta$
       & $  Ra^{3/7} Pr^{-1/7} $
       & $  Ra^{4/7} Pr^{-6/7} $
\\
         $III_\infty$
       & 
       & $\lambda_u = L/4 > \lambda_\theta$
       & $  Ra^{1/3} $
       & $  Ra^{2/3} Pr^{-1} $
\\
\hline
         $IV_l$
       & $\eps_{u,bulk}$, $\eps_{\theta,bulk}$
       & $\lambda_u < \lambda_\theta$
       & $  Ra^{1/2} Pr^{1/2} $
       & $  Ra^{1/2} Pr^{-1/2} $
\\
         $IV_u$
       & 
       & $\lambda_u > \lambda_\theta$
       & $  Ra^{1/3}  $
       & $  Ra^{4/9} Pr^{-2/3} $
\\
 \hline
 \end{tabular}
\caption[]{
The pure power laws for $Nu$ and $Re$ in the various regimes.
Table taken from \cite{gro01}.  
}
\label{tab-pure-power}
 \end{table*}

One central assumption of the GL theory
is the applicability of the scaling of the 
Prandtl-Blasius laminar BL theory. For increasing Ra and thus increasing Re this
assumption will ultimately 
break down; the BLs are expected to 
 become turbulent as well. \cite{gro00,gro02} provide an
estimate for the Rayleigh number at which 
 the breakdown occurs, based on the shear Reynolds
number $\Re_s = \lambda_u U / \nu =a  \sqrt{Re}$. For 
RB experiments using classical fluids over ``typical" Ra and Pr ranges $Re_s$ is not particularly large. This reflects 
the relatively low degree of turbulence in the interior, which also becomes
evident from flow visualizations like those done by \cite{til93,xia03,xi04}, and \cite{fun04}. For example, with Pr $\simeq 4$  one has $Re_s = 15$ when $\Ra = 10^8$ and $Re \approx 900$, and $Re_s = 190$
for $\Ra = 10^{14}$ and $Re \approx 140 000$. The dashed line in figure
 \ref{pd} is based on 
the critical value $Re_s^* \simeq 420$. Beyond $Re_s^*$
the kinetic BLs become fully turbulent
and the Prandtl-Blasius scaling is no longer applicable. 
It is not totally clear what will happen in this ultimate regime of thermal
convection. That will be discussed in the next subsection.

Note that the notion of laminar kinetic BLs in RB flow
should not be confused with
time independence or lack of chaotic behavior. The detaching thermal plumes
introduce time dependences and chaotic behavior into the kinetic BL;
but, as shown by \cite{gro04}, also then the Prandtl-Blasius
scaling laws for the thicknesses of the BLs still hold. Assuming 
a turbulent BL already for $Ra \ll 10^{14}$ as done by \cite{shr90} 
leads to a dependence of Nu on Pr that disagrees with experiments and
numerical simulations.

A detailed comparison of the GL theory with various data will be done
in sections \ref{nurenu} and \ref{nurere}.
Here we only stress that the theory has predictive power: 
The determination of the free parameters was done in the
 limited parameter range
$3\cdot 10^7 \le \Ra \le 3 \cdot 10^{9}$ and $4 \le \Pr \le 34$,
see blue stars in figure \ref{pd}. The predictions of the theory,
however, hold over a much larger domain in the Ra-Pr parameter space.

We further note that thanks to eq.\ (\ref{eps_u}) 
the knowledge of the Nusselt number allows for an estimate
of the volume averaged energy dissipation rate and derived quantities. 
E.g., when taking the conditions of the Oregon cryogenic helium experiment 
(\cite{nie00}), for $\Ra = 10^{10}$ and $\Pr=0.7$ one obtains 
either directly from  experiment or from the GL theory a
Nusselt number of 120 and with 
the experimental values
L=1m and $\nu = 5 \cdot 10^{-6}m^2/s$   an energy dissipation rate
of $\eps_u = 3 \cdot 10^{-4} m^2/s^3$. 
At $\Ra = 10^{14}$ and  $\Pr=0.7$ one obtains $\Nu \approx 2400$ 
and with $\nu = 10^{-7}m^2/s$ a value of $\eps_u = 5 \cdot 10^{-4} m^2/s^3$.
Both of these
energy dissipation rates are about three orders of 
magnitude smaller than in typical wind tunnel experiments. 
From the volume averaged energy dissipation rate eq.\ (\ref{eps_u})
one can also obtain
global estimates for the spatial coherence length $\ell$ which typically
is about 10 times the Kolmogorov length scale 
$\eta = \nu^{3/4}/\eps_u^{1/4}$.  E.g., for cryogenic helium ($\Pr = 0.7$)
at 
$\Ra = 10^7$ one obtains
$\ell/L \approx 10\eta / L =  10 \Pr^{1/2} / [ (Nu-1)^{1/4} Ra^{1/4}]
\approx 0.08
$, which is small enough to allow
 for the loss of spatial coherence and the onset of turbulence in the bulk.
In contrast, for the same Ra in water (at Pr = 4) one has 
$\ell/L \approx 0.18 $ and in glycerol (at Pr=2000) even at $\Ra = 10^9$ 
one only has   
$\ell/L \approx 0.9 $, so that there is no developed turbulence. 
In glycerol, only at  $\Ra = 10^{11}$ one obtains
$\ell/L \ltwid 0.2 $ and thus developed turbulence, 
according to this GL-model based estimate.

Finally, we note that the GL approach also has been applied to other
geometries and flows: E.g.\ \cite{eck00,eck07a,eck07b}
applied it to Taylor-Couette and pipe flow and 
\cite{tsa03,tsa05,tsa07} to turbulent electroconvection.

\subsection{Is there an asymptotic regime for large Ra? - and strict upper bounds}\label{nureup}
\cite{kra62} and later \cite{spi71} postulated an ``ultimate", or asymptotic, regime in which 
heat transfer and the strength of turbulence become independent of
the kinematic viscosity and the thermal diffusivity.
The physics of this ultimate regime is that the thermal
and kinetic boundary layers, and thus the kinematic viscosity $\nu$ 
and the thermal diffusivity $\kappa$, do not play an explicit role any
more for the heat flux. The flow then is bulk dominated.
With proper non-dimensionalization, and including logarithmic corrections due to viscous sub-layers induced by no-slip boundary conditions, 
Kraichnan's predictions for this regime read
\bea
\label{eq1}
Nu &\sim& Ra^{1/2}(\log  Ra)^{-3/2}\ Pr^{1/2}, \\
Re &\sim& Ra^{1/2}(\log Ra)^{-1/2}\ Pr^{-1/2}, \label{eq1_}
\eea 
for $Pr<0.15$, while for $0.15<Pr \le 1$ he suggested 
\bea
\label{eq2}
Nu &\sim& Ra^{1/2}(\log  Ra)^{-3/2}\ Pr^{-1/4}, \\
Re &\sim& Ra^{1/2}(\log Ra)^{-1/2}\ Pr^{-3/4}.
\label{eq2_}
\eea 
The Ra-number dependences agree with the
dependences in regimes IV$_l$ and IV$_l^\prime$ of the GL theory 
(\cite{gro00,gro01,gro02,gro04}), except for the logarithmic corrections. 
The Pr-dependence within the GL theory
in the ultimate regimes IV$_l$ and IV$_l^\prime$ is different:
\be
Nu \sim Ra^{1/2} Pr^{1/2},
\label{eq3}
\ee
\be
Re \sim Ra^{1/2} Pr^{-1/2}.
\label{eq4}
\ee 
Equation\ (\ref{eq3}) was derived first by \cite{spi71}
from a model for thermal convection in stars.

To illustrate the physical implications of the 
existence of the ultimate regime,
Andreas Acrivos (priv.\ communication) suggested the following 
Gedankenexperiment: Consider RB convection in a very large
aspect ratio sample, with the lateral dimension (say, the 
diameter D of a cylinder) much larger than the sample height L.
Now fix all dimensional parameters ($\Delta$, $\kappa$, $\nu$,
$\beta$, g, and D) and increase the sample height L, starting from
zero, but such that always still $D\gg L$, i.e., remain in the large
aspect ratio limit. 
How does the dimensional heat flux $H = \Nu \Lambda \Delta /L$ 
behave? First, $H \sim L^{-1}$, corresponding to $\Nu = 1$. With 
increasing L, the decrease will become weaker. The regime 
$\Nu \sim \Ra^{1/3}$ corresponds to the dimensional heat flux
H being independent of L. For even further increase of L
(with still $D \gg L$), the existence of the ultimate regime $\Nu 
\sim \Ra^{1/2}$ would imply that the dimensional heat flux H 
would increase again, namely with $L^{1/2}$. 
This feature may be considered
as counter-intuitive, however, our interpretation of the ultimate
 regime (if it exists) 
is that with fully developed turbulence in the bulk, 
the increasing sample height L allows for larger and larger
eddies which thus can transport more and more heat from the bottom
to the top plate.

Ever since Kraichnan's prediction in 1962, researchers have tried to 
find evidence for this regime. Various experimental efforts
will be discussed in subsection \ref{sec:large_Ra}.

There are also numerical indications of the ultimate regime: 
In order to obtain a $\Nu \sim Ra^{1/2}$ power law, 
the classical velocity and temperature boundary-conditions of the 
RB problem have been modified: 
\cite{loh03} and \cite{cal05} performed
numerical simulations for so-called ``homogeneous'' RB turbulence, in which
the top- and bottom-temperature boundary-conditions have been replaced by
periodic ones, with an unstratified temperature gradient imposed. 
The idea was to eliminate the BLs in this way.
The numerical results of \cite{cal05} -- including the found 
Prandtl number dependence -- 
are consistent with the ultimate scaling
eqs.\ (\ref{eq3})
and (\ref{eq4}),
where the Reynolds number is that of the velocity
{\it fluctuations}. 
As has been pointed out by \cite{cal06}, however, one should
 note that the dynamical equations of
 homogeneous RB turbulence allow for exponentially growing (in time)
solutions,
i.e., homogeneous RB turbulence does not have any strict upper bound
for Nu.

Such upper bounds do exist for the classical RB problem. 
Building on Howard's seminal variational formulation 
(\cite{how63,how72}), 
\cite{bus69} could prove that
$\Nu \le (Ra/1035)^{1/2}$  for any Pr. 
Later
\cite{doe96} derived a strict upper bound given by 
$\Nu \le 0.167 Ra^{1/2}-1$. They  employed the so-called ``background-method''
(\cite{doe92}). 
 The hitherto absolute best asymptotic upper
bound on Nu(Ra) comes from 
\cite{pla03}, obtaining
$Nu \le 1 + 0.02634Ra^{1/2}$, which is 20\% lower than Busse's
best estimate.
For arbritary Pr no power-law exponent of Ra smaller than 1/2 could hitherto
be obtained as an upper bound. However, for 
infinite Pr
 \cite{con99} could prove that
$\Nu \le const \times Ra^{1/3} [log\Ra]^{2/3}$. This result was improved later to
$\Nu \le 0.644\times Ra^{1/3} [log\Ra]^{1/3}$ 
by \cite{doe06}. \cite{ote02} obtained a strict upper bound for Nu for 
RB convection with constant heat flux through the plates (rather than 
with constant
temperatures of the plates), namely, $\Nu \le const \times Ra^{1/2}$ also
for this case. 
We note that the scaling laws resulting from the GL theory 
 are compatible with the upper bounds,
including those in the large-Pr limit.

\section{Experimental measurements of the Nusselt number}\label{nurenu}
\subsection{Overview}
During the last two or three decades measurements of $\Nu(\Ra)$ as a function of such parameters 
as $\Gamma$, the extent of departures from the OB approximation, 
the deliberate suppression of the large-scale circulation (LSC) by internal obstructions, 
the roughness of the confining solid surfaces,
or deliberate misalignment relative to gravity have revealed various
 aspects of the heat-transport mechanisms involved in this system.
These efforts received a significant boost when it was appreciated that liquid or gaseous helium at low temperatures offered experimental opportunities not  available at ambient temperatures 
(\cite{ahl74,thr75,ahl75,beh85,nie06b}). 
Extensive low-temperature measurements of $\Nu(\Ra)$ were initiated by the Chicago group 
(\cite{hes87,cas89,san89}), 
followed by the Grenoble group 
(\cite{cha96,cha97,roc01,cha01,roc02,roc04}) 
and the Oregon/Trieste group 
(\cite{nie00,nie00e,nie00b,nie01,nie03,nie06,nie06b}). 
Among the advantages of the low-temperature environment is the exceptionally small shear viscosity of helium gas which, at sufficiently high density,  permits the attainment of extremely large Ra. Further enhancements of the achievable Ra have be attained near the critical points of several fluids, 
including helium, where the thermal expansion coefficient diverges and the thermal diffusivity vanishes, yielding a diverging Ra at constant $\Delta$.
Here, however,
 it must be  noted that on average the increase of Ra is accompanied by an increase of Pr (see Fig.\ \ref{fig:Nred_of_largeR}, bottom) 
because  Pr diverges as well at the critical point. This makes it difficult to disentangle any influence of Ra on the one hand and of Pr on the other on this system. Another unique property of materials at low temperatures is the extremely small heat capacity and large thermal diffusivity  of the confining top and bottom plates which permit the study of temperature fluctuations at the fluid-solid interface when the heat current is held constant and led to the observation of chaos in a system governed by continuum equations
(\cite{ahl74,ahl75}). Additional advances in recent times have been due to the application of precision measurements, 
using classical liquids and gases at ever increasing Ra 
(\cite{xu00,fle02,roc02,nik05,fun05,sun05e})
and over a wide range of Pr 
(\cite{ahl01,xia02}). 

\subsection{Side-wall and top- and  bottom-plate-conductivity effects on Nu}
\label{sec:corrections}

A serious problem for quantitative measurements of Nu is the influence of the side wall 
(\cite{ahl00,roc01b,ver02,nie03}). 
The wall is in thermal contact with the convecting fluid and shares with it,  by virtue of the thermal BLs,  a large vertical temperature gradient near the top and bottom and a much smaller gradient away from the plates. Thus, the current entering and leaving the wall is larger for the filled sample than it is for the empty one. In the wall near the top and bottom
ends there is also a lateral gradient that will cause a part of the wall current to enter the fluid in the bottom half of the sample, and to leave it again in the top half. This will influence the detailed nature of the LSC 
(\cite{nie03}). 
However,  the global heat current is determined primarily by processes within the top and bottom BLs. Thus it  is insensitive to the detailed structure and intensity of the LSC and  is not influenced much by this complicated lateral heat flow out of and into the wall. Therefore the problem reduces primarily to determining the current that actually enters the fluid.
Approximate models that provide a correction for this wall effect have been proposed 
(\cite{ahl00,roc01b}), 
but these are of limited reliability when the effect is large. The cryogenic measurements have a disadvantage because the sample usually is contained by steel side walls that have a relatively large conductivity $\Lambda_w \simeq 0.2$ W/(m K), while the fluid itself has an exceptionally small conductivity of order 0.01 W/(m K), giving $\Lambda_w/\Lambda \simeq 20$. In this case the models
suggest that 
the correction is about 10\% of Nu when $\Nu \simeq 100$ ($\Ra \simeq 4\times 10^9$) and of course larger for smaller Nu. Even for $\Ra \simeq 10^{11}$ where $\Nu \simeq 280$ a correction of about 6\% is suggested. The net result is that the measured effective exponent of $\Nu(\Ra)$ is  reduced below its true value by about 0.02 or 0.03 (\cite{ahl00}). For  gases near ambient temperatures with typical thermal conductivities near 0.03 W/(m K), such as sulfur hexafluoride (SF$_6$) and ethane (C$_2$H$_6$), confined by a high-strength-steel side wall with a conductivity of 66 W/(m K) 
(\cite{ahl07}), one approaches the case of perfectly conducting lateral boundaries where subtraction of the current measured for the empty cell actually becomes a good approximation. 
Nonetheless results for Nu, although very precise, can not be expected to be very accurate. An exceptionally favorable case is that of water confined by relatively thin plastic walls 
(\cite{ahl00}), 
where $\Lambda_w/\Lambda \simeq 0.3$. In that case the side-wall correction can be as small as a fraction of a percent and may safely be ignored for most purposes. An intermediate case, for which reasonably reliable corrections can be made, is that of organic fluids confined by various plastic walls which typically have $\Lambda_w/\Lambda = {\cal O}(1)$ 
(\cite{ahl01}). 
In the case of liquid metals, which are of interest because they have very small Prandtl numbers of order $10^{-2}$ or less, $\Lambda_w/\Lambda$ is small ($\simeq 2$ for Hg and  $\simeq 0.2$ for Na as examples) and again the wall corrections are small or negligible.

A second problem involves the finite conductivity $\Lambda_p$ of the top and bottom plates 
(\cite{cha02,ver04}). 
One would like $X \equiv \Lambda_p L / (e \Lambda \Nu)$ to be very large (here $e$ is the thickness of one plate). Else the emission of a plume from the top (bottom) boundary will leave an excess (deficiency) of enthalpy in its former location, generating a relatively warm (cold) spot near the plate where the probability of the emission of the next  plume is diminished until this thermal ``hole" has diffused away by virtue of the plate conductivity. This issue was explored experimentally 
by \cite{bro05} 
by  measuring $\Nu(\Ra)$ with high precision using water ($\Lambda \simeq 0.6$ W/(m K)) as the fluid and first Al and then Cu top and bottom plates of identical shape and size (see Fig.~\ref{fig:Nred_of_R} below). The conductivities $\Lambda_{p,Cu} \simeq 400$ W/(m K) of Cu and  $\Lambda_{p,Al} \simeq 170$ W/(m K) of Al 
differ by a factor of about 2.3 and thus yield different reductions of $\Nu(\Ra)$ below the ideal value $\Nu_\infty$ for isothermal boundary conditions. The results permitted the extrapolation of $\Nu$ to  $\Nu_\infty$ by the use of the empirical formula
\begin{equation}
\Nu = f(X) \Nu_\infty;\ \ f(X) = 1 - exp(-(aX)^b)\ .
\label{eq:correct}
\end{equation}
The parameters were found to be $a=0.275$ and $b=0.39$ for $L=0.50$ m, and $f(X)$ was found  to be closer to unity for smaller $L$. At fixed $L$ both $a$ and $b$ (and thus $f(X)$)  were independent of $\Gamma$. This plate-conductivity effect is expected to be relatively small for the cryogenic and room-temperature compressed-gas experiments because typically $\Lambda_p/\Lambda = {\cal O}(10^4)$ and larger and thus $X$ is very large unless Nu becomes extremely large.  At modest Ra, say $\Ra \alt 3\times 10^{9}$, it is small also for Cu plates and organic fluids where $\Lambda_p/\Lambda = {\cal O}(10^3)$. The plate correction is a very serious problem for measurements with liquid metals where for instance $\Lambda_p/\Lambda \simeq 50$ for Hg and $\simeq 5$ for Na. It has been suggested that this problem might be overcome by using a composite plate containing a volume partially filled with a liquid of high vapor pressure. In that case the condensation and vaporization of this fluid inside the plate can yield an effective plate conductivity much larger than that of the metal alone. To our knowledge this idea has not yet been implemented.

The problem of the plate corrections was addressed again recently through
numerical simulations by
\cite{ver08,ama05}.  Results for $\Nu(\Ra)$ obtained with constant heat-flux boundary conditions (BCs)  at the lower plate and constant-temperature
BCs at the upper plate were compared with Nu(Ra) for constant-temperature BCs at both plates. As shown below in  Fig.~\ref{fig:Nred_of_largeR}a, the results for both BCs agreed reasonably well with each other and with experiment  up to $\Ra \approx 10^9$. This is also found by comparing 
2D numerical simulations with constant temperature and constant flux
BCs (\cite{joh07} and unpublished results by the same authors).  
However, 
for $Ra \agt 10^9$  (\cite{ver08,ama05}) 
the constant-temperature results for Nu were up to
20\% higher than the constant heat-flux simulations and up to nearly 30\% higher than experimental
results. This suggests that in the high-Ra experiments, even after correction for the finite plate-conductivity using Eq.~(\ref{eq:correct}), the results correspond more nearly to constant-flux at the bottom plate than to constant-temperature boundary conditions. An elucidation by experiment of this surprising result would be very welcome. 
\vfill

\subsection{The Nusselt number for $\Pr \simeq  4.38$ obtained using water as the fluid}
\label{sec:Pr=4}

\begin{figure}
\centerline{\epsfig{file=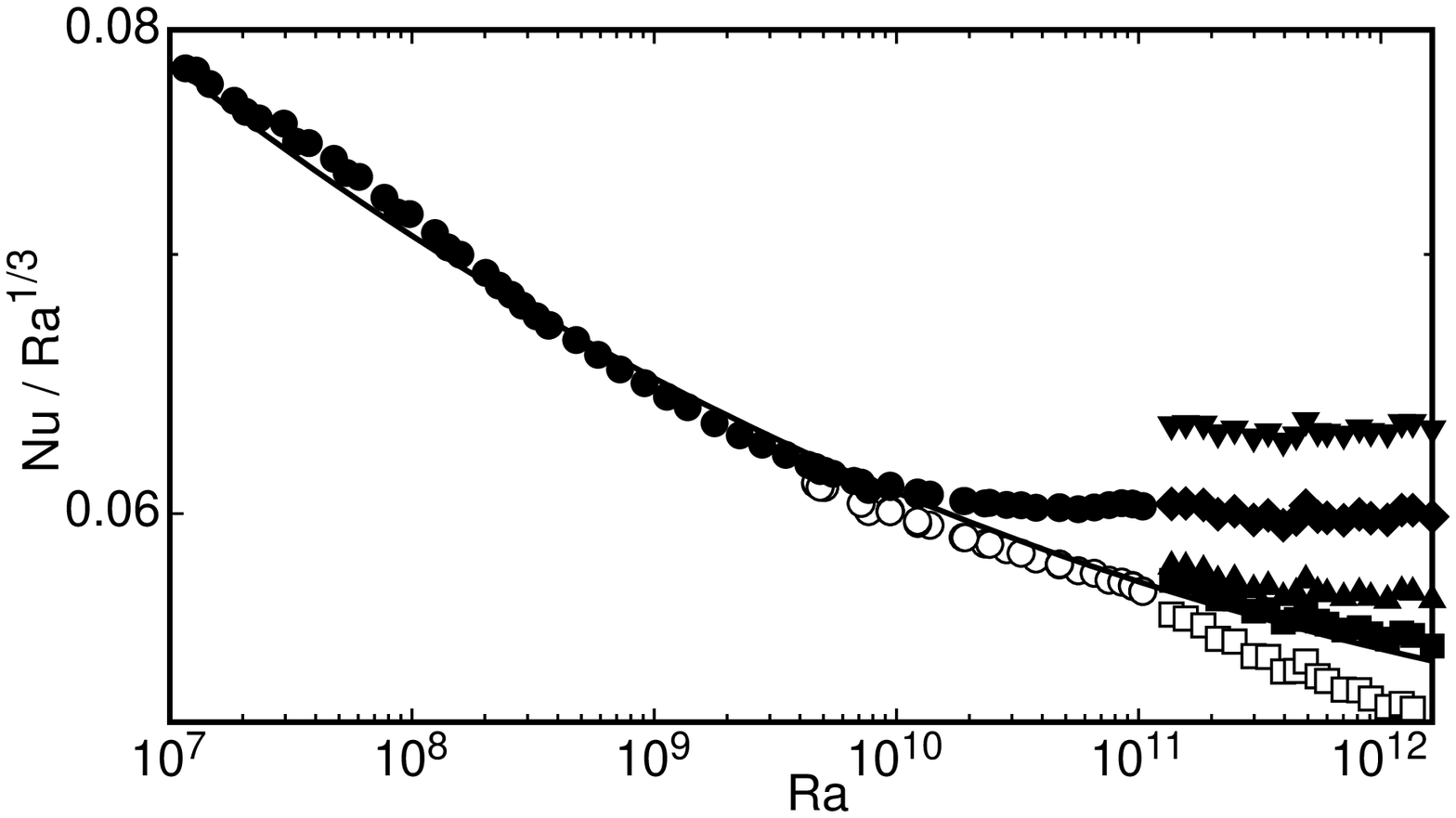,width=8cm}}
\caption{Reduced Nu for $\Gamma = 1$, obtained using water 
($\Pr \approx 4.4$) 
and copper plates, as a function of Ra. Open symbols: uncorrected data. Solid symbols: after correction for the finite plate conductivity. Circles: 
\cite{fun05}. 
Squares:  
\cite{sun05e}. 
The downwards and the upwards triangles are
upper and lower bounds on the actual Nusselt number at large Ra; the diamonds
originate from an estimate, 
see the text for details. 
The solid line is the GL prediction 
(\cite{gro01}).
}
\label{fig:Nred_of_R} 
\end{figure}

For $\Pr \simeq 4.4$ and $\Gamma = 1.00$ high-accuracy measurements of $\Nu$ 
(\cite{fun05}) 
for $10^7 \alt \Ra \alt 10^{11}$ using water and copper plates are shown in Fig.~\ref{fig:Nred_of_R} as circles. We focus on these data because for them the side-wall corrections are negligible and top- and bottom-plate corrections  based on experiments with plates of different conductivities were made (see Sec.~\ref{sec:corrections}). The wide Ra range was achieved by using three samples with different $L$. The data before the plate correction are given as open circles. Corrected data are presented as solid circles. 

For $\Gamma = 1$ and $\Pr \simeq 4$ the experiment reaching the largest Ra was conducted using Cu plates and a water sample with $L = 100$ cm and reached $\Ra \simeq 10^{12}$ 
(\cite{sun05e}). 
These data are shown as open squares in the figure. It is gratifying that they are remarkably consistent with the open circles. However, the authors used an empirical plate correction with $a = 0.987$ and $b = 0.30$  which yielded the solid squares in the figure. In an attempt to develop an estimate of the uncertainty, we applied a correction using eq.~(\ref{eq:correct}) and the parameters $a=0.275$ and $b=0.39$ obtained from the $L = 0.5$ m sample. This yielded the up-pointing triangles. This correction is too small because measurements with a $L = 0.25$ m sample and the $L = 0.50$ m sample 
by \cite{bro05} 
revealed that the correction increases with $L$. Arbitrarily assuming a power-law dependence $a = a_0 L^{x_a}$ and $b = b_0 L^ {x_b}$, an extrapolation to $L=1$ m yielded $a = 0.221$ and $b= 0.264$, and via 
eq.~(\ref{eq:correct}) led to the down-pointing triangles. We consider the up-pointing and down-pointing triangles to be estimates of lower and upper bounds on the actual $\Nu_\infty$. Arbitrarily adjusting $a$ and $b$ to the intermediate values 0.25 and 0.32 respectively yielded the solid diamonds which are consistent with the data from the $L=0.5$ m sample. New measurements in this very large cell with Al plates, which together with the Cu-plate data will yield better values of $a$ and $b$, are anxiously awaited.

The solid line in Fig.~\ref{fig:Nred_of_R} is the GL prediction 
(\cite{gro01}). It gives the shape of the experimentally found $\Nu(\Ra)$ very well for $\Ra \alt 10^{10}$. For larger Ra the data suggest $\gamma_{eff} = 1/3$ whereas the model only reaches such a value for $\gamma_{eff}$ as $\Ra \rightarrow \infty$ where the model is no longer expected to be applicable.

A detailed discussion of a number of other measurements for $Pr = {\cal O}(1)$ and $\Gamma = 0.5$ or 1 
(\cite{nie00,xu00,ahl01,cha01,fle02,roc02,nie03,nik03,roc04,nik05,sun05e}) 
is beyond the scope of this review, although we shall re-visit a few of them below in Sec.~\ref{sec:large_Ra}. We refer to previous publications  
by \cite{nie03,nik05} 
where many data sets have been compared. There is excellent agreement between several of them; however, in the range $\Ra \alt 10^{12}$ there are differences of up to 20\% or so between some of them. It is not clear whether the origin of these differences is to be found in experimental uncertainties, perhaps associated with wall or plate corrections or other experimental effects, or, as has been suggested by \cite{nie03},  in genuine differences of the fluid dynamics of the various samples. 
We find the latter explanation somewhat unlikely because, as discussed below in Sec.~\ref{sec:Nu_LSC}, the heat transport is determined primarily by 
boundary layer instabilities and is relatively insensitive to the structure of the LSC.

\subsection{The Prandtl-number dependence of the Nusselt number}

Fluids with $\Pr > 1$ are plentiful in the form of various liquids, although accurate determinations of $\Nu(\Ra)$ are in many cases problematic because the required physical properties are not known well enough. Typical gases not too 
close to the critical point have $\Pr = {\cal O}(1)$. The range $\Pr \alt 0.7$ is difficult to access because most ordinary fluids have Pr greater than or close to the hard-sphere-gas value 2/3 (see, for instance, 
\cite{hir64}). 
Liquid metals, by virtue of the electronic contribution to the thermal conductivity, have $\Pr = {\cal O}(10^{-2})$ or smaller, leaving a wide gap in the range from $10^{-2}$ to 0.7. For the liquid metals it is difficult to obtain very large values of Ra because the large thermal conductivity requires very large heat currents and tends to yield small Rayleigh numbers unless very large samples are constructed. Another problem for liquid metals  (see Sec.~\ref{sec:corrections}) is the uncertainty introduced by a large plate correction; however, side-wall corrections should be negligible.

\begin{figure}
\begin{center}
\includegraphics*[width=5.7cm]{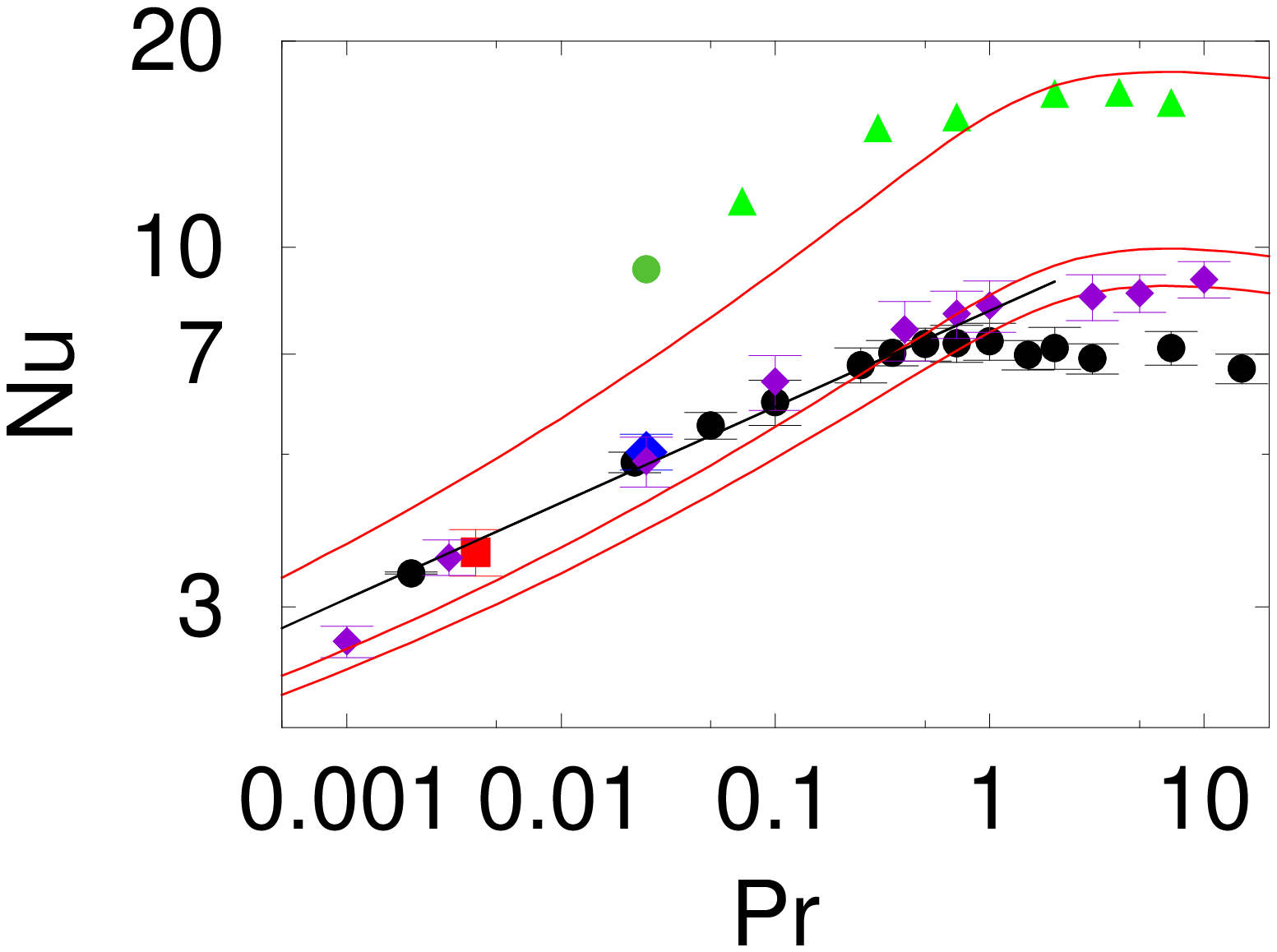}
\hspace{0.5cm}
\includegraphics*[width=5.7cm]{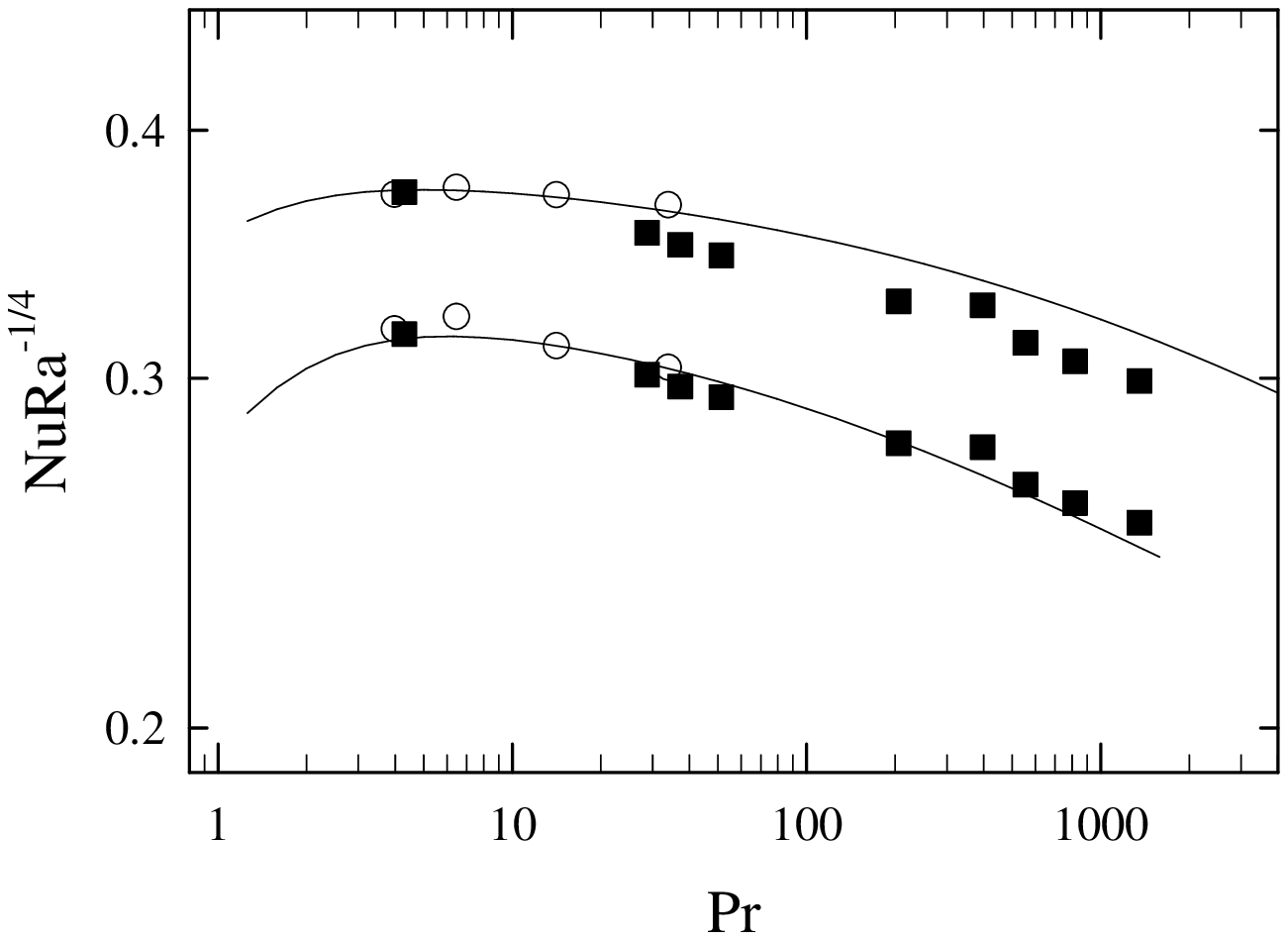}
\caption{
Top figure: $\Nu(\Pr)$ for $ \Ra= 6\cdot 10^5$ and $\Gamma = 1$
from numerical simulations by \cite{ver99} (black circles),
from experiments with mercury by \cite{ros69}  (blue diamond),
and from the experiments with sodium by \cite{hor98}  (red square).
The straight solid line is a fit to  the numerical data 
 with $\Pr < 1$ (\cite{ver99}), giving
$ \Nu= 8.1 \Pr^{0.14\pm 0.02}$. 
The exponent is in
agreement with the low-Pr expectation 1/8 of the GL theory.
The green upper triangles are the numerical data for
$\Ra = 10^7$ by \cite{ker00}, 
the dark-green circle results from the experimental
data of \cite{cio97} for the same $\Ra = 10^7$,  
and 
the violet diamonds are numerical results  for $\Ra = 10^6$ by
\cite{bre04}.
The three thick red lines are the results from the GL theory
eqs.\ (\ref{eq13}) and (\ref{eq14}) for the three 
Rayleigh numbers of the 
numerical data sets, namely
$\Ra = 6\cdot 10^5$,
$\Ra = 10^6$, and 
$\Ra = 10^7$, bottom to top.
Bottom figure:
 The reduced Nusselt number $\Nu \Ra^{-1/4}$ as a function of the Prandtl number for the two Rayleigh numbers $1.78\times 10^9$ (upper set) and $1.78\times 10^7$ (lower set) in the large-Pr regime. Open circles: 
\cite{ahl01}. Solid symbols: \cite{xia02}. Various organic fluids were used. From \cite{xia02}.}
\label{fig:Nred_of_Pr}
\end{center} 
\end{figure}                      

In spite of these difficulties, several researchers attempted low-Pr
measurements of Nu, 
in order to study the Pr dependence.
Measurements with 
 mercury
($\Pr=0.025$) were done by  \cite{ros69} 
($2\cdot 10^4 \le \Ra\le 5 \cdot 10^5$),
by \cite{tak96} and \cite{nae97} 
($10^5 \le \Ra\le  10^9$),
by 
\cite{cio95,cio96,cio97} 
($5\cdot 10^6 \le \Ra \le 2\cdot 10^9$), and by
\cite{gla99}
($2 \cdot 10^5 \le \Ra\le 8 \cdot 10^{10} $). 
\cite{hor98} made
measurements with 
liquid sodium ($\Pr=0.005$,
$Ra \le 10^6$).
Together with the results for helium gas, air (Pr=0.7), and water 
($4\le Pr\le 7$), these low-Pr data imply a strong increase of Nu with Pr at constant 
Ra, as shown in figure  \ref{fig:Nred_of_Pr}, top.
For Pr larger than about one a 
saturation sets in and Nu becomes 
 Pr-independent for some Pr-range. Recent results using helium gas at low temperatures 
(\cite{roc02})
 and covering the range $0.7 \alt \Pr \alt 21$ suggest a very mild if any 
increase with Pr. Results obtained with various organic fluids 
(\cite{ahl01,xia02}) for $\Ra= 1.78\times 10^9$ and $1.78 \times 10^7$ 
are shown in fig.~\ref{fig:Nred_of_Pr}, bottom,
 and indicate a maximum in $\Nu(\Pr)$ near $\Pr \simeq 3$, followed by a very gradual decrease of $\Nu$ with $\Pr$ that can be described by $\Nu \propto \Pr^{-0.03}$ over the Pr-range of the experiments. 

One of the successes of the GL model is that it contains most 
of the features of $\Nu(\Pr)$ observed in experiment. When Ra is not too large, it predicts $\Nu \sim \Pr^{1/8}$ 
at constant Ra 
for $\Pr \alt 1$, a maximum near $\Pr = 3$, and the very gradual decline for larger $\Pr$. 
For large Pr the GL prediction
 is shown by the solid lines in figure
\ref{fig:Nred_of_Pr}b. Although the parameters of the model had been adjusted using data for Pr up to about 30 (including the
open circles in the figure), the model agrees with the 
measurements up to  $\Pr\simeq 2000$. 
The large Pr behavior resulting from the GL theory is discussed
in more detail in \cite{gro01} and the small Pr behavior in \cite{gro08}.

\subsection{The aspect-ratio dependence of the Nusselt number}
\label{sec:Gamma}

Several experiments 
(\cite{wu92,xu00,ahl01,fle02,fun05,nik05,sun05e,nie06})
have probed the dependence of Nu at constant Ra and Pr on $\Gamma$. Using water with $\Pr \simeq 4$, it is found for $\Gamma \alt 5$ that Nu increases, albeit only very slightly, with decreasing $\Gamma$. For larger $\Gamma$ the measurements up to $\Gamma = 20$ suggest no further change, indicating that a large-$\Gamma$ regime may have been reached. 
The weak $\Gamma$ dependence suggests an insensitivity to the nature  of the LSC (see also Sec.~\ref{sec:Nu_LSC}), which surely changes as $\Gamma$ increases well beyond one, and is consistent with the determination of Nu by instabilities of the thermal BLs. --
Theoretical efforts to understand the influence of $\Gamma$ on  Nu
have been quite limited, see e.g.\
\cite{gro03} and \cite{chi06}.

\subsection{The insensitivity of the Nusselt number to the LSC}
\label{sec:Nu_LSC}

Several experiments suggest that the Nusselt number in the Ra range below the transition to the ultimate regime is remarkably insensitive to the strength and structure of the LSC.  
\cite{cio96} 
measured $\Nu(\Ra)$ with a sample of water with $\Pr \simeq 3$ in a container of rectangular cross section in which the azimuthal LSC orientation was more or less fixed.  They determined the heat flux both of the original water samples, and of the same samples after several vertically positioned screens had been installed within them.  In the absence of the screens shadowgraph visualizations showed that plumes generated at the bottom boundary layer were swept laterally just above the boundary layer by a LSC. The plumes rose vertically in  the presence of the screens, suggesting a dramatically altered and much weaker LSC. For both cases the heat current was the same within a few percent. This experiment suggests that the heat current is determined primarily by the conductance and instability of the thermal boundary layers which are not influenced significantly by the LSC, and that the plumes with their excess enthalpy will find their way to the top one way or another regardless of any LSC.   
\cite{cio96} 
also found that tilting their cells relative to gravity by an angle $\alpha$ as large as 0.06 radian, which enhances the Reynolds number of the LSC,  had no influence on the heat transport within their resolution of a few percent.

More recently 
\cite{ahl06b} 
measured $\Nu(\Ra)$ for a cylindrical water sample with $\Gamma = 1$ and $\Pr = 4.4$ as a function of the tilt angle $\alpha$ with a precision of 0.1\%. They found, for example,  a very small {\it reduction}, by about 0.4\%, for a tilt angle $\alpha = 0.12$ radian. In the same experiment the LSC Reynolds number was determined and found to {\it increase} by about 25\% for $\Ra = 10^9$ and by about 12 \% for $\Ra = 10^{11}$. If the Reynolds number had any direct influence on Nu, one would have expected an {\it increase} of Nu with Re. Again one is led to conclude that the heat transport is independent of the vigor of the LSC and thus presumably determined by LSC-independent boundary layer properties. 
This finding seems to be in conflict with the final GL results 
eqs.~(\ref{eq13}) and (\ref{eq14}), in which Nu and Re are intimately coupled
to each other.

For a $\Gamma = 0.5$ water sample 
\cite{chi04a} 
measured a reduction of Nu by about 5\% when they tilted their system by about 0.03 radian. Samples of this aspect ratio are more complex because the LSC can consist of either a single convection roll, or of a more complex structure approximated by two rolls stacked one above the other 
(\cite{ver03,xi08}). 
The authors conjecture that the tilt stabilizes the single-roll structure, and that this structure gives a smaller heat transport than the two-roll structure, thus accounting for the reduction of Nu. However, 
it seems surprising to us that for $\Gamma = 0.5$ the Nusselt number should be more sensitive to the LSC than it is for the $\Gamma = 1$ system. 

More evidence for the insensitivity of Nu to changes in the
 LSC is given by
\cite{xia97}, who
altered the LSC 
into an oscillating four-roll flow pattern by placing staggered
fingers on the side-wall and found that Nu changed very little.
\cite{xia99}
made an even stronger perturbation to the system 
by placing a baffle at the cell's mid-height, again
finding insensitivity of Nu.

In addition to the evidence of the {\it insensitivity} of $\Nu(\Ra)$ to changes in the LSC, there is good evidence for the {\it sensitivity} of $\Nu(\Ra)$ to the structure of the thermal BLs. This is  provided by an experiment of 
\cite{du00,du01} 
who covered the top and bottom plates with triangular grooves that were much deeper than the BL thickness. They found an enhancement of $\Nu(\Ra)$ by as much as 76\%, with no significant change in the dependence on Ra (however, see also 
\cite{cil99}). 
Flow visualization revealed an increase of plume shedding by the protrusions as the mechanism of the Nu enhancement.
Similar results  were found by \cite{str06} 
in their numerical simulations of RB convection over grooved plates. 

\subsection{The dependence of Nu on Ra at very large Ra}
\label{sec:large_Ra}

\begin{figure}
\centerline{\epsfig{file=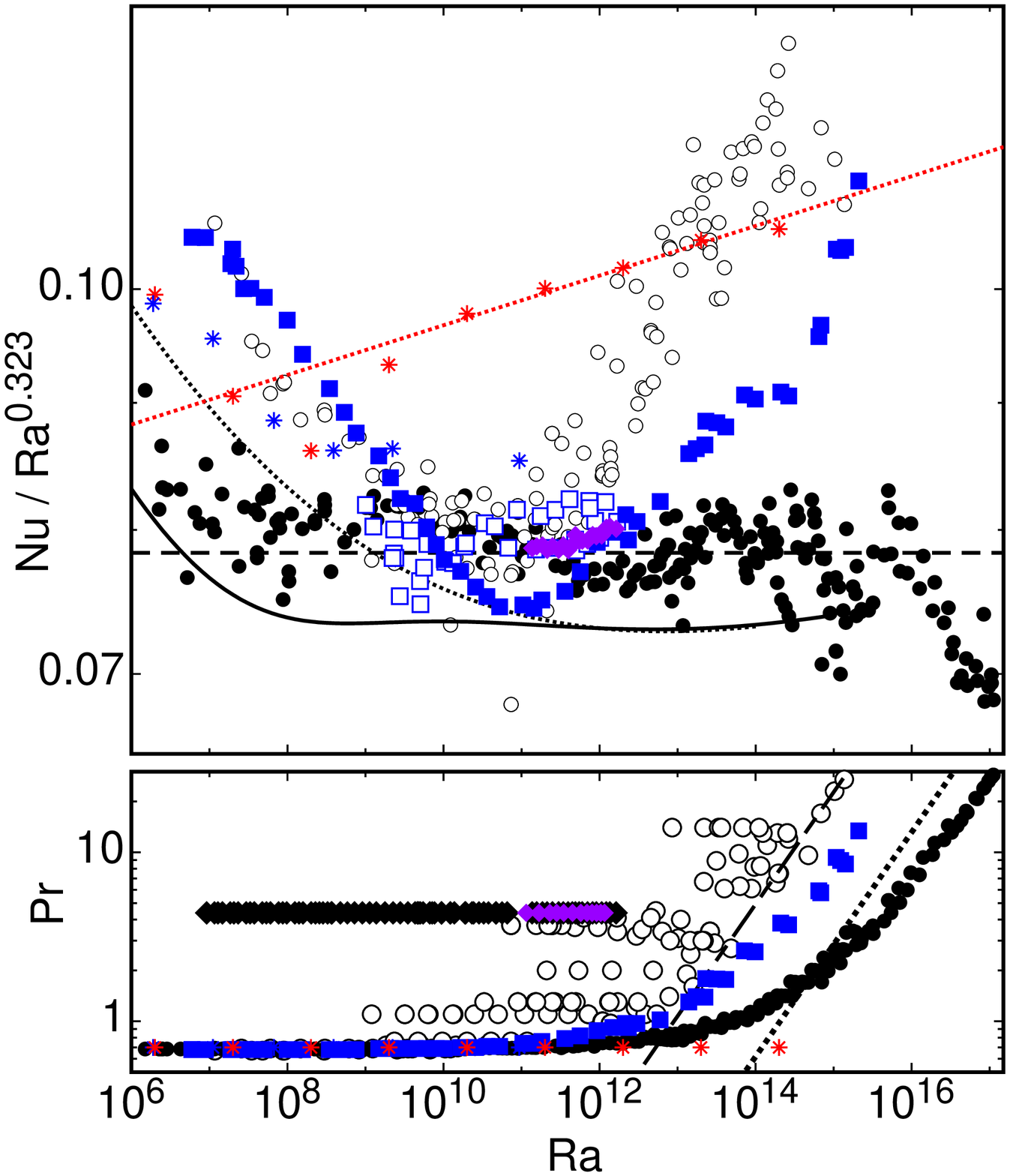,height=9cm}}
\caption{Top figure: 
 $\Nu/\Ra^{\gamma_{eff}}$ with $\gamma_{eff} = 0.323$ as a function of Ra. Black solid circles: $\Gamma = 0.5$ (\cite{nie00}) 
after a correction for plate and side-wall effects (\cite{nie06b}). 
Black open circles: $\Gamma = 0.5$ 
( \cite{cha01}).  Blue solid squares: $\Gamma = 1$ (\cite{nie03}). Purple solid diamonds:  $\Gamma = 0.67$ and 0.43 (\cite{nik05}). 
Open squares: $1 \alt \Gamma \alt 3,~\Pr = 0.7$ (\cite{fle02}). 
Red stars: numerical results for $\Gamma = 1/2$, $\Pr = 0.7$ (\cite{ama05})
for constant-temperature boundary conditions at the plates. Blue stars:  
numerical results for $\Gamma = 1/2$, $\Pr = 0.7$ (\cite{ver08})
 for constant-heat-flux BCs at the lower plate and 
constant-temperature BCs at the upper plate.
Dotted red line: a power law with $\gamma_{eff} = 1/3$ 
fitted to the red stars. 
Solid (dotted) black line: The GL prediction for $\Pr = 0.8$ ($\Pr = 29$). 
Bottom figure: 
The Prandtl numbers corresponding to the data  in the top the figure. 
In addition Pr for the measurements for $\Gamma = 1$ with water in Fig.~\ref{fig:Nred_of_R} are shown as black diamonds. The dashed and dotted lines in 
the bottom figure are estimates of the location of the transition  to the Kraichnan regime for $\Gamma = 1$, assuming critical boundary layer Reynolds-numbers $\Re_s^* = 220$ and 440 respectively  (since the parameters of the GL theory have been determined only for $\Gamma = 1$, a prediction of $ \Ra^*$ for smaller $\Gamma$ unfortunately is not available).
}
\label{fig:Nred_of_largeR} 
\end{figure}

Below the transition to the ultimate regime the Nusselt number is determined essentially by properties of the top and bottom thermal boundary layers (see Sec.~\ref{sec:Nu_LSC}). As already discussed in subsections 
\ref{nuregl} and \ref{nureup}, this is expected to change dramatically in a critical range around some $\Ra^*$, defined by the condition
that the shear across the  
laminar (albeit fluctuating) kinetic BL due to the LSC becomes so large that a transition to turbulence is induced within it. 
Note that the exact value of $\Ra^*$ depends on the strength and type of the 
turbulent noise that perturbs the BLs, but 
the transition
 is expected to happen once the shear Reynolds number $\Re_s$, based on the kinetic BL thickness, exceeds $\Re_s^* = {\cal O}(400)$. 
For $\Gamma = 1$ estimates of $\Ra^*$ based on the GL theory (\cite{gro02}) and corresponding to $\Re_s^* = 440$  and 220 are shown in Fig.~\ref{fig:Nred_of_largeR}b as dotted and dashed lines,
 respectively  (since the parameters of the GL theory have been determined only for $\Gamma = 1$, an equivalent prediction of $ \Ra^*$ for general
 $\Gamma$ unfortunately is not available). These estimates are based on the assumption that a LSC continues to exist at these very large Rayleigh numbers. If it does not, then the transition should eventually be triggered by a destruction of the kinetic BL by turbulent fluctuations rather than by a laminar (albeit fluctuating)
 flow across the plates. Understanding the regime above $ \Ra^*$ is of particular importance because it is believed by many to be the asymptotic regime that permits, in principle, an extrapolation to arbitrarily large values of $\Ra$, including those of astro- and geo-physical interest. 
 
 Experimentally it should be possible to observe the predicted transition by a dramatic change in the magnitude and/or the Rayleigh-number dependence of the Nusselt number. For $\Nu(\Ra)$ one expects a change from an effective power law with $\gamma_{eff} \simeq 0.32$ as observed below $\Ra^*$ to $\gamma_{eff} \simeq 0.4$,  which due to the logarithmic corrections is somewhat below the predicted asymptotic value $\gamma = 1/2$ (see Sec.~\ref{nureup}). Another dramatic change, according to the theory, should be the dependence on Pr. For $\Ra < \Ra^*$ the Nusselt number
 is essentially independent of Pr for $\Pr \agt 1$. For $\Ra > \Ra^*$ the Kraichnan prediction is $\Nu \sim \Pr^{-1/4}$ (see Eq.~\ref{eq2}), at least for Pr near one. However the GL theory predicts $\Nu \sim \Pr^{1/2}$ (see Eq.~\ref{eq3}), so there remains some uncertainty on this issue. Nonetheless, any significant Pr dependence would lead to a discontinuity of $\Nu(\Ra)$ of a size that would depend on Pr.

From Fig.~\ref{fig:Nred_of_largeR}, bottom, 
one sees that the measurements with water at $\Pr = 4.4$ and $\Gamma = 1$, 0.67, and 0.43  have not reached the  regime
above
$ \Ra^*$  predicted for $\Gamma = 1$. As expected, the measurements for $\Gamma = 0.67$ and 0.43 shown in Fig.~\ref{fig:Nred_of_largeR}, top, as well as those for $\Gamma = 1$ shown in  Fig.~\ref{fig:Nred_of_R}, give no indication of the BL-turbulence transition. Neither do  other measurements
for $\Pr \approx 4$, $\Gamma$ between  0.67  and  20,  and 
$\Ra$ up to $5\cdot 10^{12}$  (\cite{sun05e}). 

Measurements using cryogenic helium  
by \cite{nie00} for $\Gamma = 0.5$  are shown in Fig.~\ref{fig:Nred_of_largeR}, top,
 as solid circles. The data were corrected recently by some of the original authors
(\cite{nie06}) for side-wall and plate effects (see Sec.~\ref{sec:corrections}).\footnote{Interestingly, this side-wall correction, which was based on the model of \cite{roc01b}, yielded corrected data that are nearly identical to those that had been obtained using Model 1 of \cite{ahl00} and had been shown in Fig. 5 of that reference. The plate-effect corrections are quite small for the cryogenic data and have  little influence on the interpretation of the data.} The data are for Rayleigh numbers as  large as $10^{17}$, and for Ra up to about $10^{12}$ they are for $\Pr \simeq 0.7$. 
In the Ra-range of overlap, they are in excellent agreement with the water measurements for $\Pr = 4.4$  and $\Gamma = 0.67$ and 0.43, and with the results for compressed gases (\cite{fle02})  with $\Pr \simeq 0.7$, $1 \leq \Gamma \leq 3$, and $1\times 10^9 \alt \Ra \alt 2\times 10^{12}$, demonstrating again the insensitivity of Nu to Pr and $\Gamma$ in this Ra range, as well as a consistency between the cryogenic and room-temperature experiments. 
At large $\Ra$ the values of $\Pr$ for the \cite{nie00} data increased because of the proximity to the  critical point. As seen in Fig.~\ref{fig:Nred_of_largeR}b, the $\Gamma = 0.5$  data might have been expected to cross $ \Ra^*$ somewhere near $\Ra  = 10^{13}$ or $10^{14}$, but apparently did not do so since they reveal no change of the dependence of Nu on Ra. Two possible explanations come to mind. Perhaps the LSC was less vigorous in this experiment than it was for $\Gamma = 1$. In that  case the expected $\Re_s^* \simeq 400$ would only be reached at even higher $\Ra$. Alternatively, at the very high $\Ra$ the LSC may have deteriorated into an unrecognizable entity consisting essentially only of vigorous fluctuations as suggested by \cite{sre02}. In that case the GL estimate for $ \Ra^*$ would no longer be quantitatively applicable.  

Within their experimental uncertainty and over the very wide range $10^7 
\alt \Ra \alt 5\times 10^{15}$ the data of \cite{nie00} 
can be described by a single power law $\Nu = N_0 \Ra^{\gamma_{eff}}$ with an effective exponent $\gamma_{eff} = 0.323$ and
 $N_0 = 0.0783$. This power law agrees very well with a fit to the data of \cite{fle02}  which (over their much more narrow range of Ra) yielded $N_0 = 0.0714$ and $\gamma_{eff} = 0.327$. Both sets of measurements are inconsistent with an exponent of $1/3$, which would yield a slope like that of the dotted line in Fig.~\ref{fig:Nred_of_largeR}a. However, they are remarkably consistent with the prediction of GL, which is shown by the solid line in the figure. The drop below the power law for $\Ra \agt 5\times 10^{15}$ is unexplained. One might have attributed it to non-Boussinesq effects, but for gases near the critical point 
  these would cause an increase of $\Nu$ and not a decrease,
see \cite{ahl08} and Sec.~\ref{sec-nob}. 
Alternatively, one might look at the variation of $\Pr$ as an explanation, but in the GL model $\Nu$ is essentially independent of $\Pr$ at these large $\Ra$, cf. regime $IV_u$ in subsection \ref{nuregl}. 

An earlier set of data using helium at low temperatures and $\Gamma = 0.5$ was obtained  by \cite{cha96,cha97,cha01}. 
It  extends up to $\Ra \simeq 10^{15}$, and the results listed by \cite{cha01} are shown as open circles in Fig.~\ref{fig:Nred_of_largeR}. In the range $10^{10} \leq \Ra \leq 10^{11}$ they agree very well with the other data shown in the figure. For smaller $\Ra$ they are higher than the \cite{nie00} data. A possible reason might be found in a difference of the side-wall correction that was applied.\footnote{In fact, for $\Ra <  10^9$ the \cite{nie00} data before their side-wall correction were much closer to the \cite{cha01} data  than after this correction was made.} More interesting is the difference between the two data sets that evolves as $\Ra$ grows beyond $10^{11}$. In that regime the open circles in the figure can be described within their scatter by a power law with $\gamma_{eff} \simeq 0.38$. The authors interpret this result as corresponding to the expected $\gamma = 1/2$ in the Kraichnan regime, modified by the logarithmic corrections that are attributable to a viscous sublayer. Thus they claim to have entered the ``ultimate", or asymptotic, regime of turbulent RBC 
(\cite{cha97}). However, the  transition 
at $\Ra^*$ just above $10^{11}$ is lower than the theoretical estimates
 for the shear-flow boundary layer instability. For $\Pr=1$ and $\Ra = 3\times 10^{11}$ the GL model yields (in the $\Gamma = 1$ case) $\Re_s \simeq 100$, which is too low for a shear-induced transition to turbulence in the boundary layer. An explanation in terms of a shear-induced BL transition would require a more vigorous LSC for the $\Gamma = 0.5$ case than was measured (see Sec.~\ref{nurere}) for the $\Gamma = 1.0$ case.\footnote{\cite{cha01} measured the vertical LSC velocity component $v_z(r,z,\theta)$ in the horizontal mid-plane ($z = 0$) and at a radial position $r = L/4$ half way between the center line and the side wall. The significant size found for $v_z$ suggests the existence of well developed up-flow and/or down-flow in the mid-plane, implying that the LSC consisted of a single convection roll rather than of two rolls one above the other. The measurement yielded a Reynolds number that is not very different from that for the $\Gamma = 1$ case (see Sec.~\ref{nurere}), but it is difficult to know precisely the corresponding  shear across the viscous BLs at the top and bottom plate.}  
In any case, the data of \cite{cha01}, and the  interpretation in terms of a transition to the Kraichnan regime, differ dramatically from the measurements of 
\cite{nie00} who did not find this transition even though their data extend to higher values of $\Ra$, and were done at 
somewhat lower $\Pr$ where the shear transition should occur at even smaller $\Ra$. The reason for this difference remains a  mystery at this time, and the resolution of this apparent conflict between the two data sets is one of the major challenges in this field of research.

Yet another set of data, shown as solid squares in Fig.~\ref{fig:Nred_of_largeR}a, was obtained with low-temperature helium by \cite{nie03}, using the original apparatus of the $\Gamma = 0.5$ measurements by \cite{nie00}, but with a sample of reduced height that had $\Gamma = 1.0$. Unfortunately these results do not help to clarify the situation.  They are fairly consistent with other data in the Ra range near $10^{11}$. At smaller Ra they agree fairly well with the \cite{cha01} data, but  differ from the side-wall-corrected \cite{nie00} data. Here again one would be tempted to invoke the side-wall effect as a possible explanation. More difficult to disregard are the data for $\Ra \agt 10^{12}$, where side-wall corrections are negligible. There the data fall between the two $\Gamma = 0.5$ cryogenic data sets, thus adding to the complexity of experimental information about a possible Kraichnan transition. 

The Grenoble/Lyon group undertook several investigations in an attempt to find an explanation for the differences between the various data sets for $\Ra \agt 10^{12}$. For instance, \cite{chi04f} developed a model that attempted to explain the difference in terms of a finite plate-conductivity effect (see Sec. \ref{sec:corrections});  but measurements with relatively low-conductivity  brass plates by \cite{roc05} yielded results comparable to the high-conductivity copper-plate results. In a separate experiment \cite{roc01} made measurements using helium in a sample cell with $\Gamma = 0.5$ with walls and plates that were covered completely by grooves. The depth of the grooves was stated to be less than the thermal boundary layer thickness. Such a geometry is asserted to remove the influence of the sublayer which is responsible for the logarithmic corrections. For $\Ra \agt 10^{12}$ this experiment yielded an exponent quite close to 0.5, consistent with the expected Kraichnan value  of 1/2. However, 
\cite{nie06} pointed out that the BL thickness decreases with increasing $\Ra$ and becomes comparable to the groove depth in the $\Ra$-range of the measurements. In such a case the measurements of \cite{du00} using a sample with grooves in the plates that were deeper than the BL thickness 
indicate that the pre-factor of an effective power law describing $\Nu(\Ra)$ increases by as much as 76\% for deep grooves. Thus, it is  suggested by \cite{nie06} that  the results of  \cite{roc01} might possibly 
be due to a cross-over between rough surfaces with a groove depth less than the BL thickness to a regime where the groove depth is larger than the BL thickness. More work seems needed to resolve this issue.

An interesting  experiment related to the Kraichnan regime 
was done by  \cite{gil06}, 
taking up earlier experiments by \cite{per02} and experimentally
realizing the theoretically suggested 
homogeneous RB turbulence (\cite{loh03,cal06}).
\cite{gil06}
used a vertical channel with wide entrance and exit sections that avoided the influence of the thermal BLs on $\Nu$. They found  relationships for Nu and
Re (based on the velocity fluctuations)  
consistent with eqs.\ (\ref{eq3}) and (\ref{eq4})
when
they re-defined Ra in terms of an intrinsic $\Delta$-dependent length scale proportional to the ratio of temperature-fluctuation amplitudes and
 the vertical thermal gradient,
 instead of using a sample-geometry-dependent and $\Delta$-independent length. 
The same scaling was also  obtained by \cite{cho05,cho08} for 
 buoyancy driven turbulent exchange flow in a vertical pipe.
The flow was driven by an unstable density difference across the ends of the pipe, created using brine and distilled water. Away from either end, a fully developed region of turbulence existed with a linear density gradient. With a 
Rayleigh number based on the local density gradient 
relations consistent with eqs.\ (\ref{eq3}) and (\ref{eq4}) 
were found experimentally. 

Also {\it local} measurements of the heat flux can result in the 
1/2 exponent predicted by \cite{kra62}. 
\cite{sha08}, based on their
earlier measurements (\cite{sha03,sha04}),
determined a time-averaged local Nusselt number at fixed positions $\x$ and given by
\be
\Nu(\x) = \left< u_3 (\x , t)\theta (\x , t)\right>_t /(\kappa \Delta L^{-1})
\label{local_nu}
\ee
in a $\Gamma \approx 1$ cylindrical cell filled with water, 
in the range $ 10^8 \alt \Ra \alt 2\cdot 10^{10}$.
At half-height they found 
$\Nu( \x ) = 1.5 \Ra^{0.24 \pm 0.03}$
when $\x$ was close to the side-wall where the heat was primarily transported
by plumes, and
$\Nu( \x ) = 3.5 \cdot 10^{-4} \Ra^{0.49 \pm 0.03}$
when $\x$ was close to the cell center where the plume density was much 
less{\footnote{Note that in earlier work by \cite{chi04}
a less steep
scaling of $\Nu(\x)$ with Ra was derived at the center of the convection cell.
One possible source of discrepancy is that the association of the length
scale resulting from 
balancing buoyancy and viscous forces 
 with the thermal boundary layer thickness  done in that work does
not generally hold.}.
%
These two different
local scaling laws had been predicted
by \cite{gro04}; they correspond to the
 two independent scaling contributions 
$\eps_{\theta ,pl} = \kappa \Delta^2 L^{-2} \Nu_{side}$  
and 
$ \eps_{\theta , bg}=\kappa \Delta^2 L^{-2} \Nu_{center}$ 
to the 
thermal dissipation rate $\eps_\theta = \kappa \Delta^2 L^{-2} \Nu$ (see
equation (\ref{split3})).
The theory predicts that for $Ra \ltwid 10^{14}$ the heat 
transport via the flow close to the side-wall
dominates, but around $\Ra \approx 10^{14}$ the heat flux through the
center and thus the background fluctuations take over. 
Interestingly enough, when extrapolating the measured 
power laws for the
local Nusselt number in the center and close to the side-wall towards larger
Ra, also
\cite{sha08} obtain a crossover around $\Ra \approx 10^{14}$.
--
The spatial inhomogeneity of the local heat flux also has been confirmed 
by the numerical simulations of  \cite{shi07}. 
\section{Experimental measurements of the Reynolds numbers}\label{nurere}

\subsection{Reynolds numbers based on the large-scale convection roll}

The geometrical features and dynamics of the large-scale circulation (LSC) depend on the symmetry and aspect ratio of the sample. Here we focus on cylindrical samples with $\Gamma \simeq 1$ because for them the flow geometry is relatively simple, the dynamics is very rich, and the experimental studies are most extensive. Then the flow occurs in a near-vertical plane and yields near-elliptical stream lines, with the long axis of the ellipse slightly tilted relative to gravity (see, for instance, \cite{ver99,qiu04,sun05}), 
 as illustrated in the time-average velocity map 
Fig.~\ref{plumes}, bottom. Depending on $\Ra$, two small counter-rotating vortices positioned near the corners close to the minor axis of the ellipse may be more or less prominent. In the 2D numerical simulations of \cite{sug08} 
they can be very pronounced, depending on $\Ra$.

\begin{figure}
\begin{center}
\includegraphics*[height=5cm]{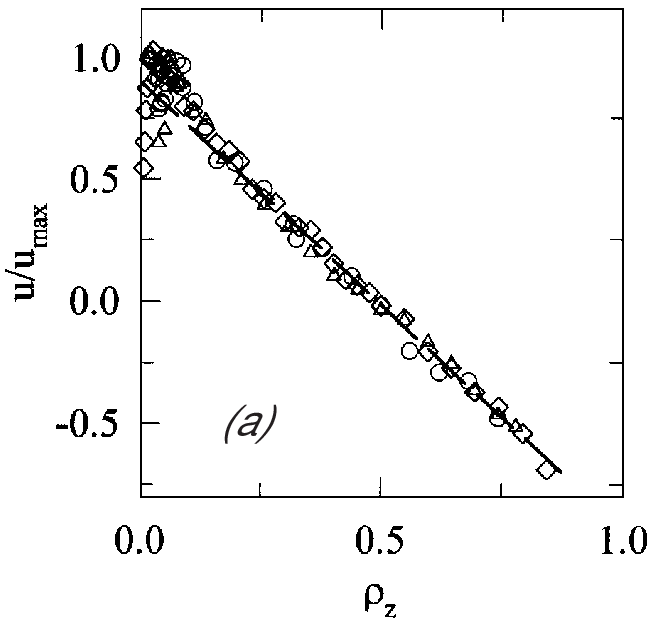}
\includegraphics*[height=5cm]{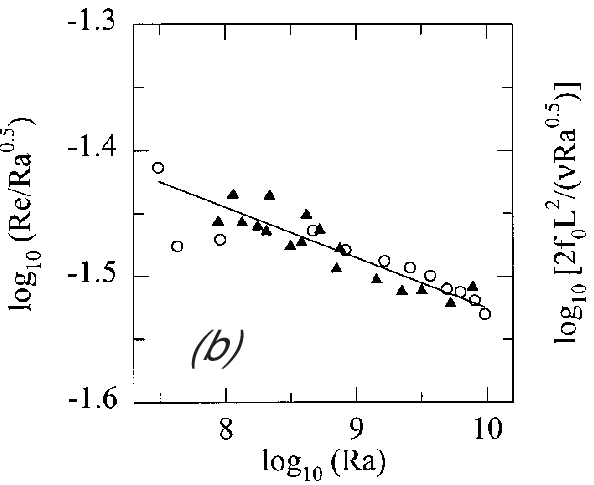}
\caption{
(a): Normalized horizontal velocity $u/u_{max}$ as a function of the vertical position $\rho_z = z/L$ (from \cite{qiu01a}). Note that the data extend only up to $\rho_z \simeq 0.85$; thus the minimum near the sample top is not shown.  (b): Reduced Reynolds numbers $\Re^u/\Ra^{1/2}$ (circles) and $\Re^\omega/\Ra^{1/2}$ (solid triangles), defined in the text, as functions of $\Ra$ (from \cite{qiu02}). 
}
\label{fig:LSC}
\end{center} 
\end{figure}                      

Often it is said that the large-scale
 convection roll is driven by the localized volumes of hot or cold fluid known as plumes that are emitted from the top and bottom thermal BLs as a result of a BL instability, and no doubt these plumes play an important role. But in addition some of the heat current is {\it conducted} across the BLs, warming or cooling the fluid adjacent to them. This in itself, without the presence of the plume inhomogeneities, would drive the flow as it does closer to the onset of convection where there are no plumes. Thus presumably the LSC derives its existence from a combination of these two heat-transport mechanisms with the relative importance of each depending on $\Pr$ and $\Ra$, and indeed it seems difficult to separate one from the other. For $\Pr \simeq 4$ it is known that, away from the BLs,  this heat current leads to a destabilizing time-averaged vertical gradient of the azimuthal average of the temperature  which is strongest near the side wall while the interior is more  nearly isothermal 
 (\cite{bro07b}). Superimposed upon this azimuthally-averaged gradient is the warm up-welling and cold down-welling current of the LSC which, when time averaged, leads to a near-sinusoidal azimuthal temperature variation with period $2\pi$ and  amplitude $\delta$ at the side wall (\cite{bro07b}), and a near-linear temperature variation along a diameter (\cite{qiu01a}).  

Although the speed of the flow shown in Fig.~\ref{fig:LSC}a varies considerably with position, one might expect that the large central roll can be described by a unique turnover time $\cal T$. Using $L$ as a relevant length scale, one can define a Reynolds number 
\be
\Re^{LSC} \equiv \frac{2L^2}{\cal T \nu}\ .
\label{eq:Re}
\ee
We would not expect $\Re^{LSC}$ to describe all aspects of the flow field; for instance the small counter-rotating vortices might require a different Reynolds number which might even have a different dependence on $\Ra$; but the main features, for instance those predicted by the GL model, might be related to $\Re^{LSC}$. Further, it has been suggested recently that the  actual path length of the  LSC circulation varies with $\Ra$ because the  shape of the flow field changes; this feature would introduce an additional $\Ra$ dependence of $\Re^{LSC}$ because the length scale, set equal to $2L$ in Eq.~(\ref{eq:Re}), would no longer be a constant (\cite{sun05d,nie03b}).

To our knowledge there are no direct measurements of $\cal T$. However, $\cal T$ was inferred from local measurements of various velocity components (see Fig.~\ref{fig:LSC}a for an example) and the assumption of a constant circulation path-length proportional to $L$. The time-averaged maximum vertical velocity  component $v_{max}$  (\cite{qiu01b,lam02})  near the region between the viscous boundary layers and the bulk of the system (sometimes known as the ``mixing zone" (\cite{cas89}) gave $\Re^{v_{max}} \propto v_{max}L/\nu$. Alternatively, the slope $\gamma_u$ in the sample interior away from the mixing zone of the time-averaged horizontal component $u$ of the LSC velocity as a function of the vertical position along the sample axis  (see Fig.~\ref{fig:LSC}a) was used to define and determine $\Re^u = \gamma_u L^2/\nu$ (\cite{qiu01b,qiu02}) (open circles in Fig.~\ref{fig:LSC}b).

An interesting property of the LSC is a torsional azimuthal oscillation mode with frequency $f_0 = \omega_0/2\pi$ (\cite{fun04,FBA08}) that can be used to define $\Re^{\omega} \equiv 2L^2f_0/\nu$. Well before the spatial nature of this mode was known, its frequency was measured in numerous single-point determinations of the temperature or the velocity 
(\cite{hes87,cas89,cil96,xin96,xin97,tak96,cio97,qiu00,qiu01a,qiu01b,qiu02,lam02,qiu04}), 
both of which have an oscillatory contribution provided the probe is not located in the horizontal mid-plane of the sample where the amplitude of this mode vanishes 
(\cite{FBA08}).  Some of the single-point measurements yielded results for $\Re^{\omega}$ that were equal to $\Re^u$  within experimental resolution 
(\cite{qiu02}), as seen in Fig.~\ref{fig:LSC}b. The reason for this equality is not known at this time. Some other experimental investigations indicated that there is a distinct difference between the $\Ra$-dependence of $\Re^{u_{max}}$ and $\Re^{\omega}$  (\cite{lam02}) and that these $\Ra$ dependences change with $\Pr$. Clearly there remain some unresolved issues.\footnote{A reliable evaluation of the experiments is complicated by the fact that experimentalists often slightly tilt their samples so as to obtain a dominant LSC orientation for their measurements. Measurements of $\Re$ are remarkably sensitive to the sample alignment relative to gravity (\cite{ahl06b}), and in a tilted sample $f_0$ can be observed even at the horizontal midplane.}

\subsection{Reynolds numbers based on plume motion}
An estimate of another $\Re$ was based on the motion of plumes. When a cold or warm plume passes a local temperature probe, then it produces a positive or negative deviation of the local temperature from the mean.
The local time-averaged vertical plume speed $v_{pl} = l/t_{pl}$ was thus determined from the peak location $t_{pl}$  of time cross-correlation functions between two temperatures measured with probes separated vertically by a small distance $l$ 
(\cite{san89,cas89,tak96,cha97,nie01,cha01}). 
The measurements yielded $\Re^{v_{pl}} \propto v_{pl} L/\nu$. 
A similar technique used thermometers mounted in the side wall at the horizontal midplane. A single temperature sensor yielded  a time auto-correlation (AC) function with a broad peak at a delay time corresponding to the plume turnover time ${\cal T}_{pl}$ (\cite{ahl06b,bro07c}).
Similarly,  two sensors on opposite sides of the sample yielded  time cross-correlation (CC) functions with broad peaks corresponding to half the plume turnover time ${\cal T}_{pl}/2$. 
These measurements gave Reynolds numbers
$\Re^{pl} \equiv (2L/{\cal T}_{pl}) (L/\nu)$. 
Quite remarkably, measurements indicate over a wide parameter range that 
$\Re^{v_{max}} = \Re^u = \Re^{pl}= \Re^{v_{pl}}$ 
within fairly small experimental errors. 
This can be interpreted to mean that the plume circulation is slaved to the LSC, or vice versa, and that all of these quantities (where they agree with each other) yield a reliable representation of $Re^{LSC}$.

The cross-correlation functions for two thermometers at the mid-plane but on opposite sides had extrema that were {\it negative}, indicting a correlation between the warm rising plumes on one side and the cold falling plumes on the other. This tends to support the ideas of \cite{vil95}, who suggested that  a plume impinging on the BL causes an instability and an associated emission of a new plume of the opposite type. However, the width of the cross-correlation function extremum, which was not very different from ${\cal T}_{pl}$, indicates that this is not a periodic process as had been originally suggested. The periodic signal, when observed in some experiments, presumably is due to the torsional oscillation discussed above and in Sect.~\ref{sec:global}.

\begin{figure}
\begin{center}
\includegraphics*[height=10cm]{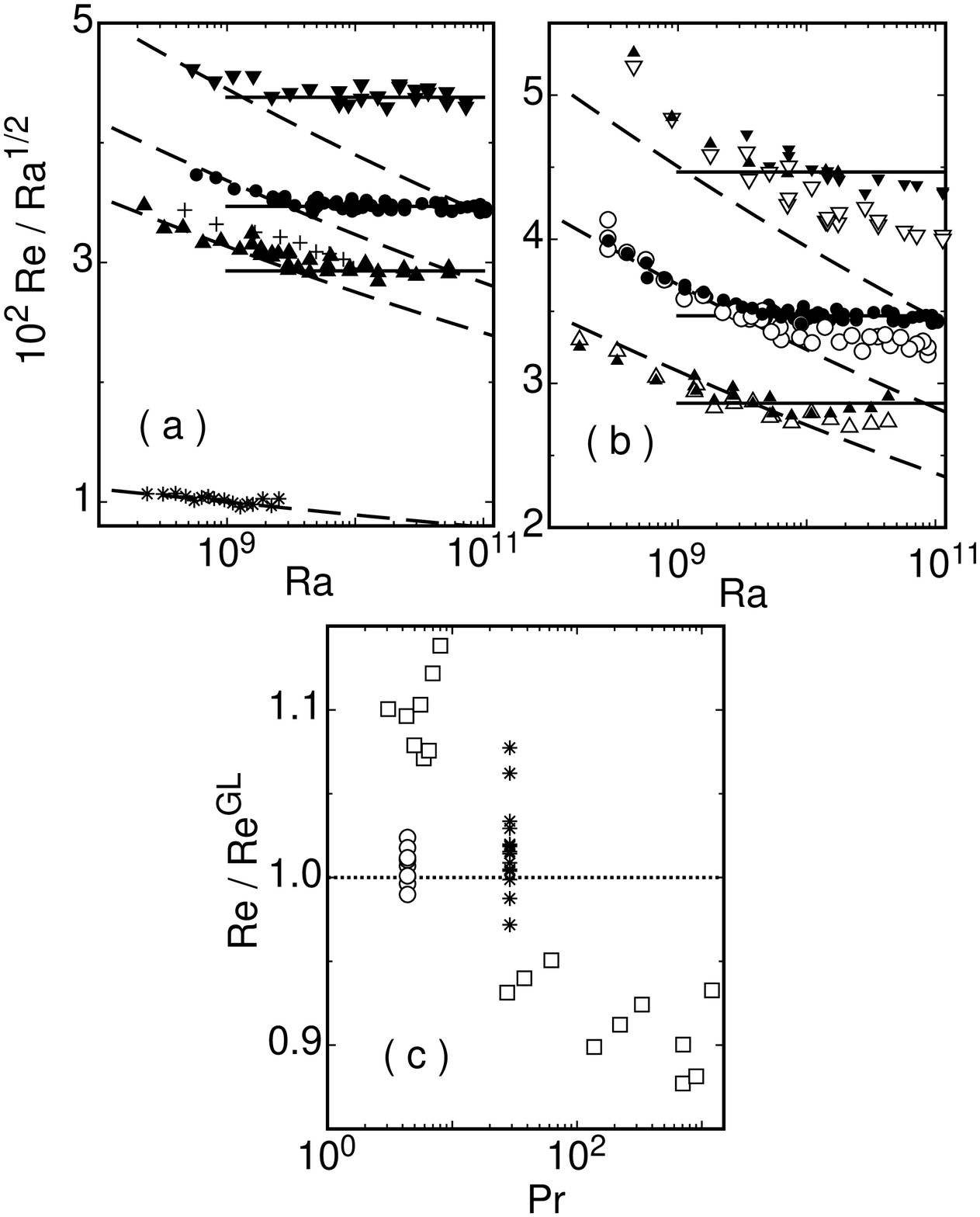}
\caption{(a) and (b): Reduced Reynolds numbers  $\Re/\Ra^{1/2}$. (a): $\Re^{pl}$ for $\Pr = 4.38$ (circles), $\Pr = 5.55$ (up-pointing triangles), and $\Pr = 3.32$ (down-pointing triangles) from \cite{bro07c}.  Stars: $\Re^\omega$ for $\Pr = 28.9$  from \cite{bro07c}.   Plusses:  $\Re^u$ from \cite{qiu02}, $\Pr \simeq 5.4$. 
Dashed lines (from top to bottom): GL predictions for $\Pr = 3.32, ~4.38, ~5.55,$ and 28.9. (b): $\Re^{pl}$ (solid symbols) and $\Re^{\omega}$ (open symbols) for $\Pr = 4.38$ (circles), $\Pr = 5.68$ (up-pointing triangles) and $\Pr = 3.26$ (down-pointing triangles). Dashed lines (from top to bottom): GL predictions for $\Pr = 3.26, ~4.38, ~5.68$. (c): The ratio of $\Re$ to the GL prediction as a function of $\Pr$ for $\Ra \leq 3\times 10^9$. 
Open circles: $\Re^{pl}$,  $\Pr = 4.38$. Stars:  $\Re^{\omega}$, $\Pr = 29.8$. Squares: $\Re^{\omega}$ from  \cite{lam02}, $5.6 \leq \Pr \leq 1206$.
}
\label{fig:Re}
\end{center} 
\end{figure}                      

For modest $\Ra$, say $\Ra \alt 3\times 10^9$ when $\Pr \simeq 4$, the GL prediction (\cite{gro02}) is in  good agreement with experimental results 
(\cite{qiu02,bro07c}) 
for $\Re^{u}(\Ra,\Pr)$, $\Re^{\omega}(\Ra , \Pr)$, and $\Re^{pl}(\Ra, \Pr)$ 
(see Fig.~\ref{fig:Re}). In that parameter range all three measured quantities, so far as they have been determined,  agree with each other. In that regime both the prediction and the experiment can be well
described  by a power law $\Re \propto \Ra^{\gamma_{eff}}$ with a very slightly $\Ra$-dependent effective exponent $\gamma_{eff} \simeq 0.44$ (there also are some notable differences between the predictions and some other measurements (\cite{lam02})).
However, for $\Ra \agt 3\times 10^9$ the measurements reveal a relatively sudden transition to a new state of the system, with a $\Re^{pl}$ that is larger than $\Re^{\omega}$, which in turn is larger than the GL prediction.
 These results agree quite well with recent experimental results for $\Re^{pl}$ obtained by \cite{sun05d} in the range $7\times 10^{10} \ltwid Ra \ltwid 2\times 10^{12}$. If the GL model still correctly predicts the $\Ra$-dependence of $\Re^{LSC}$ in this parameter range, then the experimental results for $\Re^{pl}$ and $\Re^\omega$ suggest the existence of a new LSC state in which the plumes and the LSC are  no longer slaved to each other and where the twist oscillation of the LSC is no longer synchronous with the LSC turnover time (see also \cite{xi06}). The observation that $\Re^{pl} > \Re^{LSC}$ would then indicate that the plumes rise and fall  more rapidly than the background flow of the LSC. However, 
\cite{nie03} and \cite{sun05d} suggested that it is
 more likely that the LSC evolves into a more complex flow structure where its dynamics can no longer be described quantitatively by a uniquely-defined Reynolds number. It is unclear at present whether the difference between this state and the one at smaller $\Ra$ will be found in the geometry of the flow, in the nature of the viscous boundary layers that  interact with it, or in the nature and frequency of plume shedding by the thermal boundary layers adjacent to the top and bottom plates. 

The dependence of $\Re^{\omega}$ on $\Pr$ has been investigated over a wide range of $\Pr$  by \cite{lam02}. The data for $\Ra \alt 3\times 10^9$ are shown as squares in Fig.~\ref{fig:Re}c. Their agreement with the GL prediction is not as good as one might wish. However, at $\Pr \simeq 4$, they also differ from the measurements of  $\Re^{pl}$ by \cite{bro07c}. More work over a wide $\Pr$ range seems desirable.

\section{Nu(Ra,Pr) and Re(Ra,Pr) in direct numerical simulations}\label{nuredns}

Direct numerical simulations (DNS) of Rayleigh-B\'enard flow
have several advantages in comparison to experiments: (i) Any local or 
global quantity can be ``measured'' without interfering with the probe or 
having restricted accessibility. (ii) The boundary conditions
can be chosen as in the idealized RB case, i.e., with exactly zero heat
flux through the side-walls and constant temperatures at the top and bottom
plates. Therefore, a modeling of side-wall and plate corrections (see section~\ref{sec:corrections}) is not 
necessary. (iii) The material properties of the fluid can be chosen at
will. E.g., one can study fluids which exactly obey the Boussinesq 
approximation, or fluids which show temperature 
dependence of one material property only.
By doing simulations with such fluids, 
theoretically suggested mechanisms can be tested.

However, obviously there are also major disadvantages of DNS, 
the main one being the limitations in the obtainable sizes of Ra and Pr. For fixed $\Pr \approx 1$
the required CPU time increases roughly as $ \Ra^{3/2} \log \Ra$.
The spatial resolution requirements for thermal convection have been defined by \cite{gro82,gro83}. 
In some simulations good small-scale resolution, which is particularly crucial in the BLs, 
is sacrified for higher Ra. 

In this DNS part of the review we again will focus on global properties
such as Nu and Re and the dynamics of the wind, omitting DNS results for the
small scales.

The RB simulations by \cite{delu90,wer91}, and \cite{wer93} were
restricted to two dimensions (2D) and employed free-slip (but impermeable) 
velocity boundary conditions at the side-walls, to make efficient use of
pseudo-spectral methods. Note that free-slip side-wall BCs lead
to a different large-scale circulation (LSC) pattern than no-slip side-wall BCs.
In particular, the counter rolls in the corners of the cell are suppressed.
With this method,
\cite{delu90} carried out simulations up to $\Ra = 1.6
\cdot 10^8$ for $\Pr = 1$, finding an effective Nu vs Ra power-law exponent
consistent with 2/7 at the largest Ra.
\cite{wer93} extended these 2D simulations to $\Pr = 7$.

3D simulations in the early 90s -- also employing spectral methods --
could only achieve relatively small Ra,
e.g.\ 
$\Ra = 6.5 \cdot 10^6$ 
in the work by \cite{sir89} and \cite{bal89} or
$\Ra = 6.3 \cdot 10^5$ 
in the work by \cite{chr92}; 
both are too small to make statements on 
exponents for power-law dependences on Ra, and also because there are still
coherent flow structures. 

One of the first 3D RB simulations from which such exponents
could be extracted was that by \cite{ker96}. He employed lateral periodic
BCs and achieved $\Ra = 2 \cdot 10^7$
on a 288 x 288 x 96 grid. For Pr=0.7 he found an effective power law
$\Nu \sim Ra^{0.28}$.
Later, \cite{ker00} extended these simulations to Prandtl numbers
in the range $0.07 < \Pr < 7$, finding $\Nu \sim \Ra^{0.25}$ for the
smallest Pr and effective exponents 
consistent with 2/7 for Pr $\ge 0.7$. 
For the Prandtl dependence of $\Nu$ at fixed $\Ra = 10^7$ and $\Pr \ltwid 0.7$
(see figure \ref{fig:Nred_of_Pr}, top), 
\cite{ker00} give an effective exponent of 0.12.
A more recent example of a spectral RB code is the work of 
\cite{har03}, achieving $\Ra = 10^7$. In that work the focus was 
on the flow organization in RB convection with sidewise periodic boundary 
conditions. 

A second class of DNS for RB convection is that of Lattice-Boltzmann 
(LB) simulations (\cite{ben94b}).
In one of the first large 
LB RB simulations \cite{ben98} achieved
$\Ra \approx 3.5 \cdot 10^7$ and found  $\gamma_{Nu} = 0.283\pm 0.003$
for the effective Nu vs Ra power-law exponent (they used $\Pr=1$ and free slip on vertical walls).


The third class of simulations contains DNS based on finite-difference or finite-volume.
\cite{ver99}, building on their earlier work (\cite{ver97,cam98}),
 employed such finite-difference simulations to 
obtain effective power-law exponents for both Nu and Re vs Ra for several 
Pr for Ra up to $2\cdot 10^7$ (see figure \ref{fig:Nred_of_Pr}a).
The advantage of this method is that the correct no-slip BCs at the
side-wall can easily be implemented and that simulations can also be
performed efficiently in the cylindrical geometry as used in many experiments,
so that a direct comparison with the experimental results is possible.
A further advantage of  finite-difference or finite-volume methods 
is that the 
spatial resolution can easily be refined at will in the BLs, using a coarser grid in the bulk.
For the small Prandtl number $\Pr = 0.025$ and $\Gamma = 1$
the effective power law $\Nu \approx 0.119 \Ra^{0.25}$ 
was found in the regime $5 \cdot 10^4 \le \Ra \le 10^6$.
For $\Pr = 0.7$ \cite{ver99} obtained
the effective power law $\Nu \sim \Ra^{0.285}$. These authors also 
explored the Pr dependence of Nu over the range  $0.0022 \le \Pr \le
15$ at fixed $\Ra = 6 \cdot 10^5$, finding 
$\Nu \approx 8.5 \Pr^{0.14}$ 
as an effective power law. 
For the Reynolds number they found 
$\Re \sim \Pr^{-0.73}$ for $\Pr < 1$ 
and $\Re \sim \Pr^{-0.94}$ for $\Pr > 1$. 
Similarly, \cite{bre04} performed finite volume integrations of the
Boussinesq equations, finding 
$\Nu \sim Pr^{0.182\pm 0.012}$ and $\Re \sim Pr^{-0.607\pm 0.012}$ for
$10^{-3} \le \Pr \le 1$ and
$\Nu \sim Pr^{0.032\pm 0.003}$ and $\Re \sim Pr^{-0.998\pm 0.014}$ for
$1 \le \Pr \le 10^2$, for no-slip boundary conditions and
$\Ra = 10^6$.

\cite{ver03} and \cite{str06} extended the earlier
 simulations to the remarkably high $\Ra = 2 \cdot 10^{11}$
for a slender $\Gamma = 1/2$ cell and $\Pr = 0.7$. Beyond $\Ra = 10^9$
the numerical data are consistent with $\Nu \sim Ra^{1/3}$. The focus of those papers is on 
the flow organization in the $\Gamma = 1/2$ cell: Beyond $10^{10}$ and 
for $\Pr \approx 0.7$ the single convection roll can break up
into two smaller counter-rotating rolls, each approximately of aspect 
ratio one.
\cite{str06} in addition showed that the thermal properties of the
side-walls can stabilize the large-scale convection roll.

At present the largest-$\Ra$ DNS of RB flow are the ones
by \cite{ama05} and by \cite{ver08}, achieving $\Ra = 2 \cdot 10^{14}$ for
$\Gamma = 1/2$ and $\Pr = 0.7$. The focus here is 
on the difference between constant-temperature and constant-flux boundary
conditions at the top and bottom plates, which already have been discussed in 
subsection \ref{sec:corrections}. Similarly, side-wall corrections
(\cite{ver02}) and plate corrections (\cite{ver04}) have been studied, 
which also already
have been discussed in the same above subsection.

Another advantage of finite-difference simulations is that 
complicated geometries like those with rough walls can 
be treated, see \cite{str06b}.
The results -- an enhanced heat flux consistent with $\Nu \sim \Ra^{0.37}$
over the range $2 \cdot 10^9 \le \Ra \le 2\cdot 10^{11}$ --  
are in reasonable agreement 
with the
experimental results of \cite{qiu05}, who found a power law exponent of
0.35 in the range $10^8 \le \Ra \le 10^{10}$, but in conflict with
the earlier mentioned results by \cite{du00}, who found an unchanged 
effective power law exponents for the rough wall case, but a larger prefactor. 
Similarly,
\cite{ver03b} found that by manipulating
the velocity BC at the plates, the viscous BL could be affected and
also the absolute value of Nu, but its power-law exponent was rather robust against
such manipulations. 
Clearly, more research is necessary to clarify this matter.

The numerical simulations of \cite{shi06,shi07,shi08} 
also are based on a finite
volume scheme; they focus on the role of plumes and the flow
organization in 
RB flow with $\Pr = 0.7$ and $\Pr=5.4$ and $\Ra$ up to $2\cdot 10^{9}$. 
For $\Ra = 2\cdot 10^{10}$ large-eddy simulations (LES) were done.
In this subgrid modeling the
length scales in the dissipative and diffusive 
regime are under-resolved.

Finally, when full spatial resolution of the turbulence
field was abandoned even further, obviously much larger Ra could be achieved.
An example for such a calculation is the so-called 
transient Reynolds averaged Navier-Stokes (RANS) method, which 
\cite{ken02} applied to RB convection. These authors find
$\Nu \sim \Ra^{0.31}$ between $\Ra= 10^5$ and $10^{15}$. For the 
last two $\Ra$ decades some reorganization of the plumes
and a slightly enhanced Nusselt number were observed, but given
the progressively decreasing spatial resolution of the numerical
scheme at these high Ra the implications of this finding are presently 
unclear and a detailed discussion of RANS simulations and LES  of RB flow is beyond the scope 
of this review.

Given the enormous CPU power needed to achieve large Ra,
one may wonder whether two-dimensional simulations would not 
be  sufficient to
reflect at least some aspects of 
the dynamics of the three-dimensional RB problem. 
This point has been analyzed in detail by 
\cite{sch02,sch04} whose conclusion is that for $\Pr \ge 1$ 
various properties observed in numerical 3D 
convection (and thus also in experiment) are indeed well reflected in  2D simulations. 
This in particular holds for the BL profiles and for the Nusselt number.
Also \cite{sug07,ahl08,sug08} employed 2D numerics to study the non-Oberbeck-Boussinesq deviations of the
Nusselt number and of the bulk (central) temperature from that in the Oberbeck-Boussinesq cases, 
see section \ref{sec-nob}.

\section{Boundary layers} \label{sec:bl}

\subsection{Relevance of boundary layers and challenges}

Boundary layers describe the temperature and flow fields in the vicinity of the plates and walls.
They are
 characterized by their time-averaged profiles in the direction perpendicular to the respective solid boundary, i.e.,
 in the $z$-direction off the bottom and top plates or in the $x$-direction off the side walls.
As the top and bottom 
boundary layers contribute the main resistance for the heat transfer 
through the cell and thus dominantly determine the Nusselt number, 
they deserve special attention. Indeed, nearly all theories 
of Nu(Ra,Pr) in RB convection are in essence boundary-layer theories:
This holds for the now classical mixing-layer theory of \cite{cas89},
the turbulent-BL based scaling theory of \cite{shr90} and extensions
thereof (e.g.\ \cite{chi97}), the turbulent BL type 
theories of \cite{dub01,dub02b}
and of \cite{hoe05,hoe06}, and the theory of \cite{gro00,gro01,gro02,gro04} which
scaling-wise builds heavily  on the laminar, though time dependent 
Prandtl-Blasius BL theory as elaborated in subsection \ref{nuregl}. 
It is therefore of prime importance to {\it directly} study the 
BLs, to see whether the assumptions on which these theories are based are
fulfilled at least to a reasonable approximation.

Unfortunately, directly studying the BLs in the large-Ra regime
is equally challenging
experimentally, numerically, and theoretically. Experimentally,
the required spatial resolution is very difficult to achieve. Even
for the $L \approx 6.5$ m high Ilmenau RB barrel 
both the thermal and the kinetic BL thicknesses
are only a few millimeters when e.g. $\Ra \approx 10^{12}$, see \cite{pui07b,pui07}. 
In addition, both laser-Doppler velocimetry (LDV) and hot-wire anemometry
are notoriously difficult to employ in regions with large temperature
fluctuations and a small mean-flow velocity; both complications occur in the BLs. 
Numerically, large Rayleigh numbers are hard to reach, 
in particular with sufficient spatial resolution, see section \ref{nuredns}. Even state-of-the-art 
DNS such as those of \cite{ver08} have only a few gridpoints in the BLs. Theoretically, 
no generally accepted BL theory for flow over a strongly heated surface 
exists. Even in the truely laminar case an analytic theory for the two-way 
coupling 
of the 
temperature to the velocity field is missing
(i.e., taking the temperature as an {\it active} scalar), 
see e.g.\ \cite{sch00}. And even if the two-way coupling is suppressed
as done in the interesting numerical work by
\cite{chi01a} and \cite{chi02}, no exact analytical 
results exist for the thermal BL thickness
and thus for the heat transfer through the shear flow over the heated plate.

The analysis of the BLs is further complicated by their extreme
complexity, not only in time, where there is plume detachment (see figure \ref{plumes}),
but also in space. Recent experimental studies such as those by 
\cite{lui98,wan03,may06}, and \cite{sun08} and 
numerical studies such as those by \cite{sug08} give increasing evidence that
the  BL thicknesses as well as the 
profiles of the mean quantities and of the fluctuations 
depend on the position along and above the plate(s),
not only relative to the walls, but also relative to the fluctuating
large convection roll (see section \ref{nurere}). It is this main convection roll 
which creates the viscous BL because of the no-slip condition at the top and
bottom plates.

The aim of this section is to give an overview of  what has been found experimentally,
theoretically, and numerically about the BL thicknesses.

\subsection{Thermal boundary layers}
The thermal boundary layer thickness can be defined in several ways.
From an experimental point of view, it is easiest to
 time-average the temperature profile
at fixed lateral positions $(x,y)=*$ for various $z$ and to extract
a thermal boundary layer thickness $\lambda_\theta (*)$
from the resulting local profile.
From the 
theoretical viewpoint
area-averages on $x,y$ (in addition to the time-averaging) 
are of main interest. Indeed, it is  the thickness 
$\lambda_{\theta}$ of the 
{\em area-averaged} temperature profile in the $z$-direction which enters theory 
while comparison with experiment so far refers to locally 
measured BL thicknesses $\lambda_{\theta}(*)$.
Note that strictly speaking the global thickness $\lambda_{\theta}$ is 
not equal to
$\langle \lambda_{\theta}(*) \rangle_A$ in general; the reason is that the thickness is a (nonlinearly defined)
property of a given profile. Then 
$\lambda_{\theta}$ in particular is a property of the area-averaged
profile and not the average of properties $\lambda_{\theta}(*)$ of the respective local profiles. Along large plates 
the local BL width may well represent the area-averaged global one; but for RB samples with $\Gamma$ of order 1 
one has to expect significant differences, even different scaling of the global and the local thicknesses, 
since the area-averaged profiles comprise all 
inhomogeneities of the temperature and flow fields along the plates, including backflows, the regions between them, 
and the near wall ranges. These all will locally have quite different profiles. The more homogeneous the flow,
the more similiar
 the global width will be to the average of the local widths. Experimental analysis of all these 
details is still a big challenge.        

The (global) thermal BL thickness $\lambda_\theta$ can be defined in several ways.
The most popular way is to define $\lambda_\theta$ through the
slope of the area-averaged time-mean temperature profile at the plates: 
Take the tangent of 
 the area-averaged mean temperature $\theta(z)$ at the plate.  
That distance between the plate and the vertical position
where this tangent crosses the bulk (or center) temperature $T_c$ is then called 
$\lambda_\theta^{sl}$. The center temperature $T_c$ is equal to the mean temperature 
$T_m = (T_t + T_b)/2$ if the Oberbeck-Boussinesq approximation holds because of up-down-symmetry. 
The very notion of $T_c$ already refers to an area-averaged profile. 

Locally, i.e., for fixed $(x,y) \equiv *$, the slope of the time-averaged profile, as well as the temperature 
for $z \rightarrow \infty$ and thus the local width $\lambda_{\theta}^{sl}(*)$,  will depend on the 
horizontal position $*$ where the local time-mean temperature profile is taken. 

One of the first measurements of  temperature and  also velocity
profiles in RB cells was done by \cite{til93}, namely in water ($\Pr = 6.6$)
at fixed  $\Ra = 1.1 \cdot 10^9$ and at a fixed lateral position $*$.
\cite{bel93} extended these measurements to the $\Ra$-range
$5\cdot 10^5 \le  \Ra \le 10^{11}$ in compressed 
gas (air) at room temperature ($\Pr = 0.7$), 
but still at  fixed lateral position.
\cite{lui98} measured the lateral dependence of
 $\lambda_\theta^{sl} (*)$ 
on the  
positions $x$ (in the mean LSC direction) and $y$ (perpendicular 
to the preferred LSC direction) in a cylindrical water filled RB cell with $\Gamma = 1$ in the regime 
$\Ra = 2 \cdot 10^8 - 2\cdot 10^{10}$. They report variations of the BL width by nearly
a factor of 2, depending on the lateral location where the profile was measured. \cite{wan03} found 
similar results for a cubic cell.
Even the scaling exponent of $\lambda_\theta^{sl} (*)$ with
$\Ra$ depends on the position and varies between $-0.35$ and
$-0.28$. \cite{lui98} found the thermal BL to be thinnest 
close to the center of the plates; there 
$2\lambda_\theta^{sl} (0,0)/L  = (0.23\pm 0.02) \Ra^{-0.285\pm 0.004}$ 
holds.

One cannot draw conclusions about the dependence of Nu on Ra from such measurements of local
$\lambda_\theta^{sl}(*)$ at a particular lateral location. It is the length scale resulting 
from the {\it laterally averaged} BL temperature profile 
which is connected with the Nusselt number through the exact relation 
\be
\Nu = {|\partial_z \left< \theta (x,y, z=0 \ \hbox{or} \, \, z=L)\right>_A |
\over {\Delta L^{-1}} } = 
{ L \over 2 \lambda_\theta^{sl} }
\label{nu-lambda}
\ee
which follows from equation (\ref{nu}).

A different definition for a thermal boundary-layer thickness is the
one employed by classical (laterally homogeneous) laminar BL theories 
 (\cite{pra04,bla08},
see also \cite{mek61,sch00,cow01}). 
This thickness, known as $\lambda_\theta^{99\%}$, is defined 
as the vertical
 distance from the plate to the point where $99\%$ of 
 the temperature 
difference between plate and 
  mean center temperature is achieved. 
E.g.\ for the bottom plate $\lambda_\theta^{99\%}$ is  the distance 
where the temperature reaches $T = T_b - 0.99 \Delta_b$. 
This definition is in analogy to the common definition of the thickness $\delta$ of the 
kinetic BL, defined by the distance where, say, 99\% of the maximum bulk velocity is achieved.
The profile-based thickness $\lambda_\theta^{99\%}$ can be calculated within
the truly laminar Prandtl-Blasius BL theory, which implies lateral homogeneity and thus yields local 
and area-averaged widths that are the same. One obtains
\be
\frac{\lambda_\theta^{99\%}}{L} =  \frac{C(\Pr)}{\Re^{1/2} \Pr^{1/3}}
\label{pohlhausen-thickness}
\ee
with a function $C(\Pr)$ given by \cite{mek61}. For large Prandtl numbers
$C(\Pr)$ becomes constant, whereas for small $\Pr$ one finds $C(\Pr) \propto \Pr^{-1/6}$.
We note that $\lambda_\theta^{99\%}$ can display different
scaling behavior than  $\lambda_\theta^{sl}$, both as a function
of Ra and of Pr, depending on the parameter-space regime. To our knowledge this issue has not yet been
explored systematically, neither for $\lambda_\theta^{99\%}$, nor for its local counterpart $\lambda_{\theta}^{99\%}(*)$
which depends on the lateral position $*$ and refers to the local, only
time-averaged, profiles. 

A third way to define a thermal BL thickness, suggested by \cite{bel94}, is to take the 
position of the maximal temperature fluctuations at the edge of the thermal BL. 
We call the respective BL thicknesses defined in this way $\lambda_\theta^\sigma$ and $\lambda_\theta^\sigma(*)$.
\cite{pui07} show that thermal BL thicknesses defined in this way
and measured in the Barrel of Ilmenau using air at one atmosphere as the fluid, behave differently from $\lambda_\theta^{sl}$. 
However, since BL thicknesses are determined largely by diffusive processes,
 it cannot be ruled out that a mixture like air (where both mass diffusion 
and heat diffusion play a role and where marginal stability depends both 
on concentration and temperature gradients) might in this respect behave
 differently from a pure fluid.

\begin{figure}
\begin{center}
\includegraphics*[height=5.5cm]{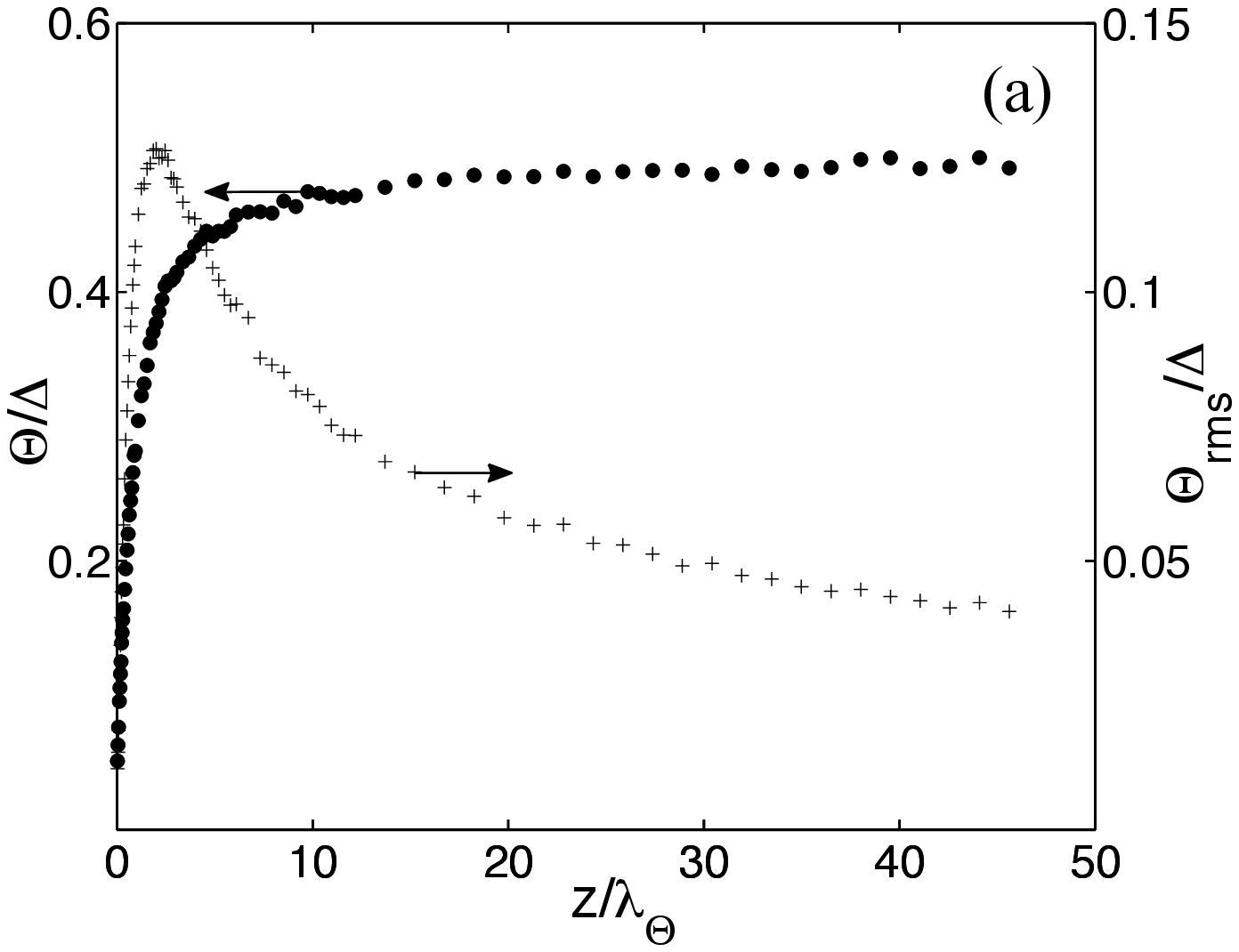}
\vspace{0.5cm}
\includegraphics*[height=5.5cm]{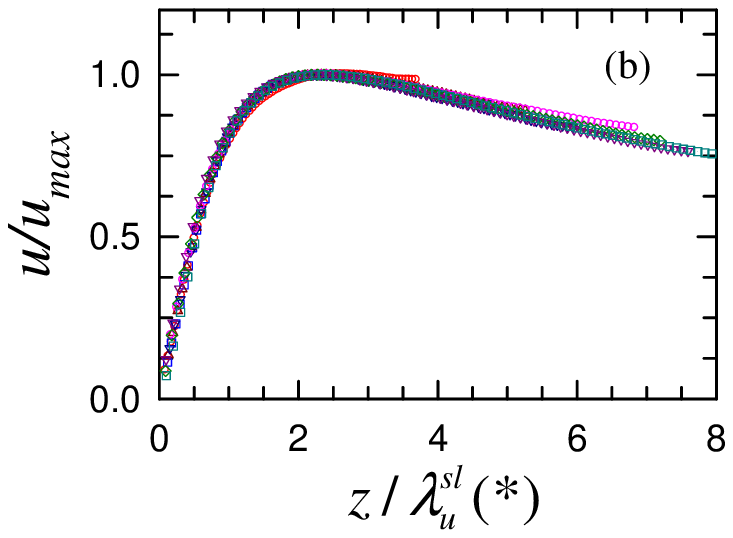}
\caption{
(a) Local, only time-averaged, temperature and temperature-fluctuation profiles
at $\Ra = 7.7 \cdot 10^{11}$ and $\Gamma = 1.13$ as functions of the height
z (normalized by $\lambda_\theta^{sl}$), 
measured below the center of the top plate.
The fluid is air at atmospheric pressure, the cell height is 6.3 m. 
From \cite{pui07}. 
(b)
Rescaled mean velocity profiles $U(z)$ above the center of the plate 
for 
$\Ra = 1.25 \cdot 10^9$ (shortest data set) up to 
$\Ra = 2.02 \cdot 10^{10}$ (longest data set) for RB flow in water.
$u$ is normalized by $u_{max}$ and $z$ by the 
kinetic BL thickness $\lambda_u^{sl}(*)$. 
Figure taken from \cite{sun08}. 
}
\label{bl-profiles} 
\end{center}
\end{figure}

\begin{figure}
\begin{center}
\includegraphics*[height=5.5cm]{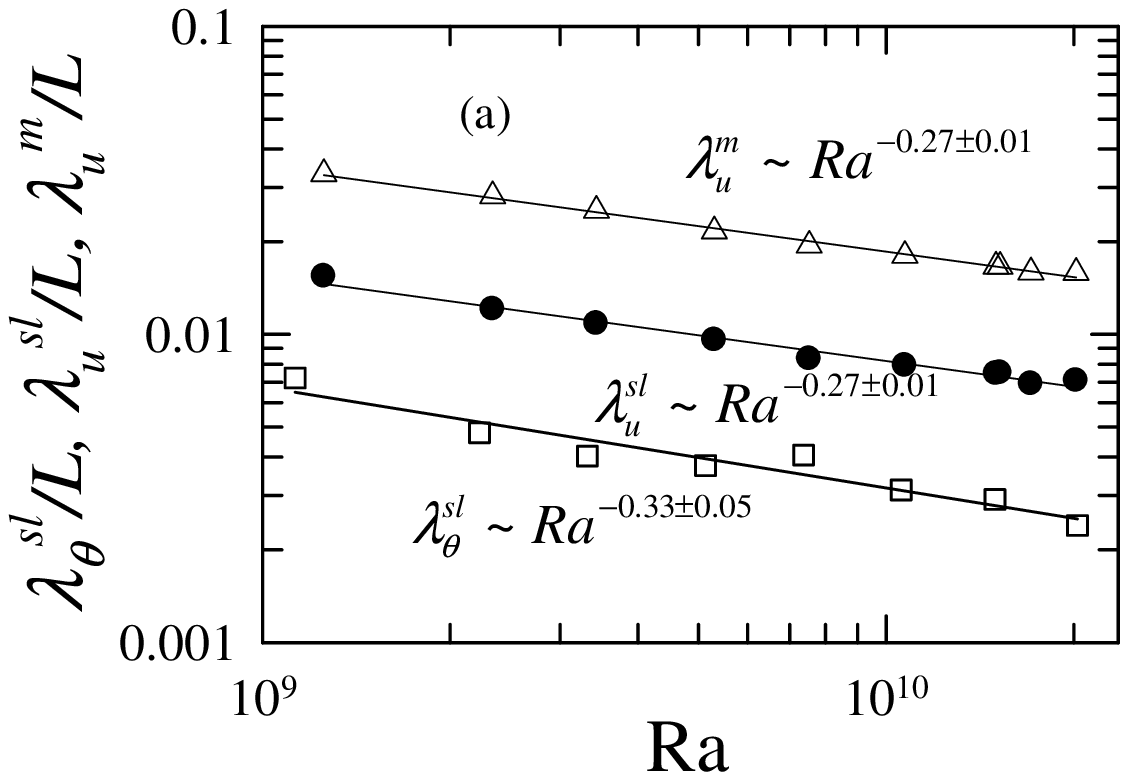}
\vspace*{0.9cm}
\includegraphics*[height=5.5cm]{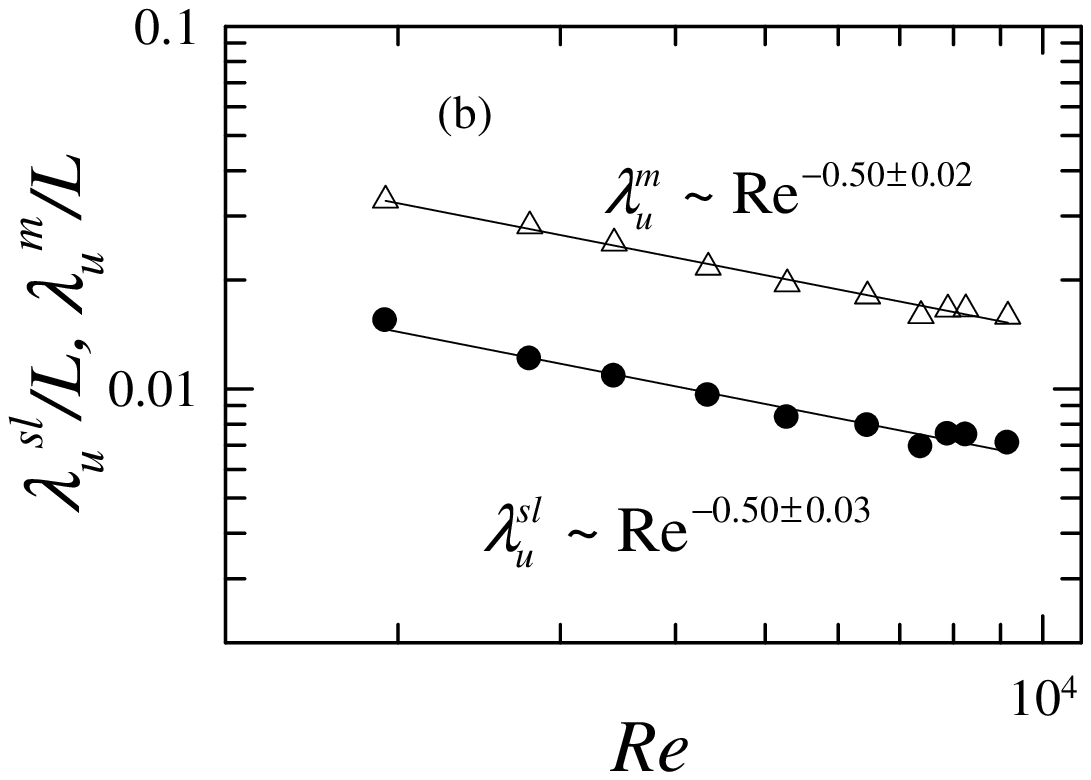}
\caption{
Scaling of the kinetic BL thicknesses $\lambda_u^{sl}(*)$ and $\lambda_u^m(*)$ 
as functions of (a) Ra and of (b) Re, taken above the center 
of the lower plate.
 Taken from \cite{sun08}.
In (a) also the scaling of the thermal BL thickness $\lambda_\theta^{sl}
 (*)$ is shown.
It has been deduced from the temperature profile above the center 
of the lower plate. 
The lines are power-law fits.
}
\label{bl-thicknesses} 
\end{center}
\end{figure}

Profile measurements, and thus the determination of BL thicknesses,
are obviously less difficult if the absolute length scale $L$ is larger. That is
why the about 6.5 m high barrel of Ilmenau is particularly suited for these purposes. 
\cite{pui07} measured local time-averaged temperature profiles
up to $\Ra = 7.7 \cdot 10^{11}$, for an 
example see figure \ref{bl-profiles}a. Outside a small linear regime near the plate
they obtain a  power law $\theta (z) \sim z^\alpha$ with $\alpha \approx 0.5$, 
only weakly dependent on Ra and $\Gamma$. 
For even larger $z$ the temperature saturated at its bulk value. The 1/2 power law for  intermediate heights 
supports neither the logarithmic scaling laws for turbulent BL theories nor the temperature profile
for a one-way coupled 
temperature field (i.e., taking the temperature as a passive scalar)
predicted by the 
(laminar) Prandtl-Blasius approach. However,
the power law $\lambda_\theta^{sl}(*)/L \propto Ra^{-0.254}$ found by \cite{pui07} below the center 
of the top plate is not inconsistent with what follows from the scaling-wise laminar Prandtl-Blasius approach.
\cite{sun08} obtained $\lambda_\theta^{sl}(*)/L \propto Ra^{-0.33\pm 0.05}$
from temperature-profile measurements above the center of the plate
in water with Ra up to $ 2 \cdot 10^{10}$.

Thermal BL thicknesses also have been determined from DNS. For a $\Gamma = 1$
RB cell \cite{ver99} obtained for the area-averaged width 
$\lambda_\theta^{sl} = 3.1 Ra^{-0.29}$ for $Pr= 0.7$ over the range $5\cdot 10^5 \le \Ra \le 2\cdot 10^7$ 
and $\lambda_\theta^{sl} = 2.8 Ra^{-0.25}$ for $Pr= 0.022$ in the regime $5\cdot 10^4 \le \Ra \le 10^6$.
When going towards the larger Ra regime $2\cdot 10^6 \le \Ra \le 2\cdot 10^{11}$
\cite{ver03} find $\lambda_\theta^{\sigma} \sim Ra^{-0.31}$ for $Pr= 0.7$ and $\Gamma = 1/2$.
For even  larger $\Ra$ up to $ 2\cdot 10^{14}$ \cite{ver08} report 
$\lambda_\theta^{\sigma} \sim Ra^{-1/3}$.

The profiles of the rms temperature fluctuations are much less settled,
see e.g.\ \cite{lui98,fer02,wan03,pui07,sun08}. This issue will not be discussed in the present review.

As stated already in section \ref{nuregl}, thermal {\it plumes} can be 
viewed as detached pieces of the thermal BL. 
Since their discovery in RB convection by \cite{zoc90} they have
 been studied so intensively, -- experimentally, numerically,
and theoretically -- that a review of its own on this subject
would be justified. A number of recent papers were devoted to them 
(\cite{the98,the00,zho02,bre04,fun04,har04,xi04,put05,put05b,shi06,zho07,put08,shi08}).  
Their nature was discussed most recently by \cite{FBA08}. Here we only make a few remarks. They seem to 
originate as one-dimensional excitations of the marginally stable BL. As they are born out of the BL, 
they become oriented by the LSC with their long axis in the direction of the flow, and then are swept 
by the LSC toward the side wall. Along their way they separate from the BL and progress vertically near 
the wall, developing their famous mushroom top (\cite{zoc90,xi04,put05,zho07}) in this process. Although 
they are often referred to as `sheet-like", implying an extension in two spatial dimensions,
the term ``line-like" would more appropriately describe their length scale of order $L$ in one dimension 
and of order the thermal boundary-layer thickness 
in the other two.

\subsection{Kinetic boundary layers}

The kinetic BL thicknesses $\lambda_u$ are determined from the velocity profiles or the velocity-fluctuation profiles. Also $\lambda_u $ can be defined in various ways, 
starting either from time- and area-averaged quantities or only locally 
time-averaged ones. 
An additional alternative here is to average the 
velocity vectors themselves or to take the rms velocities, component wise or the full magnitude. 
Given the velocity profile of interest, its thickness may be defined
(i) via the slope of the velocity profile in the $z$-direction (for the BL
above or below the bottom or top plate)
at a given position $*$ on the plate, 
known as $\lambda_u^{sl}(*)$,  
(ii) through the distance $\lambda_u^m(*)$ to the local maximal mean velocity,
(iii) through the distance $\lambda_u^\sigma(*)$ to the maximal 
velocity rms fluctuations, etc. Again, all these quantities depend on the lateral position $* = (x,y)$, 
unless area-averages are considered. 

The first direct systematic measurements of velocity profiles in RB convection
as a function of Ra where done by \cite{xin96} and \cite{xin97}, using a novel light-scattering technique
in a cylindrical cell. They found $\lambda_u^{sl}(*)/L \sim  Ra^{-0.16}$ 
from the velocity profile above the center of the lower plate. 
At the side-walls at half height as functions of $x$ instead of $z$ they measured $\lambda_u^{sl}(*)/L \sim Ra^{-0.25}$.
\cite{qiu98,qiu98b} extended these measurements to convection in cubic cells, finding the same scaling 
exponents as in the cylindrical case. By using various organic liquids, \cite{lam02} could also explore
the Pr dependence over the range $6 \le \Pr \le 1027$
and $2\cdot 10^8 \le \Ra \le 2 \cdot 10^{10}$, finding 
$\lambda_u^{sl}(*)/L \sim \Pr^{0.24} \Ra^{-0.16}$ above the 
center plate position. The small $\Ra$-exponent of the effective 
power law is
remarkable, as for a Prandtl-Blasius area-averaged BL profile one would expect an exponent of
$-0.25$, just as measured at the side-walls, roughly corresponding to 
the Prandtl-Blasius scaling $\lambda_u^{sl}/L \sim  \Re^{-1/2}$ with (approximately) 
$\Re \sim \Ra^{1/2}$. We will come back to this point later.
Note that all these measurements were still done above a single 
position over the plate.

In 2003, particle image velocimetry (PIV) measurements 
revolutionarized the experimental analysis of the velocity field
in thermal convection (\cite{xia03,sun05,sun05a,sun06}), 
including the analysis of the kinetic BLs (\cite{sun08}).
This allowed not only for the direct identification of various Reynolds numbers, cf.\ Sect. \ref{nurere}, 
but also for that of various kinetic BL thicknesses such as $\lambda_u^{sl}(*)$,
$\lambda_u^{m}(*)$, or $\lambda_u^{\sigma}(*)$, even at different positions $*$ in the cell.

For $10^9 \le \Ra \le 10^{10}$ and $\Pr = 4.3$ in a rectangular cell
with $\Gamma = 1$, \cite{sun08} obtained time-averaged velocity profiles
in the center above the bottom plate. They found that if the vertical lengths are rescaled by $\lambda_u^{sl}(*)$
and the velocities by the maximal velocity, then the shape of the local velocity profile does not depend on 
$\Ra$; in this sense it is universal in that regime, see figure \ref{bl-profiles}b.
Moreover, \cite{sun08} find $\lambda_u^{sl}(*)/L\sim \lambda_u^{m}(*)/L \sim \Ra^{-0.27\pm 0.01}$
(see figure \ref{bl-thicknesses}a), in contrast to the much weaker Ra dependence reported earlier 
(\cite{xin96,xin97,qiu98,qiu98b}. The origin of this discrepancy is not clear at this point. However,
we note that the newly found scaling 
$\lambda_u^{sl}(*)/L\sim \lambda_u^{m}(*)/L \sim \Ra^{-0.27\pm 0.01}$
is consistent with the Prandtl-Blasius expectation. As a double check,
\cite{sun08} plotted $\lambda_u^{sl}(*)/L$ and  $\lambda_u^{m}(*)/L$ against
the independently determined Reynolds number, finding the Prandtl-Blasius
scaling $\lambda_u^{sl}(*)/L \sim \lambda_u^{m}(*)/L \sim Re^{-0.50 \pm 0.03}$, see
figure \ref{bl-thicknesses}b. 
In numerical simulations of RB flow for $\Pr = 1$ and $\Gamma = 1$ 
\cite{bre04} find $\lambda_u^{sl}/L \sim \Re^{0.44\pm 0.02}$,
but a weaker Re-dependence for $\lambda_u^{m}/L$. 
Numerical simulations by \cite{ver99}
 for the same $\Gamma = 1$ and $\Pr =0.7$  
gave an area-averaged profile thickness exponent consistent with the Prandtl-Blasius
BL theory, namely $\lambda_u^{sl} /L = 0.95 Ra^{-0.23}$.
For lower $\Pr = 0.022$ \cite{ver99} obtained
$\lambda_u^{sl}/L = 0.1 Ra^{-0.18}$.

The PIV BL study of \cite{sun08} also yielded the wall shear
stress $\tau_w$, the viscous sublayer lengthscale $y_w = \nu(z=0)/u_\tau$, 
the skin-friction velocity $u_\tau = (\tau_w / \rho )^{1/2}$, 
and the skin-friction coefficient $c_f = \tau_w / (\rho U_{max}^2)$.
(All these quantities have been measured above the plate center, but here
for simplicity we suppress the $*=(x,y)$-dependence in the notation.)
They found 
$\tau_w/(\rho \nu^2 /L^2)  \sim Re^{1.55}$,
$y_w/L \sim Re^{-0.91}$, 
$u_\tau/(\nu/L) \sim Re^{0.80}$, 
and 
$c_f \sim Re^{-0.34}$.
The respective Prandtl-Blasius scaling exponents for a laminar BL over
a flat plate are
$3/2$, $-1$, $3/4$, and $-1/2$. Apart from the last one all are in fair
agreement with the measurements. It is hard to judge the importance
of the discrepancy between $-0.34$ and $-0.50$ for the scaling of
the skin-friction coefficient, in particular because the last three
data points  around $\Re \approx 10^4$ show a steeper dependence
than $-0.34$, see figure 22d of \cite{sun08}. Moreover, 
in the numerical simulations of \cite{ver03} 
($\Gamma = 1/2$, $\Ra$ up to $2\cdot 10^{11}$)
the friction coefficient $c_f$ does show a Reynolds-number scaling-exponent consistent with $-1/2$. 
(That work did not find any scaling
for $\lambda_u^\sigma$ as function of Ra, presumably because of a
reorganization of the flow in the $\Gamma = 1/2$ cell with increasing Ra.)
Also in the hitherto largest numerical RB simulations of 
\cite{ama05} $c_f \sim Re^{-1/2}$ is found, see figure 6b of that paper.
Next, also the experimental analysis of 
\cite{cha01} finds a $-1/2$ scaling of a surrogate of the friction factor with 
$\Re$ up to $\Ra = 10^{11}$. However, at $\Ra = 10^{11}$, corresponding
to $\Re = 8 \cdot 10^4$, the data of \cite{cha01} give evidence for a transition 
towards a weaker dependence. \cite{ver03} find the same transition 
in this friction factor surrogate, but not in the actual friction factor
$c_f$. 

Finally, we would like to mention the interesting numerical study by
\cite{yu07}, who analyse the heat transfer and boundary-layer thicknesses
in {\it laminar} and  steady convection with $\Ra$ up to $10^8$, 
finding basically the same Ra-scaling laws for
Nu, $\lambda_u$, and $\lambda_\theta$ as in classical RB convection.
Their conclusion is that turbulence does not play a decisive role for 
the heat transfer.

In summary, it is fair to state that, although some of the various velocity BL profiles
differ from the idealized 
Prandtl-Blasius truly laminar profile due to the permanently ongoing plume emission,  the 
{\it scaling} of the kinetic BL thickness with Ra, Re, and Pr 
is consistent with the laminar Prandtl-Blasius
theory at least up to $\Ra = 10^{11}$, but presumably also beyond.
There is no indication of any transition towards a different kinetic BL 
thickness-scaling even at the largest Ra realized up to now.
Moreover, for $\Pr \gtwid 1$ the thermal BL is nested in the kinetic one,
whereas for $\Pr \ltwid 1$ it is the other way round.

The scaling of the rms velocity fluctuations (see e.g. \cite{sun08,pui07b} for recent work on this issue)
will not be discussed in this review.

\section{Non-Oberbeck-Boussinesq effects}\label{sec-nob}

The problem of RB convection is commonly analyzed within the
so-called Oberbeck-Boussinesq (OB) approximation (\cite{obe79,bou03}), in which  the fluid properties are assumed to be temperature independent, 
apart from
the density for which the
 linear temperature dependence 
 eq.\ (\ref{beta-def})
is assumed.
Under normal conditions, i.e.,\ ``small''  temperature
differences $\Delta$ between the bottom and top plates,
this approximation is rather good. However, 
in order to achieve ever larger values of Ra for given $L$ and fluid 
properties $\beta$, $c_p$, $\kappa$, and $\nu$, 
the temperature difference $\Delta$ between top and bottom plate 
quite frequently was increased to such an extent that
the OB approximation had to be expected to fail. Non-Oberbeck-Boussinesq (NOB) effects, i.e.,
 deviations of various properties including Nu and the center temperature $T_c$ from the OB case,
 then had to be expected at the largest Ra in several experiments (\cite{cas89,cha97,ash99,nie00,nie03,nie06b}). Particularly problematic were measurements in the vicinity of liquid-gas critical points, where $Ra$ tends to become exceptionally large but where NOB effects are to be expected at relatively small $\Delta$. 
To be able to interpret these high-Ra data -- in particular with respect
to the question of whether there is an intrinsic large-Ra transition in the
flow -- it is therefore of prime importance to understand the physical nature, the signatures, and 
the size of the NOB effects. In addition to this practical consideration, there is much interesting physics to be learned from their study. 

NOB effects in  high-Ra convection were measured and analyzed 
first using helium
gas by \cite{wu91a}, and then using water and glycerol by \cite{zha97}.
These authors confirmed experimentally that the temperature dependence of the fluid properties
leads to a symmetry breaking between the top and the bottom of the sample. 
The temperature drops $\Delta_b$ and $\Delta_t$
across the bottom and top boundary layers become different, 
$\Delta_b \neq \Delta_t$, and so do the thicknesses of the
thermal BLs, $\lambda_{\theta,b}^{sl} \neq \lambda_{\theta ,t}^{sl}$. 
Both phenomena are associated with a corresponding shift of the temperature $T_c = T_b - \Delta_b 
= T_t + \Delta_t$ in the center (bulk) of the cell
away from the arithmetic mean temperature $T_m = (T_b + T_t)/2$ 
of the bottom and top plate temperatures $T_{b}$ and $T_t$. 
Moreover,
 one would expect NOB observables to include deviations of Nu and Re from their OB values if $\Delta$ becomes large. 
Surprisingly, \cite{wu91a} and \cite{zha97} found that $\Nu$ is remarkably insensitive to NOB effects. To our knowledge NOB effects on Re have yet to be observed experimentally.

In order to assess the validity of the OB approximation, \cite{nie03}
suggested that the following three measures are basically equivalent:
Busse's weighted sum of fractional deviations 
$(X_t - X_b)/X_m$ for the relevant
material properties X, the ratio $\chi = \Delta_b /\Delta_t$ 
(introduced by \cite{wu91a}) of the temperature drops
across the bottom and top BLs, and the relative change of the density 
$\beta \Delta$. The last criterion is  the simplest, and 
\cite{nie03} suggested on empirical grounds that OB conditions can reasonably be expected 
to prevail when $\beta \Delta$ is less than about 0.1 to 0.2.
We will see below that the situation is more complicated and that there
is a plethora of different NOB effects.

NOB effects for RB convection in
 water were measured and analyzed systematically by
\cite{ahl06}. Of the relevant fluid properties, the kinematic viscosity $\nu$ had the largest 
temperature dependence in this case. These authors compared Nu for three samples with different L,
but the {\it same} Ra, all with $\Gamma = 1$. The same Ra was realized in different ways: 
Either in a cell with larger L but
 $\Delta$ so small that NOB effects should be negligible, 
or in a cell with smaller L but larger $\Delta$ up to 40 K 
that had strong NOB effects. For the largest $\Delta$  the deviation of the center temperature
$T_c$ from the mean temperature $T_m$ was only about 1.8 K (see fig.\ \ref{fig:nob}c), 
corresponding to $\chi \simeq 0.85$. 
\cite{ahl06} calculated the increase of $T_c$ theoretically
by developing an extended Prandtl-Blasius BL theory, which takes the T-dependence 
of $\nu$ and $\kappa$ into account.

Surprisingly, for these water measurements
again the NOB effects on Nu were tiny, see
figure \ref{fig:nob}a: less than 1.4\% even for 
$\Delta = 40K$. To account for this finding, 
\cite{ahl06} derived an extension of the exact 
relation eq.\ (\ref{nu-lambda}) for the Nusselt number
to the NOB case, proving the generally valid
exact relation 
\be 
\label{nu_ratio}
\frac{Nu}{Nu_{OB}} = 
\frac{2 \lambda_{\theta, OB}^{sl}}{\lambda_{\theta ,b}^{sl} 
+ \lambda_{\theta , t}^{sl}}
 \cdot \frac{\kappa_b \Delta_b + \kappa_t \Delta_t}
{\kappa_m \Delta } \equiv F_{\lambda} \cdot F_{\Delta} \ .
\ee
As can be seen, in each of the two factors  $F_{\lambda}$
and  $F_{\Delta}$ the symmetry-breaking different bottom and top BL properties 
enter additively and thus tend to 
compensate each other, leading to a cancellation of the linear contributions to their temperature dependences. Thus the remaining NOB effects on Nu originate only from the quadratic and higher-order variations  of the material properties
with temperature.  

With the help of relation (\ref{nu_ratio}) \cite{ahl06} could identify the origins 
of the NOB corrections for water: For this fluid it is mainly the temperature 
dependence of the thermal diffusivity $\kappa$ that is responsible
for the NOB correction of  Nu, while the NOB correction of the center 
temperature $T_c$ has its main origin
in the temperature dependence of the kinematic viscosity $\nu$.
Instead in glycerol, which displays an extreme dependence of $\nu$ on the temperature, 
the NOB corrections for $T_c$ are much larger,
both in experiment (\cite{zha97}) and in the
two-dimensional numerical simulations of \cite{sug07}.
 
Theoretical calculations of $T_c$ from the extended Prandtl-Blasius BL theory 
of \cite{ahl06}
for various 
water-like fluids with hypothetical temperature dependences of the 
material parameters were in good agreement with the corresponding numerical
results  of \cite{sug07,sug08}.
\cite{ahl07} applied the extended Prandtl-Blasius theory of \cite{ahl06}
to NOB effects in gases, finding good agreement for
$T_c$ with experiments using pressurized ethane gas, 
in which the material properties strongly depend on temperature.
For this case a decrease of $T_c$ relative to $T_m$ and an increase of Nu in comparison 
to the OB case was found. Note that both effects
are in the opposite directions as compared to the NOB effects in water or glycerol.

In experiments by \cite{ahl08} 
the measurements using ethane were extended to the region near the critical point, see fig.~\ref{fig:nob} (b) and (d). 
On first sight the results for $T_c$ seem surprising. NOB effects in liquid-like ethane 
($\rho > \rho_c$ where $\rho_c$ is the critical density) caused an increase of $T_c$ relative to $T_m$, 
whereas in gas-like ethane ($\rho < \rho_c$) NOB effects made $T_c$ smaller than $T_m$.
The physical reason for this qualitative difference was found in the opposite sign of the bottom-top asymmetry of
the buoyancy strength 
due to the opposite temperature dependence of $\beta(T)$ for the two cases. In the liquid-like case the buoyancy, proportional to  $\beta (T) = \beta (T_m) + (T-T_m) \beta '(T_m)$,
is larger at the bottom and smaller at the top, supporting the uprising warmer 
bottom plumes more than the down-coming colder top plumes.
This brings predominantly hotter material into the bulk, leading to $T_c  > T_m$. 
For gas-like ethane the buoyancy is larger at the cooler top, favoring down-going cold 
 over uprising warm plumes. 
This in turn brings more cooler material into the bulk, resulting in $T_c < T_m$. 
It is the sign of the slope $\beta'$ of 
$\beta$ at $T_m$ that is the relevant quantity for this type of NOB correction.

Of course, the extended Prandtl-Blasius BL theory fails when the thermal expansion coefficient 
$\beta$, which is disregarded in that theory by construction as it 
treats the temperature as a passive scalar, shows an extreme temperature dependence 
such as close to a criticial point. The above explanation in terms of the 
strong temperature dependence of $\beta$, which leads to broken buoyancy symmetry, as an additional cause of NOB 
effects on $T_c$ was verified numerically by \cite{ahl08}. 

Even though the two cases $\rho < \rho_c$ and  $\rho > \rho_c$ discussed above had $T_c$ displaced away from $T_m$ in opposite directions, they both yielded an enhancement of Nu above the OB value, i.e., in the same direction
despite opposite deviations of $T_c$ from $T_m$.
This insensitiveness from the sign of $T_c-T_m$ 
 can be 
 understood from Eq.~(\ref{nu_ratio}). 
Physically, the BLs act as two thermal resistances in series, since
Nusselt number deviations only depend on the sum of both, 
and it does not matter much  
which of them  is reduced  and which one is enhanced.

\begin{figure}
\begin{center}
\includegraphics*[height=8cm]{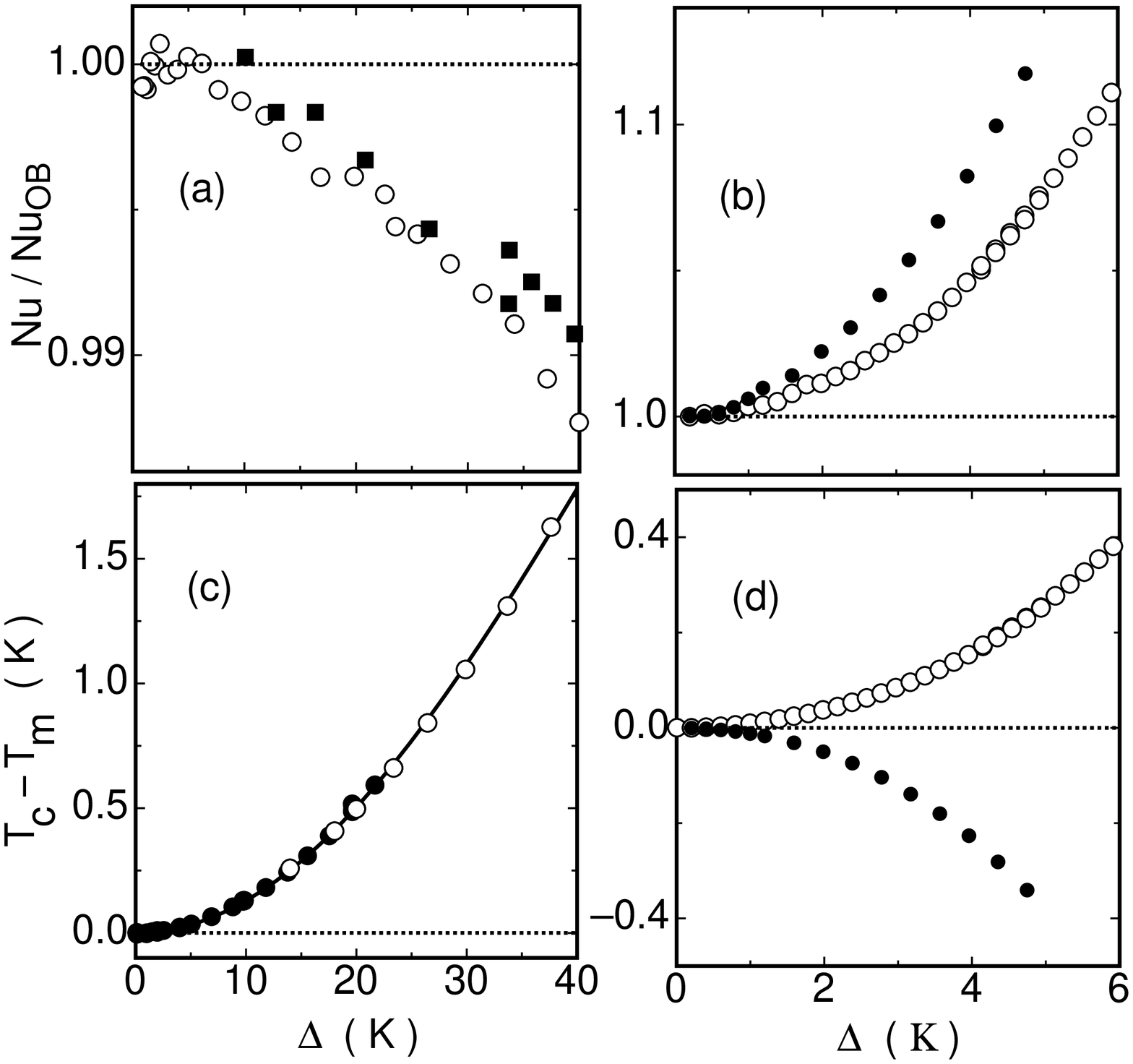}
\caption{ 
(a) and (b): The ratio $Nu/Nu_{OB}$ as a function of $\Delta$ for fixed $T_m$. (c)
and (d): $T_c - T_m$ as a function of $\Delta$, also fixed $T_m$. (a) and (c): Water at $T_m =
40.0^\circ$C, $\Gamma = 1.0$. Solid circles: $L =  49.7$ cm. Open circles:
$L = 24.8$ cm. Solid squares: $L = 9.2$  cm.
(b) and (d): ethane at a pressure of 55.17 bar for $\Gamma = 0.5$. Open
symbols: $T_m = 35.0^\circ$C below the critical isochore temperature $T_\phi = 38.06^\circ$C 
at this pressure. Solid symbols: $T_m = 41.0^\circ$C, above $T_\phi$. 
}
\label{fig:nob} 
\end{center}
\end{figure}

The flow organization due to NOB effects in water 
and other liquids was studied numerically
by \cite{sug08}, again with 2D RB simulations. 
It was confirmed that buoyancy, in particular the temperature 
dependence of the 
thermal expansion coefficient $\beta$, 
is the main origin of the NOB effects on the center-roll Reynolds-number,
which roughly behaves like $Re / Re_{OB} \approx ({\beta (T_c)
/\beta (T_m)})^{1/2}$. Reynolds-number measurements using water (\cite{ahl06}) were 
still unable to resolve NOB effects within their resolution of one or two percent 
even though $\Delta$ was as large as 38 K.

\section{Global wind dynamics}
\label{sec:global}

\subsection{Experiment}

\begin{figure}
\centerline{\epsfig{file=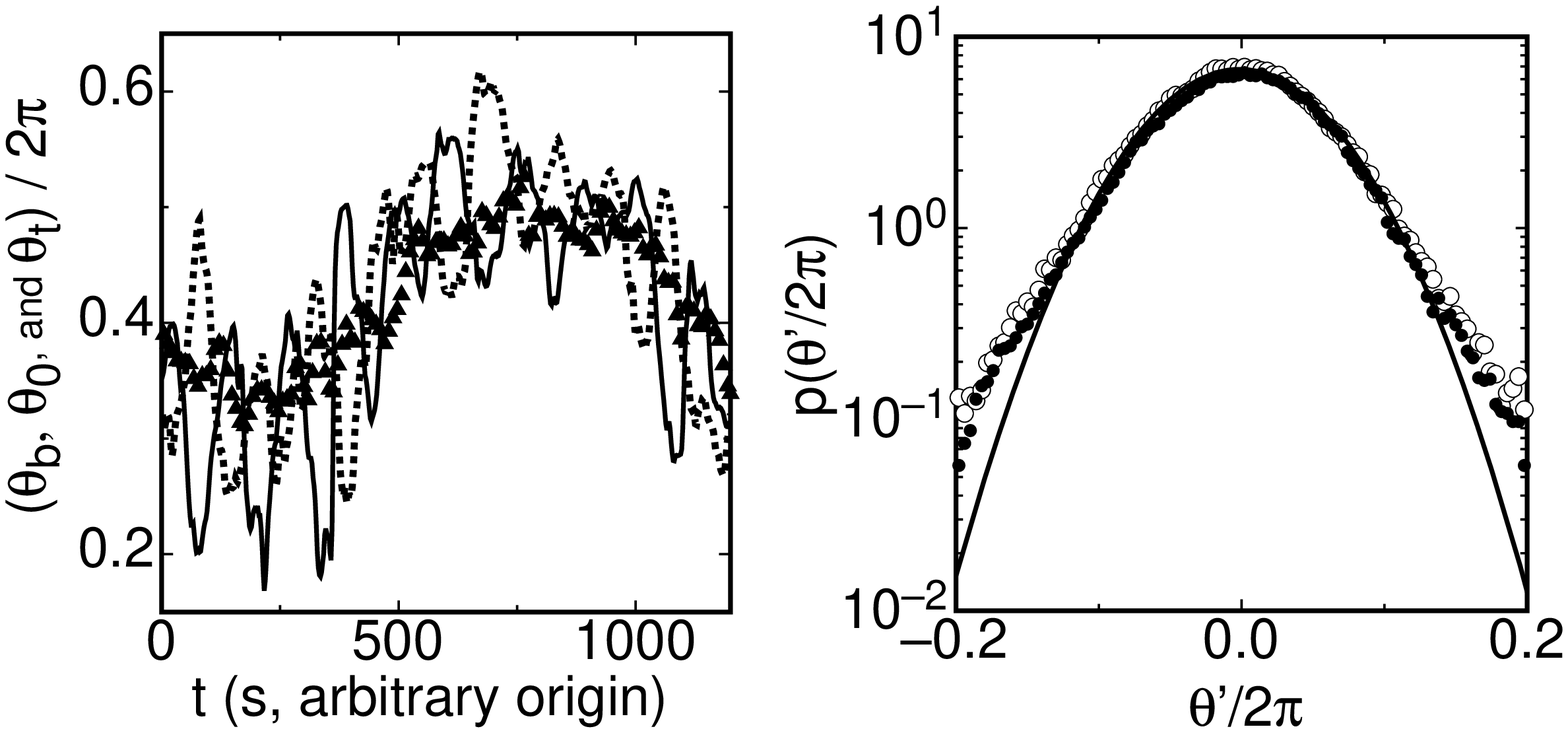,height=4.2cm}
}
\caption{Left: The LSC orientation as a function of time at $z=-L/4$ ($\theta_b$, dotted line), $z = 0$ ($\theta_0$, triangles), and $z=L/4$ ($\theta_t$, solid line). Right: The probability distribution $p(\theta'/2\pi)$ as a function of $\theta'/2\pi$. Here $\theta' = \theta_t - \theta_0$ (open circles) or $\theta' = \theta_b - \theta_0$ (solid circles). Solid line: a Gaussian distribution. Adapted from \cite{FBA08}. 
}
\label{fig:twist} 
\end{figure}                      

For cylindrical samples with $\Gamma = 1$ the LSC circulation plane breaks the rotational invariance of the cell. 
This leads to interesting dynamics which includes oscillations of the circulation plane
(\cite{hes87,cas89,cil96,tak96,cio97,doo00,qiu00,qiu01a,qiu01b,nie01,sre02,qiu02,qiu04,res06,xi06,xi07}). 
These are caused by a torsional mode in which the orientation of the upper half of the LSC undergoes azimuthal oscillations 
(\cite{fun04,res06}) 
that are out of phase with those of the lower half 
(\cite{fun04,FBA08}). 
This is illustrated in the left part of Fig.~\ref{fig:twist}, which shows the LSC azimuthal orientations (characterized by the angle) $\theta_b$ at the vertical position $z = -L/4$, the azimuthal orientation 
$\theta_0$ at $z = 0$, and the azimuthal orientation 
$\theta_t$ at $z = +L/4$ (the origin of the $z$-axis is taken here at the cell center).  These orientations were determined by measuring the azimuthal temperature variation of the side wall at the three vertical positions. Even casual inspection shows that $\theta_t$ and $\theta_b$ oscillate, out of phase with each other and perhaps with random amplitudes, about $\theta_0$. Quantitative analysis using correlation functions confirms this qualitative result.

The origin of this twisting mode is not known. However, important insight is gained from the probability-distribution functions  $p(\theta_{b,t}')$ shown in the right part of Fig.~\ref{fig:twist}. Here $p(\theta_{b,t}')$ is the probability of the azimuthal displacement $\theta_{b,t}' \equiv \theta_{b,t} - \theta_0$, relative to $\theta_0$, of $\theta_b$ or $\theta_t$.  If the mode had its origin in a Hopf bifurcation, then one would expect it to have a characteristic finite amplitude $A$ which would lead to two peaks of $p(\theta_{b,t}')$ at or near $\theta' = \pm A$. In contrast to this, the experimental result for $p(\theta_{b,t}')$ is a near-Gaussian distribution with only one maximum centered at $\theta_{t,b}' = 0$. Such a distribution is characteristic of a damped oscillator driven by a broad-band noise source (see, for instance, 
\cite{git05}). This driving is attributed to the action of the small-scale turbulent fluctuations on the large-scale flow. 

\begin{figure}
\centerline{\epsfig{file=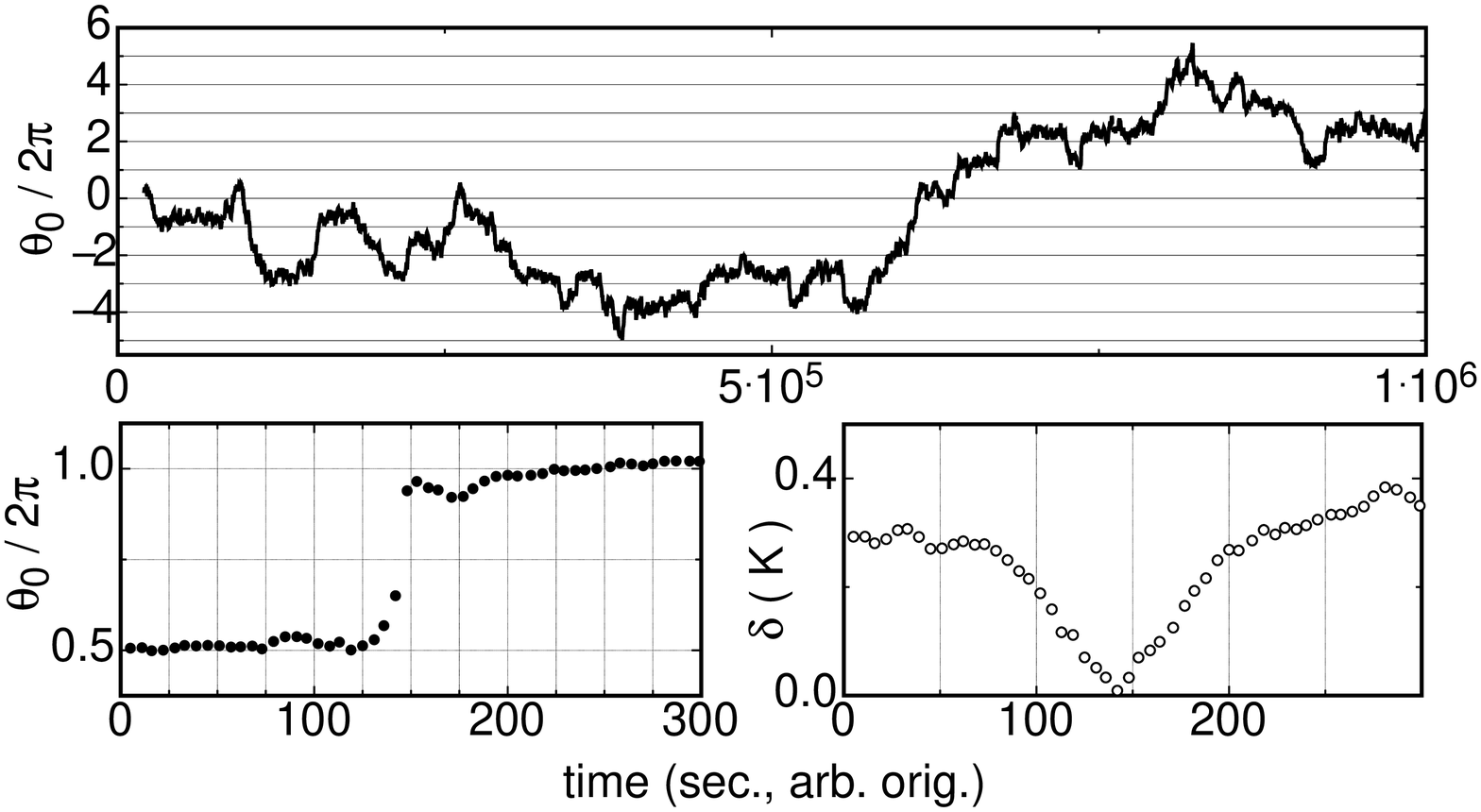,width=9.2cm}}
\caption{Top: A time series of $\theta_0(t)$ spanning about 11 days (adapted from \cite{bro06}). Bottom left: the LSC orientation $\theta_0(t)$, bottom right: the LSC azimuthal temperature amplitude $\delta(t)$ at the side wall, both 
during a cessation (adapted from  \cite{bro05b}).
}
\label{fig:diff_cess} 
\end{figure}                      

The twisting mode does not influence the LSC orientation $\theta_0$ at $z=0$ because there it has a node. However, the LSC breaks the rotational symmetry of a cylindrical sample and its circulation plane must somehow choose an azimuthal orientation.  This orientation has been found to undergo spontaneous diffusive meandering 
(\cite{sun05,xi06,bro06,bro06b}) 
in the sense that its mean square azimuthal displacement is proportional to the elapsed time. This diffusion is illustrated in the top of Fig.~\ref{fig:diff_cess} by a long time series of $\theta_0(t)$. It is also attributed to the action of the small-scale fluctuations on the large-scale flow. Measurements of its diffusivity have yielded results for the intensity of the fluctuating force 
(\cite{bro06,bro06b}).

Very recently \cite{xi08b} and \cite{zho09} discovered a ``sloshing''
 mode in a $\Gamma = 1$ cylindrical sample that occurs in addition to the torsional mode and that was missed by the analysis method employed by \cite{fun04} and \cite{FBA08}. This mode consists of an in-phase horizontal displacement of the entire LSC. The sloshing mode is said to be responsible for the oscillations seen in some of the local temperature measurements that had been attributed by some authors to periodic plume emission from the thermal boundary layers.

The LSC also undergoes re-orientations both by azimuthal {\em rotations} 
(\cite{cio97,bro06}), 
and by {\em cessations} in which the LSC slows to a stop and restarts in a random new orientation (\cite{bro05b,bro06,xi07}). A cessation is illustrated in the lower part of Fig.~\ref{fig:diff_cess}. The left part reveals a sudden change in orientation. The right one shows that the temperature amplitude $\delta$, and thus presumably the LSC itself, vanishes or comes close to zero at the time of the cessation. The angular change $\Delta \theta_c$ during a cessation has a nearly uniform probability distribution, indicating that, once the flow stops, it re-organizes itself in a random new orientation. The angular change $\Delta \theta_r$ of rotations  has a power-law distribution, with small changes more likely  then large ones. The time interval $\tau_1$ between successive events for either cessations or rotations is Poisson distributed, showing again that successive events are uncorrelated. 

On longer time scales, Earth's Coriolis force (at a latitude of 34$^\circ$ in the Northern hemisphere) was found to cause a net clockwise (as seen from above) rotation of the LSC orientation on average once every 3 days, and to align the LSC in a preferred orientation close to West. The  measured probability distribution $p(\theta_0)$ agreed quantitatively with a model calculation of the Coriolis-force interaction that involved no adjustable parameters 
(\cite{bro06b}). The net clockwise rotation also was consistent with this model. 

The LSC in other geometries and aspect ratios adds to the richness of this phenomenon but is beyond the scope of this review. We mention briefly that DNS revealed  two counter-rotating near-circular rolls, stacked vertically one above the other (\cite{ver03}) as the dominating feature of the LSC in cylindrical samples with  $\Gamma \alt 0.5$. Recent experimental evidence 
provided by \cite{xi08} for $\Gamma = 0.5$ and 0.33 indicates random temporal successions of one-roll and two-roll states, with the two-roll state becoming more prevalent as $\Gamma$ is decreased.
\cite{xi08c} found a strong aspect ratio dependence 
when comparing 
the azimutal motion, reorientation, cessation, and reversal 
of the large-scale circulation 
in cylindrical RB samples with 
$\Gamma = 2.3$, 1, and 0.5. 
In samples of square or rectangular cross section the LSC is locked in a predominant orientation along a diagonal (\cite{day01}). 
This is now understood in terms of the pressure gradients that arise when the rotational invariance of a cylindrical  sample is broken (\cite{BA08d}).
\cite{zho07b} studied temperature and velocity oscillations in 
a rectangular cell, too, finding that the temperature oscillations scale
differently with Ra than the velocity oscillations, which are affected
by the cell geometry. 
Of great interest, but largely unexplored, is the flow structure that will be found in large-$\Gamma$ systems where numerous convection rolls potentially can co-exist next to each other. Recent papers addressing this issue numerically include that by \cite{har05}. Recent experimental work going in this direction includes the measurements by \cite{nie06}.  
 

\subsection{Models}

Stochastic models of flow reversal have been proposed by \cite{sre02}, and by \cite{ben05}. They treated  diffusion of the LSC strength in a potential well, but there was no physical motivation for the shape of the potential that was used  and the model parameters were chosen phenomenologically. They also did not address the azimuthal dynamics of the LSC. 
Two other models describe the LSC with deterministic ordinary differential equations that have  chaotic  solutions (\cite{fon05,res06}). They were derived by retaining some arguably relevant aspects of the Navier-Stokes equations and making various approximations. The model of \cite{fon05} 
is based on assumptions about the life times $\tau_p$ of plumes. It is estimated that plumes with a sufficiently large $\tau_p$ will be carried over by the LSC  to the far side where their buoyancy tends to act {\it against} the prevailing flow. This physical process can lead to cessations. Since this model does not contain an azimuthal mode, it can not describe the rich azimuthal dynamics of the physical system. 

Recently a model consisting of two stochastic ordinary differential equations, one each for the circulation strength $U$ and the  azimuthal circulation-plane  orientation $\theta_0$,  was developed by \cite{bro07,BA08a}. 
Starting with the Navier-Stokes (NS) equation, it is argued that $U$ is driven by the buoyancy term and hindered by the dissipation in the viscous boundary layers near the walls. A phenomenological stochastic driving term representing the interaction between the small-scale turbulent fluctuations and the large-scale circulation is added to the deterministic model. For $\theta_0$ the only driving is the turbulent fluctuations, but the component of the nonlinear term in the NS  equation that describes the rotational inertia of the LSC provides damping and couples the two equations. Assuming that the experimentally accessible azimuthal temperature amplitude $\delta$ is proportional to $U$ near the side wall, the model becomes
\begin{equation}
\dot\delta = \frac{\delta}{\tau_{\delta}} - \frac{\delta^{3/2}}{ \tau_{\delta}\sqrt{\delta_0}} + f_{\delta}(t)\ , \qquad
\label{eq:lang_delta}
\ddot\theta_0 = - \frac{\dot\theta_0\delta}{\tau_{\dot\theta}\delta_0} + f_{\dot\theta}(t)
\label{eq:lang_theta}
\end{equation}
with the coefficients
\begin{equation}
\delta_0 \equiv \frac{18\pi \Delta  \sigma \Re^{3/2}}{R} \  ,\ \ 
\tau_{\delta} \equiv \frac{L^2}{18\nu \Re^{1/2}} \ ,\ \ 
\tau_{\dot\theta} \equiv \frac{L^2}{2\nu \Re}\ .\\
\label{eq:tau_theta}
\end{equation}
The intensities of the stochastic forces $f_{\delta}(t)$ and $f_{\dot\theta}(t)$ are obtained from model-independent  measurements of the diffusivities of $\delta$ and $\dot \theta_0$.  

The potential corresponding to the $\dot \delta$ equation  
has a $\delta^{5/2}$ rather than the usual quadratic nonlinearity. However, this does not change its qualitative structure, which has an unstable fixed point at $\delta = 0$ and a stable one at $\delta = \delta_0$. The driving will cause diffusion in the vicinity of the stable fixed point, with occasional excursions to $\delta = 0$ corresponding to cessations. During a cessation (when $\delta$ is small) the damping term on the rhs of 
the $\ddot \theta_0$ equation, and thus the angular momentum of the LSC, become small, and it becomes easy for the stochastic driving to cause relatively large angular changes $\Delta \theta_c$. The model agrees well with many experimental results for the LSC dynamics (\cite{BA08a}), including a time interval between  cessations of 1 to 2 days, a near-uniform  distribution $p(\Delta \theta_c)$ for cessations, and the  dependence of $\delta_0$ on $\Re$ and $\Ra$ expressed by the relation $(\delta_0/\Delta ) \times (\Ra/\Pr) \propto \Re^{3/2}$. There is, however, a significant disagreement between the model and the measured {\it tail} of the probability distribution $p(\delta)$ at small $\delta$, for $\delta \alt \delta_0/2$. It has  been suggested (\cite{BA08a}) that this is caused by the neglect of thermal conduction across the thermal BLs which 
is expected to become important when $\delta$ is small.

Very recently the above model was extended by \cite{BA08d}
by including various perturbations that break the rotational invariance of the sample. These include (a) the effect of Earth's Coriolis force, (b) an elliptic rather than circular horizontal cross section of the sample, (c) a tilt of the sample axis relative to gravity, and (d) a small horizontal temperature gradient at the top or bottom  plate. It turns out that (a) and (b) only influence the $\ddot \theta_0$ equation  and  not the $\dot \delta$ equation. Perturbations of this type, although they introduce a preferred orientation of the LSC, leave the Reynolds number  and the frequency of cessations largely unchanged. However, perturbations like (c) and (d) which affect $\dot \delta$ influence Re and suppress cessations. A tilt or elliptic eccentricity of sufficient magnitude creates a new oscillatory mode of the LSC which is different from the torsional oscillation in that the phase of the oscillation is uniform along the height of the sample. For the tilted sample this mode was found in recent experiments (\cite{BA08d}).

\section{Issues for future research}\label{future}

As seen in this review,  since Siggia's 1994 article in Annual Reviews of Fluid Mechanics  tremendous progress of the understanding of the turbulent Rayleigh-B\'enard system has been achieved by experiment, theory, and numerical simulation. However, it has also become clear that our
 understanding is far from complete. In the following we summarize what we consider as  major issues for future research.

Presumably the most important  challenge is  to clarify whether and, if so, where in parameter space the ultimate state of convection exists. Estimates suggest that  (for Pr near 1) a  transition to such an ultimate state should 
occur at  Rayleigh numbers around $10^{13}\ {\rm to}\ 10^{14}$. Beyond that transition the Nusselt number should increase more rapidly with Ra than below it.  Though the Grenoble experiments suggest such a transition near $Ra = 10^{11}$, neither the Oregon/Trieste experiments nor numerical simulations do so. The reason for the discrepancy is unresolved at present.  It is of utmost importance to clarify this issue in order to allow extrapolations of 
the heat transfer to the very large Ra regime of geo-physical and astro-physical interest. Perhaps related to this issue is whether a coherent LSC continues to exist at very large Ra, or whether it is totally overwhelmed by fluctuations.
Several experimental efforts are planned to try to answer these question.

Rather than focusing on global quantities such as the heat flux,
which may be very difficult to control in the large-scale setups necessary
in an  ultimate regime, a promising complementary 
strategy may be to focus on a detailed analysis of the top and bottom
BLs, whose structure should reflect such a transition. Here the challenge
is the opposite: Extremely small structures must be spatially resolved. Even for
an about
 seven-meter high sample (such as the Barrel of Ilmenau) the 
thermal boundary layer is 
only about 1mm thick when $\Ra = 10^{14}$. Helpful strategies may be to try to trigger the 
transition in the BL, by controlled roughness or even by moving parts.

Another 
major question to be clarified is the role of the plate 
corrections: Though major efforts have been undertaken to correct for the
finite plate heat-conductivities, beyond $\Ra = 10^9$ 
there still seems to be a discrepancy of 20\% or more
 between experimental measurements of Nu and numerical 
data for constant plate-temperature boundary-conditions, 
see figure \ref{fig:Nred_of_largeR}. 
In fact, this question may even be related to above one on the existence
of the ultimate regime. 

Also the three-dimensional 
dynamics of the large scale convection roll will need 
further attention and analysis. 
As we have seen in section \ref{sec:global}, it is rather rich,
including torsional modes, rotation, cessation, and sloshing. 
Here the key question is: 
What features of the LSC can be described through a 
deterministic model somehow based on the Oberbeck-Boussinesq dynamics or
some force balance and what features need stochastic elements for a
description? Routes for further research on this subject will include 
experiments and numerical simulations with modified geometries 
as e.g.\ cylindrical samples with a horizontal axis.

Another issue for further
 research is the exploration of the aspect-ratio dependence of the flow.
Though the $\Gamma$-dependence of Nu is found experimentally 
to be remarkably weak, a full understanding of the RB system
requires further theoretical efforts also in that direction.
For the characterization of the 
 corresponding flow organizations 
detailed PIV measurements could play a crucial role.
These will  also shed light
on the Ra and Pr dependences of the various Reynolds numbers that one can define,
and their connections. 

From the theoretical viewpoint, the GL theory has provided a good guideline
for the understanding of Nu(Ra,Pr) and Re(Ra,Pr) and even allowed 
various predictions, but clearly the theory has its limitations:
\begin{itemize}
\item 
It builds on the Prandtl-Blasius BL theory for the
temperature as a  {\it passive} scalar.
The buoyancy term  
is skipped by construction in
the Prandtl BL equations, but obviously it is of high 
importance for thermally driven flow, leading to plume detachment 
from the thermal BLs.
This detachment mechanism  as a time dependent BL separation
    process needs further study and analysis of its parameter dependences.
The GL theory only represents the global effect of the plumes and
the resulting self-organizing
buoyancy-driven flow,
namely the large scale wind.
\item The three-dimensional (torsional) dynamics of the large-scale wind 
is not taken into consideration
 by the GL theory (or any other model).
 Although it was established experimentally that this oscillation 
is a stochastically  driven mode, its origin is presently unknown.
\item Experiments suggest that Nu and Re are not as intimately
related as the relations (\ref{eq13}) and (\ref{eq14}) of the GL theory
suggest. For instance, by tilting the system Re can be enhanced considerably 
whereas Nu changes very little. 
\item 
Figure \ref{fig:Nred_of_R} implies that there are 
more sudden transitions between the various regimes 
in reality than within the GL theory.
\item 
Finally, in the $\Nu \propto \Ra^{1/3}$ regime the 
experimentally found co-existence of a
  measured Reynolds number $ \propto Ra^{1/2}$, 
achieved in
  Santa Barbara, and the GL prediction $\Re\propto 
\Ra^{4/9}$ for the global
wind 
   has to be clarified and resolved.
\end{itemize}
More theoretical work is necessary to improve on the GL theory. 
This unfortunately may only 
be possible by sacrificing  its conceptual appeal and simplicity.

An important further extension of the present theoretical understanding
of RB convection, but of course also of the numerical simulations and the
experiments, is the extension towards {\it rotating} RB convection.
With rotation, the (Ra,Pr)-parameter-space is extended to 
(Ra,Pr,Ro), where the Rossby number Ro is defined as $\Ro = 
(2\Omega)^{-1} \sqrt{\beta g \Delta /L}$, i.e., as ratio of the
time scale given by the rotation rate $\Omega$ and the one
given by buoyancy. The obvious questions to address
are: What is the dependence of the Nusselt number on the control parameters
Ra, Pr, and Ro, i.e., what is Nu(Ra,Pr,Ro)? And how do the large
scale convection roll and the 
Reynolds number react to the rotation, i.e., what is Re(Ra,Pr,Ro)?
When will the large scale convection roll break down? 
How are the top, bottom, and side-wall kinetic and thermal 
BLs modified through the rotation? Only a small fraction of the
parameter space (Ra,Pr,Ro) has hitherto been explored.  
Given that rotating turbulence is known for its
counter-intuitive features (e.g.\ the Taylor-Proudman effect), 
we expect many surprising results ahead of us. We do not
want to give an extensive literature review on rotating RB here,
but good starting points are the book by \cite{gre90} or the classical
article by \cite{ros69} himself.

Future developments undoubtedly will also include the extension of  
the current state of the art of RB convection  
to complex fluids, to fluids containing bubbles or suspensions of particles, 
to fluids undergoing phase transitions, 
and so on.
But clearly these topics are beyond the scope of the present review.

Fifteen years after Siggia's Annual Review of Fluid Dynamics article on RB convection,
and inspite of the huge progress achieved during this time,
we still close with a similar statement as he did: The current rate 
of experimental and numerical advances will again soon antiquate this summary. 

\vspace{0.5cm}

\noindent
{\it Acknowledgement:} We would like to thank all our coworkers and colleagues
for their contributions to our understanding 
of this great problem, for the many stimulating 
discussions we had the priviledge to
enjoy over  the years, and for their many valuable comments on this 
manuscript.  -- The work of G.A.\ was supported in part by  US National Science Foundation Grant DMR07-02111 and by the Alexander von Humboldt Stiftung.
D.L.\ acknowledges FOM for the continuous support.


\end{document}